\documentclass[11pt,a4paper]{article}

\usepackage{jheppub}

\usepackage[utf8]{inputenc}
\usepackage[T1]{fontenc}
\usepackage[english]{babel}
\usepackage{microtype}

\usepackage{mathtools}
\usepackage{physics}
\usepackage{bbold}
\usepackage[scr=boondox]{mathalfa}

\usepackage{subcaption}

\usepackage{tikz}
\usetikzlibrary{
    shapes,
    decorations,
    angles,
    patterns,
    calc,
    arrows.meta,
    positioning,
    cd
}

\usepackage{array}
\usepackage{tabularx}

\usepackage[dvipsnames]{xcolor}

\usepackage{cleveref}

\newcommand{\q}{\mathrm{q}}
\newcommand{\qbar}{\overline{\q}}

\renewcommand{\dd}{\mathrm{d}}

\newcommand{\e}[1]{\mathrm{e}^{#1}}

\newcommand{\Z}{\mathbb{Z}}
\newcommand{\T}{\mathbb{T}}
\newcommand{\Un}{\mathbb{1}}

\newcommand{\Tp}{\ensuremath{\mathrm{T}_{79}}}
\newcommand{\Tt}{\ensuremath{\mathrm{T}_{6789}}}

\title{D-branes and orientifolds on quasicrystalline orbifold backgrounds}

\author[a]{Antony Wendels}

\affiliation[a]{CPHT, CNRS, École polytechnique, Institut Polytechnique de Paris,\\
Palaiseau, France}

\emailAdd{antony.wendels@polytechnique.edu}

\abstract{
We construct and analyse orientifold descendants of a six-dimensional type~IIB $\mathbb Z_{12}$ quasicrystalline orbifold. The parent background is defined by a Narain-lattice automorphism that is not crystallographic on the geometric torus. We determine the closed-string spectrum and twisted-sector orbit structure, construct a compatible unoriented projection, and compute the Klein bottle, annulus and Möbius strip amplitudes. Tadpole cancellation selects boundary orbits that admit a partially T-dual description as D7-branes at angles and fixes the Chan-Paton gauge group and massless spectrum. The resulting supersymmetric six-dimensional vacuum satisfies the irreducible anomaly constraints, its reducible anomaly polynomial factorizes through the Green-Schwarz mechanism, and its anomaly lattice obeys the Kim-Shiu-Vafa string-probe bound. We also construct a Brane Supersymmetry Breaking descendant with a perturbatively tachyon-free rigid branch, and a freely acting five-dimensional orientifold whose two decompactification limits reproduce a geometric type~IIA~$\mathbb Z_3$ orientifold and an antisymmetric type~IIB~$\mathbb Z_4$ orientifold. These examples show that quasicrystalline closed-string backgrounds can admit controlled perturbative open-string completions with non-trivial boundary and crosscap sectors.
}


\begin{document}

\maketitle

\section{Introduction}

String compactifications on Calabi-Yau manifolds, orbifolds and orientifolds have long provided a fruitful laboratory for investigating the landscape of consistent vacua~\cite{Candelas:1985en}. Calabi-Yau compactifications furnish a broad and geometrically rich class of supersymmetric models, while orbifolds~\cite{Dixon:1985jw,Dixon:1986jc,Giaccari_2023} provide exactly solvable limits in which many features of the spectrum and interactions can be analysed explicitly. Orientifolds, in turn, enlarge this framework by combining closed-string backgrounds with unoriented sectors, D-branes and O-planes, thereby giving access to vacua with gauge sectors localised on branes and to intrinsically open-string phenomena. At the same time, asymmetric orbifolds~\cite{Narain:1986qm,Narain:1990mw} show that string theory also admits genuinely non-geometric phases, in which the worldsheet construction remains perfectly consistent even though no conventional target-space description exists. Such non-geometric constructions have repeatedly proven useful for uncovering corners of the string landscape that remain invisible from a purely geometric viewpoint. Among them, quasicrystalline orbifolds occupy a particularly interesting place.

Quasicrystalline compactifications were introduced in~\cite{Harvey:1987da} and have recently been systematically revisited in~\cite{Baykara:2024vss}. Recent applications of these constructions include new non-supersymmetric tachyon-free string vacua~\cite{Baykara:2024tjr}. The relevant symmetry groups of $\T^4$ sigma models preserving the $\mathcal{N}=(4,4)$ algebra were classified in~\cite{Volpato_2014}, and the corresponding six-dimensional type II string islands, including the $\Z_{12}$ model considered below, were further analysed in~\cite{baykara2026stringislandsdiscretetheta}.
Their defining feature is that the orbifold action is induced by a symmetry of the Narain lattice which does not descend from a symmetry of the underlying bulk torus. As a result, these vacua do not admit an ordinary geometric interpretation. Moreover, since such symmetries exist only at special points in moduli space, quasicrystalline compactifications typically come with only a few geometric moduli.

Another important corner of the string landscape is formed by orientifold vacua. These constructions involve the worldsheet parity operator and require the introduction of new extended objects, namely orientifold planes and D-branes. They are by now well understood in the case of geometric orbifolds, see for instance~\cite{Sagnotti:1987tw,PradisiSagnotti:1989,BianchiSagnotti:1990,Angelantonj:2002ct} and references therein, and have also been investigated for asymmetric orbifolds~\cite{Bianchi_2000,Angelantonj:2000xf,Blumenhagen:1999md,Blumenhagen:2000fp}. In these models the consistency problem is richer than in the purely closed-string case, since one must simultaneously control the Klein bottle, annulus and Möbius strip amplitudes, the associated tadpole conditions, and the cancellation of low-energy anomalies. These constructions give rise to genuine gauge sectors localised on space-time-filling branes, similarly to heterotic string vacua. 
Orientifolds also provide a natural framework for supersymmetry breaking in open-string sectors. A particularly important mechanism is Brane Supersymmetry Breaking (BSB), in which R-R tadpole cancellation requires branes whose supersymmetry is incompatible with that preserved by the orientifold planes. Supersymmetry then remains linearly realised in the closed-string sector but is broken on the branes. The ten-dimensional $USp(32)$ model of Sugimoto~\cite{Sugimoto:1999tx} provides the canonical example, while compact realisations and generalisations were subsequently constructed in lower dimensions~\cite{Antoniadis:1999xk,Angelantonj:1999ms,AldazabalUranga:1999,mourad2017updatebranesupersymmetrybreaking,dudas2026supersymmetrybreakingfieldsstrings}.

Compared with heterotic and type II closed-string constructions, where modular invariance strongly constrains the form of the one-loop partition function, orientifold vacua require an additional dynamical analysis. In practice, the consistency of the model is controlled by tadpole cancellation, anomaly cancellation~\cite{Gimon_1996,Aldazabal_1999,Bianchi:1999}, and, more generally, by the consistency of the open-string sectors, including probe-brane constraints and the unitarity of the corresponding worldvolume CFT~\cite{Kim_2019,Uranga_2001,Angelantonj:2020pyr,leone2024newcommentssixdimensionalorientifold}. For this reason, extending a given asymmetric closed-string background to an orientifold vacuum is a non-trivial problem.

A natural question is therefore whether the quasicrystalline sector also admits a consistent orientifold completion. This issue is particularly subtle in six dimensions. In known orientifolds of antisymmetric but crystallographic orbifolds, the orientifold projection can produce non-vanishing twisted contributions to the Klein bottle amplitude and therefore requires non-standard (magnetised) open-string sectors in order to cancel the resulting tadpoles~\cite{Blumenhagen:1999md, Bianchi_2000, Angelantonj:2000xf}. The T-dual description in terms of magnetised type-I branes is discussed in \cite{Blumenhagen:2000wh}. This already shows that the passage from a closed-string asymmetric orbifold to an orientifold background is not a formal step, but rather a genuinely dynamical problem.

In this paper we address this question for a concrete six-dimensional quasicrystalline orbifold background. We start from the type IIB quasicrystalline $\Z_{12}$ asymmetric orbifold whose generator is an automorphism of the Narain lattice but is not crystallographic on either the geometric torus or its dual. We construct an unoriented descendant, compute the Klein bottle amplitude, identify the crosscap tadpoles, and determine the open-string sectors required for their cancellation. The relevant boundary states admit a useful T-dual interpretation in terms of $D7$-branes at angles. Combining the Klein bottle, annulus and Möbius amplitudes, we derive the tadpole conditions, the resulting Chan-Paton gauge group, the massless spectrum and the factorized six-dimensional anomaly polynomial. We also construct non-supersymmetric vacua and a freely acting quasicrystalline orientifold. In this way, we provide an orientifold realisation of a six-dimensional quasicrystalline orbifold background.

Beyond the construction itself, these models are of interest for two related reasons. First, they enlarge the set of explicit six-dimensional orientifold vacua in a direction that is already genuinely non-geometric at the level of the parent closed-string background. Second, they provide a concrete setting in which techniques familiar from orientifolds can be brought to bear on quasicrystalline compactifications, thereby linking two lines of development that have so far remained essentially separate. More broadly, they offer a new example of how the non-geometric sector of string theory refines our understanding of the space of consistent vacua.

The paper is organised as follows. In \Cref{sec:orbifold-conventions} we introduce the Narain-lattice, orbifold and orientifold conventions used throughout the paper. In \Cref{sec:Z12-parent} we construct the six-dimensional type IIB $\Z_{12}$ quasicrystalline parent theory, determine the orbit structure of its twisted sectors and extract its closed-string spectrum. In \Cref{sec:probe-branes} we classify the boundary orbits relevant for the subsequent orientifold construction and derive their annulus amplitudes, emphasising the partially T-dual description in terms of branes at angles. In \Cref{sec:orientifold-descendants} we construct the unoriented theory, compute the Klein bottle, annulus and Möbius strip amplitudes, solve the tadpole conditions and determine the supersymmetric open-string spectrum and its consistency conditions. The same section also presents the Brane Supersymmetry Breaking descendant and the freely acting five-dimensional orientifold. Technical conventions, additional probe sectors and auxiliary amplitudes are collected in the appendices. We summarise the physical implications and open questions in \Cref{sec:discussion-conclusions}. Unless otherwise stated, we work in units in which $\alpha'=1$.

\section{Orbifold and orientifold conventions}\label{sec:orbifold-conventions}

We first recall the Narain description of toroidal compactifications. A type II compactification on $\T^4$ is specified by an even self-dual lattice $\Gamma_{4,4}$ of left- and right-moving momenta~\cite{Narain:1985jj,Narain:1986am}. Equivalently, it is described by a metric $G$ and a Kalb-Ramond background $B$ on the torus. The T-duality group acts as automorphisms of the abstract Narain lattice, while only those automorphisms preserving the chosen left-right splitting define genuine symmetries of the worldsheet CFT. The symmetry groups of $\T^4$ sigma models preserving the $\mathcal N=(4,4)$ algebra were classified in~\cite{Volpato_2014}. Orbifolding by such a symmetry gives an exact CFT, even when the symmetry is not induced by an ordinary geometric action on the torus~\cite{Narain:1986qm,Narain:1990mw}.

Orbifold compactifications provide one of the simplest exact worldsheet descriptions of string vacua beyond flat ten-dimensional space-time. In six dimensions they are especially useful, since they furnish solvable limits of $K3$ compactifications and allow one to keep full control of the twisted sectors, of the multiplicities associated to fixed points, and of the one-loop partition functions. For the present work they also play a second role: they provide the closed-string starting point from which the corresponding orientifold models will later be constructed.

In order to discuss orientifold vacua on quasicrystalline backgrounds, it is useful to begin with the general orbifold construction in six dimensions, independently of whether the orbifold action admits a geometric interpretation on the bulk torus. The geometric $\T^4/\Z_N$ models provide the standard reference point, while asymmetric and quasicrystalline orbifolds arise by relaxing the requirement that the left- and right-moving sectors be acted on in the same way. In this sense, the geometric orbifolds reviewed below should be viewed as the simplest representatives of a broader class of exactly solvable worldsheet backgrounds.

Let $g$ generate a cyclic group $\Z_N$. On the worldsheet, orbifolding amounts to enlarging the closed-string Hilbert space by sectors twisted by powers of $g$, and then projecting onto $\Z_N$-invariant states. If $\mathcal H_\alpha$ denotes the sector twisted by $g^\alpha$, the fields satisfy~\cite{Dixon:1985jw,Dixon:1986jc}
\begin{equation}
X^I(\tau,\sigma+2\pi)=g^\alpha \cdot X^I(\tau,\sigma)\,,
\end{equation}
while the orbifold projection keeps only the states invariant under the lifted action $\widehat g$ on the CFT Hilbert space. This construction applies equally well to geometric orbifolds, for which the action is left-right symmetric, and to asymmetric orbifolds, where the left and right movers are twisted differently~\cite{Narain:1986qm,Narain:1990mw}. In the latter case the orbifold generally has no conventional target-space interpretation, although it still defines a perfectly consistent worldsheet theory provided modular invariance and level matching are satisfied. 

For a general left-right action
\begin{equation}
g=(\theta_L,v_L;\theta_R,v_R)\,,
\end{equation}
the rotation matrices $\theta_{L,R}$ are conveniently encoded in twist vectors $\phi_{L,R}$, while $v_{L,R}$ denote possible shifts on the Narain lattice. In the following, we will not consider the possible shifts and will focus on orbifolds with pure rotational action and no additional shifts. A freely acting shift will be introduced separately in the five-dimensional extension of \Cref{sec:orientifold-descendants}. This general framework will be the one relevant for quasicrystalline orbifolds, where the symmetry acts on the Narain lattice but does not descend from an ordinary symmetry of the torus itself. 

We now specialise to orbifolds of type IIB on $\T^4$ where we write $\T^4 = \T^2_1 \times \T^2_2$ with complex coordinates
\begin{equation}
Z^1=X^1+U_1X^2\,, \qquad Z^2=X^3+U_2X^4\,,
\end{equation}
where $U_i$ denotes the complex structure of the $i$-th two-torus. One can also turn on a Kalb-Ramond background field on the torus. Then there is a non-trivial Kähler structure $T_i$ on each torus where $\Re(T_i)=B_i$ denotes the corresponding Kalb-Ramond field on the $i$-th two-torus. 

The $\Z_N$ orbifold action is generated by
\begin{align}
\theta_L &: (Z^1_L,Z^2_L)\ \longmapsto\ \bigl(e^{2i\pi \phi_{L} /N} Z^1_L, e^{-2i\pi \phi_{L} /N} Z^2_L\bigr)\,,\\
\theta_R &: (Z^1_R,Z^2_R)\ \longmapsto\ \bigl(e^{2i\pi \phi_{R} /N} Z^1_R, e^{-2i\pi \phi_{R} /N} Z^2_R\bigr)\,.
\label{eq:generalZN_action}
\end{align}
The opposite phases in the two complex planes guarantee that half of the type IIB supersymmetries are preserved.
Geometric, or symmetric, orbifolds are generated by rotations that act crystallographically on the torus lattice and identically on left- and right-moving coordinates, i.e. $\phi_L=\phi_R$. In six dimensions, they are the familiar $\T^4/\Z_N$ examples with $N=2,3,4,6$ describing singular limits of K3 compactifications. In these models the crystallographic condition constrains the complex structure of the torus, while the Kähler moduli remain continuous parameters. Asymmetric orbifolds~\cite{Narain:1986qm,Narain:1990mw} are all the other cases where $\phi_L\not=\phi_R$, but may still define exact orbifold CFTs since the action is a symmetry of the Narain lattice. Among them, a special subclass is formed by antisymmetric orbifolds, for which $\phi_L = - \phi_R$. In this case the orbifold action is generated by rotations that act crystallographically on the dual torus lattice and on the Narain lattice, but not on the torus lattice itself. They can be interpreted as T-duals of geometric orbifolds~\cite{Bianchi_2000,AngelantonjBlumenhagen:1999,Blumenhagen:1999md,Blumenhagen:2000fp}, and therefore admit a geometric interpretation in terms of singular limits of K3 surfaces. In these models the crystallographic condition constrains the Kähler structure of the torus, while the complex structure remains a continuous parameter. This should be distinguished from genuinely quasicrystalline orbifolds~\cite{Harvey:1987da,Volpato_2014,Baykara:2024tjr,Baykara:2024vss,baykara2026stringislandsdiscretetheta}, whose generators act crystallographically on the full Narain lattice but not on either the torus lattice or the T-dual torus lattice separately. Such models have no conventional geometric interpretation in terms of the bulk torus in any of these simple duality frames, and typically retain only a small number of geometric moduli since both complex and Kähler structures are constrained.

At the level of closed strings, the oriented type IIB spectrum on these orbifolds is universal: one obtains the six-dimensional $\mathcal{N}=(2,0)$ supergravity multiplet together with 21 tensor multiplets, distributed differently between the untwisted and twisted sectors depending on $N$. For instance, the $\Z_2$ model has 16 fixed points, the $\Z_3$ model 9, while the $\Z_4$ and $\Z_6$ cases contain both genuine $\Z_4$ or $\Z_6$ fixed points and orbits associated to the $\Z_2$ or $\Z_3$ subgroups. This fixed-point structure controls the multiplicities of the twisted Hilbert spaces and will later play a central role in the orientifold analysis. 

The untwisted sector contains the usual Narain zero modes, whereas twisted sectors have fractional oscillator moding and a finite degeneracy of twisted ground states. For geometric subgroups these ground states can be associated with fixed loci on the torus or on an appropriate T-dual torus. The corresponding twisted sectors are localised at the orbifold singularities and carry the data associated to the corresponding collapsed cycles in the $K3$ resolution. For genuinely quasicrystalline sectors, however, the degeneracy should be understood intrinsically at the level of the Narain CFT and need not admit a geometric fixed-point interpretation.

The one-loop amplitudes are of particular interest in order to study the massless spectrum of the theory. They are conveniently written in terms of the standard $\Z_N$ blocks. For the left-moving sector, in the $\alpha$-twisted sector with insertion $g^\beta$, with~$\alpha,\beta=0,\ldots,N-1$,
\begin{align}\label{eq:ZN-blocks}
    T[^\alpha_\beta] =& \Tr_{\mathcal{H}_\alpha\, L}(P_{GSO}\theta_L^\beta q^{L_0-c/24}) \\
    =& \sum_{a,b=0,\frac12} \frac{s_{a,b}}{2} \left[\frac{1}{\eta^{4}}\right] \left[\frac{d_{\alpha,\beta}\,\eta^2}{\theta[^{1/2+\alpha/N}_{1/2+\beta/N}]\theta[^{1/2-\alpha/N}_{1/2-\beta/N}]}\right] \left[\frac{\theta^2[^a_b]}{\eta^2}\right] \left[\frac{\theta[^{a+\alpha/N}_{b+\beta/N}]\theta[^{a-\alpha/N}_{b-\beta/N}]}{\eta^2}\right]\,. \notag
\end{align}
where $s_{a,b}=(-1)^{2a+2b+4ab}$ encodes the GSO projection for type IIB, while the normalisation factor is \begin{equation}
    d_{\alpha,\beta} = 
    \begin{cases}
        -4 \sin(\pi \beta/N)^2 & \text{ if } \alpha=0 \text{ and } \beta \not= 0\,,\\
        1 & \text{ otherwise.}
    \end{cases}
\end{equation}
In the expression of~\Cref{eq:ZN-blocks}, computed in the light-cone gauge, the first square bracket contains the contribution of the four space-time bosons, the second one the contribution of the non-zero modes of the internal bosons, the third one the contribution of the four space-time worldsheet fermions and the last one the contribution of the internal worldsheet fermions. The right-moving blocks are defined similarly.
It is often more convenient to diagonalise the orbifold action and work with the eigencharacters $\chi[^\alpha_\beta]$, defined through
\begin{equation}
T[^\alpha_\beta]=\sum_{\beta'=0}^{N-1}
e^{2\pi i \beta\beta'/N}\,\chi[^\alpha_{\beta'}]\,.
\end{equation}
This is the notation that will be used throughout the paper, since it is the most convenient one both for reading the spectrum and for passing to the unoriented descendants. The massless states contained in these characters transform in the standard bulk $SU(2)_L\times SU(2)_R$ representations as
\begin{align}
    \chi[^0_0]&\sim (2,2)\oplus 2(2,1)\,, \\
    \chi[^0_1]\sim \chi[^0_{N-1}]\sim \chi[^{\alpha \not = 0}_0]&\sim 2 (1,1) \oplus (1,2)\,.
\end{align}

In the oriented closed-string theory, the perturbative expansion is organised by closed orientable Riemann surfaces. At one loop, the only vacuum diagram is the torus. If D-branes are introduced, open strings add worldsheets with boundaries, and the oriented one-loop vacuum amplitude also contains the annulus. In an unoriented theory the set of allowed surfaces is enlarged: one must also include non-orientable worldsheets and worldsheets with crosscaps~\cite{Sagnotti:1987tw,Angelantonj:2002ct,Pradisi_1995,Pradisi_1995_open,FioravantiPradisiSagnotti:1993}. At one loop, the vacuum amplitudes are therefore the torus, the Klein bottle, the annulus and the Möbius strip. The conventions used in this paper for these amplitudes are summarised in \Cref{sec:6d_amplitudes}.

The torus amplitude is constrained by modular invariance of the closed-string CFT. The remaining amplitudes are not modular-invariant objects by themselves, but their double-cover tori have modular properties that relate two equivalent descriptions. In the loop channel they are one-loop vacuum amplitudes of closed or open strings. In the tree channel they are tree-level exchanges of closed strings between extended objects: crosscaps for orientifold planes and boundaries for D-branes. Consistency requires these two interpretations to agree. This is the basic reason why the loop channel expressions are strongly constrained by their tree channel interpretation, in particular by tadpole cancellation and by the positivity and factorization of the corresponding closed-string couplings. We denote tree channel amplitudes with a tilde. For the Klein bottle and annulus the tree channel is obtained by an $S$ modular transformation on the double cover, whereas for the Möbius strip it is obtained by the standard $P$ transformation.

The generic shape of the torus amplitude for a $\Z_N$ orbifold is
\begin{equation}
    \mathcal{T} = \frac1N \sum_{\alpha,\beta=0}^{N-1}\Tr_{\mathcal{H}_\alpha}(P_{GSO}\theta^\beta \q^{L_0-c/24} \qbar^{\overline{L}_0-c/24}) = \frac1N \sum_{\alpha,\beta=0}^{N-1} n[^\alpha_\beta] T[^{\phi_L \alpha}_{\phi_L \beta}] \overline{T}[^{- \phi_R \alpha}_{- \phi_R\beta}] \Lambda[^\alpha_\beta]~,
\end{equation}
where $\Lambda[^\alpha_\beta]$ is the lattice sum over the torus directions unaffected by the orbifold action and $n[^\alpha_\beta]$ is the degeneracy factor counting the number of fixed points in the $(\alpha,\beta)$ sector. Non-trivial momentum and winding contributions occur only for $(\alpha,\beta)=(0,0)$ and the lattice sum is
\begin{align}
    \Lambda[^{0}_{0}] =& \Gamma_{2,2}^1\Gamma_{2,2}^2, \\
    \mathrm{with~} \Gamma_{2,2}^i=& \sum_{m,w\in\Z^2}\q^{\frac14(m+(g_i-B_i) w)^T g_i^{-1}(m+(g_i-B_i) w)}\qbar^{\frac14(m-(g_i-B_i) w)^T g_i^{-1}(m-(g_i-B_i) w)},
\end{align}
where $g_i$ is the metric and $B_i$ is the B-field on the $i$-th two-torus and $m$ and $w$ denote, respectively, the momentum and winding numbers. For later convenience, here we separate the Narain lattice sums of the two two-tori. All other lattices appearing in this paper can be found in~\Cref{app:lattices}.

All this orbifold technology applies to both geometric and non-geometric cases. What changes is not the basic CFT construction, but the nature of the symmetry that defines the quotient. Once this language is in place, one can deform the construction to genuinely asymmetric actions and eventually to quasicrystalline ones, where the bulk torus picture breaks down but the orbifold CFT remains well defined. This is precisely the setting needed for the orientifold vacua studied in the rest of this work.

Let us note here that we will encounter two different kinds of T-duality throughout this article. In order to avoid confusion, we shall distinguish two different T-duality operations on $\T^4=\T^2_{(67)}\times \T^2_{(89)}$.
First, we denote by $\Tp=\text{T}_{x^7}\text{T}_{x^9}$ the partial T-duality along one direction of each two-torus. This duality maps the symmetric geometric orbifold description, for which $\phi_L=\phi_R$, to the antisymmetric description, for which $\phi_L=-\phi_R$. In particular, it changes the sign of the orbifold action on the right-moving modes, $\phi_R\to-\phi_R$, thus relating symmetric and antisymmetric orbifolds. Moreover, it sends the standard orientifold projection $\Omega$ to $\Omega \mathcal I_{79} (-1)^{F_L}$, with $\mathcal I_{79}:x^7\to-x^7, x^9\to-x^9$ and $F_L$ the left-moving fermion number, and maps the $(O9,O5)$ and $(D9,D5)$ systems to two corresponding sets of $O7$-planes and $D7$-branes. It is therefore the appropriate duality for passing between the geometric and antisymmetric pictures~\cite{Blumenhagen:1999md,blumenhagen2000newclasssupersymmetricorientifolds}. On the torus geometry it exchanges complex and Kähler structures.

Second, we denote by $\Tt=\text{T}_{x^6}\text{T}_{x^7}\text{T}_{x^8}\text{T}_{x^9}$ the full internal T-duality along all four compact directions. This second transformation plays a different rôle: rather than changing the symmetric or antisymmetric nature of the orbifold action, it exchanges the two kinds of space-filling branes and orientifold planes, $D9 \leftrightarrow D5$ and $O9 \leftrightarrow O5$. On the torus geometry it sends $U$ to $-1/U$ and $T$ to $-1/T$.

\section{The six-dimensional \texorpdfstring{$\Z_{12}$}{Z12} quasicrystalline orbifold}\label{sec:Z12-parent}
\subsection{Narain-lattice construction}

The construction of the quasicrystalline orbifold is detailed in~\cite{Harvey:1987da,Baykara:2024vss}. In 6d, there exist only $\Z_N$ quasicrystalline orbifolds with $N=5,8,10,12$~\cite{Volpato_2014,baykara2026stringislandsdiscretetheta}. In the following we focus on the $\Z_{12}$ case with $\phi_L=1$ and $\phi_R=7$. The $\Z_{12}$ symmetry group here decomposes into $\Z_3\times \Z_4$.

The $\Z_3$ subgroup is generated by $g^4$ and is therefore symmetric since
\begin{align}
    \frac{2\pi i}{12}\times4(\phi_L,\phi_R) = \frac{2\pi i}{3}(1,1)\,.
\end{align}

The $\Z_4$ subgroup is generated by $g^3$ and is therefore antisymmetric since
\begin{align}
    \frac{2\pi i}{12}\times3(\phi_L,\phi_R) = \frac{2\pi i}{4}(1,-1)\,.
\end{align}
Consequently, the complex structure is fixed by $\Z_3$ while the Kähler structure is fixed by $\Z_4$. We therefore take for both two-tori
\begin{equation}
    U=\e{i\pi/3} \qquad \text{and} \qquad T=i\,.
\end{equation}
Let us notice that this geometry is self \Tt-dual.

\subsection{Twisted sectors and fixed-point orbits}

The torus partition function is obtained by summing over all twisted sectors and by projecting onto states invariant under the full $\Z_{12}$ action. In practice, this requires two pieces of data. First, one has to determine the degeneracy of each twisted sector, namely the number of fixed points, or more generally fixed Narain lattice sectors, associated with the relevant power of the orbifold generator. Second, one has to determine how these degenerate ground states are permuted by the full $\Z_{12}$ generator, since the invariant combinations are organised into orbits of this action. For the genuinely quasicrystalline sectors, the term fixed point should be understood in the CFT sense of a degenerate twisted ground-state sector associated with the Narain-lattice automorphism, rather than as a fixed locus in the geometric torus.

The sectors $\alpha=3,9$ are controlled by the $\Z_4$ subgroup generated by $g^3$.
In the present frame this subgroup is not geometric on the original torus, but is geometric on the dual torus. Its twisted ground states are therefore associated with the four fixed points of the $\Z_4$ action on the dual torus. Under the full $\Z_{12}$ action these four fixed points decompose into one singlet and one triplet.

The sector $\alpha=6$ is controlled by the $\Z_2$ subgroup generated by $g^6$.
This subgroup acts geometrically on the original torus and has the usual sixteen fixed points of a $\T^4/\Z_2$ action. These sixteen fixed points are not all invariant separately under the full $\Z_{12}$ generator: rather, they split into one singlet, one triplet and two sextuplets.

The sectors $\alpha=4,8$ are controlled by the $\Z_3$ subgroup generated by $g^4$.
This subgroup also has a geometric realisation on the original torus. Its nine fixed points decompose, under the full $\Z_{12}$ action, into one singlet and two quadruplets. 

The remaining twisted sectors have unit degeneracy. The sectors $\alpha=2,10$ belong to the geometric $\Z_6$ subgroup generated by $g^2$ and are localised at the origin. By contrast, the sectors $\alpha=1,5,7,11$ are genuine $\Z_{12}$ sectors: they are fixed only in the full Narain-lattice sense and do not correspond to non-trivial fixed-point sets on either the torus or the dual torus. Their degeneracy is therefore one.

The twisted-sector degeneracies are summarised in \Cref{tab:Z12-degeneracies}.
\begin{table}[htb]
\centering
\begin{tabular}{c|c|c|c|c}
Sector(s) & Subgroup & Fixed locus & Degeneracy & $\Z_{12}$ orbits \\
\hline
$\alpha=3,9$
& $\Z_4$
& dual torus
& $4$
& $1+3$
\\
$\alpha=6$
& $\Z_2$
& torus
& $16$
& $1+3+6+6$
\\
$\alpha=4,8$
& $\Z_3$
& torus
& $9$
& $1+4+4$
\\
$\alpha=2,10$
& $\Z_6$
& origin of the torus
& $1$
& $1$
\\
$\alpha=1,5,7,11$
& genuine $\Z_{12}$
& Narain lattice only
& $1$
& $1$
\end{tabular}
\caption{Degeneracies of the twisted sectors and their decomposition into orbits under the full $\Z_{12}$ action.}\label{tab:Z12-degeneracies}
\end{table}

\subsection{Torus amplitude and closed-string spectrum}
The torus partition function takes the following form
\begin{align}
    \mathcal{T}
    =& \frac{1}{12} |T[^0_0]|^2 (\Gamma_{2,2}^1\Gamma_{2,2}^2)' + \frac{1}{12}\Bigg[\sum_{\alpha, \beta=0}^{11} T[^\alpha_\beta]\overline{T}[^{5\alpha}_{5\beta}] + 
    8\sum_{\beta=0}^2T[^4_{4\beta}]\overline{T}[^8_{4\beta}]+ 
    8\sum_{\beta=0}^2T[^8_{4\beta}]\overline{T}[^4_{4\beta}] \notag\\
    &+ 3\sum_{\beta=0}^3|T[^{3}_{3\beta}]|^2
    + 3\sum_{\beta=0}^3|T[^{9}_{3\beta}]|^2
    + 3\sum_{\beta=0}^3|T[^6_{3\beta}]|^2
    + 12\sum_{\beta=0}^1|T[^6_{6\beta}]|^2 \Bigg],
\end{align}
where the prime on $(\Gamma_{2,2}^1\Gamma_{2,2}^2)$ means that the zero-momentum and zero-winding contribution has been removed.

In order to read the massless spectrum, let us write this amplitude in terms of the eigencharacters $\chi[^\alpha_\beta]$ which contain only invariant states under the orbifold action. Let us also introduce new eigencharacters $\xi_k[^\alpha_\beta]$ that diagonalize the $\Z_k$ action on the $\alpha$ twisted sector with $k=2,3,4,6$. They describe how the $\Z_{12}$ eigenstates are organised in representations of the $\Z_k$ subgroups.
\begin{align}
    \xi_k[^\alpha_\beta] &= \sum_{\gamma=0}^{12/k-1}\chi[^{\alpha}_{k\gamma+\beta}],
\end{align}
for $\alpha=0,1,\ldots,11$ and $\beta=0,1,\ldots,k-1$. The massless contributions of these characters, in standard bulk $SU(2)_L\times SU(2)_R$ representations, are
\begin{align}
    \xi_k[^0_0]&\sim (2,2)\oplus 2(2,1)\,, \\
    \xi_k[^0_1]\sim \xi_k[^0_{k-1}]\sim \xi_k[^{\alpha \not = 0}_0]&\sim 2 (1,1) \oplus (1,2)\,.
\end{align}

With these notations, the torus amplitude can be rewritten in the following compact form
\begin{align}
    \mathcal{T} &= \frac{1}{12}|T[^0_0]|^2 (\Gamma_{2,2}^1\Gamma_{2,2}^2)' + \sum_{\alpha,\beta =0}^{11}\chi[^\alpha_{\beta}]\overline{\chi}[^{5\alpha}_{5\beta}] \\
    &+ \sum_{\alpha=1}^3\sum_{\beta=0}^3 |\xi_4[^{3\alpha}_\beta]|^2 + 2\sum_{\alpha=1}^2\sum_{\beta=0}^2 \xi_3[^{4\alpha}_\beta]\overline{\xi_3}[^{-4\alpha}_{-\beta}] + 2(|\xi_2[^6_0]|^2+|\xi_2[^6_1]|^2) \notag.
\end{align}
On the first line one can read the untwisted contribution together with the twisted-sector contributions associated with the singlet orbit. The second line collects the contributions of the twisted sectors belonging to non-trivial orbits under the relevant subgroups of $\Z_{12}$.
\begin{table}[htbp]
\centering
\small
\renewcommand{\arraystretch}{1.15}
\begin{tabularx}{\textwidth}{c|c|c|X}
Sector(s) & Orbit structure & Projected tensors & Interpretation \\
\hline
Untwisted & $-$ & $1$ & Untwisted tensor multiplet; the supergravity multiplet is counted separately \\
\hline
$\alpha=1,5,7,11$ & $1$ for each sector & $4$ & Genuine $\Z_{12}$ singlet sectors \\
\hline
$\alpha=2,10$ & $1$ for each sector & $2$ & $\Z_6$ singlets at the origin \\
\hline
$\alpha=3,9$ & $1+3$ for each sector & $4$ & One singlet orbit and one triplet orbit in each sector \\
\hline
$\alpha=4,8$ & $1+4+4$ for each sector & $6$ & One singlet orbit and two quadruplet orbits in each sector \\
\hline
$\alpha=6$ & $1+3+6+6$ & $4$ & One singlet, one triplet and two sextuplet orbits \\
\hline
Total twisted & & $20$ & $11$ singlet, $4$ quadruplet, $3$ triplet and $2$ sextuplet orbits \\
\hline
Total & & $21$ & $\mathcal N=(2,0)$ supergravity plus $21$ tensor multiplets \\
\hline
\end{tabularx}
\caption{Distribution of tensor multiplets in the oriented $\Z_{12}$ model. The entries in the second column describe the decomposition of the twisted ground-state degeneracies into orbits under the full $\Z_{12}$ action. Each orbit supplies one invariant combination after the orbifold projection.}
\label{tab:Z12-oriented-spectrum}
\end{table}

From here we can read the massless spectrum of the oriented theory. We have the six-dimensional $\mathcal{N}=(2,0)$ supergravity multiplet together with 21 tensor multiplets. The twisted sectors therefore provide $20$ invariant tensor multiplets: $11$ associated with singlet orbits, $4$ with quadruplet orbits, $3$ with triplet orbits and $2$ with sextuplet orbits. The spectrum is summarised in \Cref{tab:Z12-oriented-spectrum}. By looking at the massless modes surviving the orbifold projection, one can also determine the moduli of the theory. The orbifold freezes the geometric complex-structure and Kähler moduli, but five untwisted scalar fields survive: the dilaton and four R-R scalars, including the axion.

\section{Probe branes and boundary sectors}
\label{sec:probe-branes}

\subsection{Boundary orbits under the quasicrystalline action}

Since we are dealing with type IIB backgrounds, we restrict attention to BPS D-branes of odd dimension~\cite{Polchinski:1995mt}. In the six-dimensional compactification, the relevant boundary sectors can be organised according to their internal boundary conditions: branes localised in the compact space, branes wrapping the full compact space, and branes wrapping a product of one-cycles on the two two-tori.

Probe branes provide a useful diagnostic of the quasicrystalline background~\cite{DouglasMoore:1996}. In the loop channel, their annulus amplitudes determine the open-string spectrum and the Chan-Paton projection induced by the orbifold action. In the tree channel, the same amplitudes identify the closed-string modes sourced by the corresponding boundary states~\cite{Cardy:1989ir,Bill__2001}. This second interpretation will be important in the orientifold construction, where the boundary sectors have to be matched with the crosscap tadpoles. 

The quasicrystalline generator acts non-trivially on boundary conditions. Consequently, the orbit of a brane under the full $\Z_{12}$ action may contain branes with different internal dimensionalities, positions and Wilson lines. The appropriate description depends on the type of boundary sector. For the $Dp/D(p+4)$ orbit, the partial \Tp-dual frame is useful because it turns the mixed localised/wrapped system into $D(p+2)$-branes at angles. For genuine $D(p+2)$-branes, however, no such reinterpretation is needed: they are already branes at angles, which is their most convenient description. For type IIB D-branes, $p$ is odd, so we write $p=2k+1$.

We work throughout in the frame in which the branes wrap one-cycles of each torus since it is the most intuitive and convenient one~\cite{Berkooz_1996,Blumenhagen:2002wn}. Indeed, in this frame, the action of the orbifold generator on the boundary conditions can be easily described. Let us describe it for one two-torus of complex structure $U$ and Kähler structure $T$ but of course it applies to both two-tori. Let us package the two real open-string moduli into~\cite{Bianchi:1991eu,Blumenhagen_2007,Berg_2012}
\begin{equation}
    u=x_{\parallel}+T y_{\perp}\,,\qquad
    u\in\mathbb C/(\Z+T\Z)\,,
\end{equation}
with the Wilson line $x_{\parallel}$ along the cycle wrapped by the brane and the position $y_{\perp}$ transverse to it. Let us also denote by $(a,b)$ the wrapping numbers of the brane along the two one-cycles of the torus. Then this brane makes an angle $\theta$ with respect to the horizontal axis, where the wrapping number and the angle are related by
\begin{equation}
    \cot(\theta)=\frac{a+b\Re(U)}{b\Im(U)}\,.
\end{equation}
Now, let us consider an orbifold action characterised by the twist vector $\phi_{L,R}$. We can extract the symmetric and antisymmetric parts of this action as
\begin{equation}
    \Sigma=\frac{\pi}{N}(\phi_L+\phi_R) \qquad \text{and} \qquad \Delta=\frac{\pi}{N}(\phi_L-\phi_R)\,.
\end{equation}
The symmetric part $\Sigma$ acts as a rotation of the brane support. Thus it sends $\theta$ to $\theta+\Sigma$. The antisymmetric part $\Delta$ acts as a rotation in the dual torus and therefore sends $u$ to $e^{i\Delta}u$. Finally,
\begin{equation}
    g.(u,\theta)=(e^{i\Delta}u,\theta+\Sigma)\,.
\end{equation}
There is a small subtlety with order two actions. Indeed, since a $\Z_2$ action is both symmetric and antisymmetric, it imposes the following identification
\begin{equation}
    (u,\theta+\pi)\sim(-u,\theta)\,.
\end{equation}

In what follows, we shall label a brane orbit by the moduli and wrapping numbers or equivalently by the angle, of a chosen representative. The remaining elements of the orbit are then obtained by acting with the orbifold generator according to the transformation law described above. For the case under consideration in this paper $\Sigma=2\pi/3$ and $\Delta=-\pi/2$.

\subsection{\texorpdfstring{$Dp/D(p+4)$}{Dp/D(p+4)} orbits and their \texorpdfstring{\Tp}{T79}-dual branes-at-angles description}

Let us first consider a probe $Dp$-brane filling $p+1$ non-compact directions and localised in the four internal directions. Its image under the full $\Z_{12}$ generator contains a $D(p+4)$-brane wrapping the internal space. This follows from the factorization of the generator at the level of left- and right-moving coordinates,
\begin{equation}
    \frac{2\pi i}{12}(\phi_L,\phi_R)=\frac{2i\pi}{12}(1,1)+i\pi (0,1)\,.
\end{equation}

The first factor is a symmetric rotation, while the second factor acts as a full internal {\Tt-duality} on the boundary conditions. It exchanges Dirichlet and Neumann conditions along the four compact directions and therefore maps a localised $Dp$-brane to a wrapped $D(p+4)$-brane. The remaining images are generated by the symmetric $\Z_6$ subgroup generated by $g^2$.

\begin{table}[htb]
\centering
\begin{tabular}{c|c|c|c}
Initial position & Invariant group & Orbit size \\
\hline
generic point & trivial & $12$ \\
$\Z_2$ fixed point & $\Z_2$ & $6$ \\
$\Z_3$ fixed point & $\Z_3$ & $4$ \\
origin & $\Z_6$ & $2$
\end{tabular}
\caption{Orbit sizes for localised $Dp$-brane probes under the $\Z_{12}$ action.}\label{tab:Dp-orbits}
\end{table}

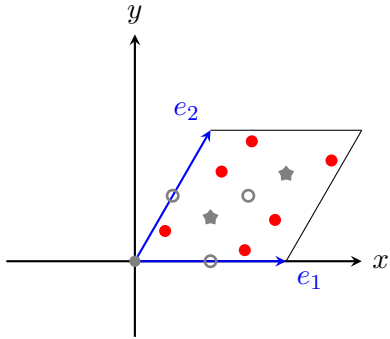
\begin{figure}[htb]
    \centering
    \begin{tikzpicture}[scale=2,>=stealth]
          % Axes
          \draw[->, thick] (-.85, 0) -- (1.5, 0) node[right] {$x$};
          \draw[->, thick] (0, -0.5) -- (0, 1.5) node[above] {$y$};

          % Vectors
          \draw[->, thick, blue] (0,0) -- (1,0) node[below right] {$e_1$};
          \draw[-] ({cos(60)}, {sin(60)}) -- ({1+cos(60)}, {sin(60)});

          \draw[->, thick, blue] (0,0) -- ({cos(60)}, {sin(60)}) node[above left] {$e_2$};
          \draw[-] (1,0) -- ({1+cos(60)}, {sin(60)});

          % Fixed points
          \node[circle, fill=gray, inner sep=1.5pt] at (0,0) {};
          \node[star, star points=5, fill=gray, inner sep=1.5pt] at ({1/3+cos(60)/3}, {sin(60)/3}) {};
          \node[star, star points=5, fill=gray, inner sep=1.5pt] at ({2/3+2*cos(60)/3}, {2*sin(60)/3}) {};
          \node[circle, draw=gray, fill=none, inner sep=1.5pt, line width=1pt] at ({1/2+cos(60)/2}, {sin(60)/2}) {};
          \node[circle, draw=gray, fill=none, inner sep=1.5pt, line width=1pt] at ({cos(60)/2}, {sin(60)/2}) {};
          \node[circle, draw=gray, fill=none, inner sep=1.5pt, line width=1pt] at (1/2,0) {};

          %D5s
          \node[circle, fill=red, inner sep=1.6pt] at (0.2,0.2) {};
          \node[circle, fill=red, inner sep=1.6pt] at ({1+0.2*(cos(60)-sin(60))},{0.2*(cos(60)+sin(60))}) {};
          \node[circle, fill=red, inner sep=1.6pt] at ({1+0.2*(cos(120)-sin(120))},{0.2*(cos(120)+sin(120))}) {};
          \node[circle, fill=red, inner sep=1.6pt] at ({1+cos(60)+0.2*(cos(180)-sin(180))},{sin(60)+0.2*(cos(180)+sin(180))}) {};
          \node[circle, fill=red, inner sep=1.6pt] at ({cos(60)+0.2*(cos(240)-sin(240))},{sin(60)+0.2*(cos(240)+sin(240))}) {};
          \node[circle, fill=red, inner sep=1.6pt] at ({cos(60)+0.2*(cos(300)-sin(300))},{sin(60)+0.2*(cos(300)+sin(300))}) {};
    \end{tikzpicture}
    \caption{Example of a $Dp$-brane orbit at a generic position on the torus lattice. The $\Z_3$ fixed points are marked by gray stars, the $\Z_2$ fixed points by empty gray circles, and the $Dp$-brane images by red dots. The orbit also contains the same number of $D(p+4)$-brane images, with Wilson lines rather than position moduli; these are displayed as a similar configuration in the fully \Tt-dual lattice. At a generic position the full orbit contains twelve images. At a $\Z_3$ fixed point the orbit has size four, at a $\Z_2$ fixed point it has size six, and at the origin it has size two. Here is displayed one of the two $\T^2$, the second one behaving the same.}\label{fig:Z12Dps}
\end{figure}

The size of the orbit is determined by the invariant group of the initial brane position (see \Cref{tab:Dp-orbits}). A generic point has trivial invariant group and gives an orbit of twelve images (see \Cref{fig:Z12Dps}). A point fixed by the $\Z_2$ subgroup gives an orbit of six images, a point fixed by the $\Z_3$ subgroup gives an orbit of four images, and the origin, fixed by the geometric $\Z_6$ subgroup, gives an orbit of two images. In all cases, the orbit contains equal numbers of localised $Dp$-branes and wrapped $D(p+4)$-branes.

The $D(p+4)$-brane images carry Wilson lines rather than position moduli. This becomes manifest after the full internal \Tt-duality, which exchanges $Dp$- and $D(p+4)$-branes and maps Wilson lines to positions. For the $Dp/D(p+4)$ annulus computations, however, the partial \Tp-duality is the most convenient description: in that frame the boundary sectors are represented by $D(p+2)$-branes wrapping a one-cycle on each two-torus.

Let us recall that, in the \Tp-dual frame, the complex and Kähler structures are exchanged, while the right-moving twist changes sign. Accordingly, the open modulus introduced above is written in this frame as $u=a_\parallel+U\,\phi_\perp$. Since the exchange $U\leftrightarrow T$ is accompanied by $\phi_R\to-\phi_R$, the two combinations $\Sigma$ and $\Delta$ are exchanged. Therefore the orbifold action on the brane data becomes ${(u,\theta)\mapsto \left(\e{i\Sigma}u,\theta+\Delta\right)}$.

For the purpose of this article let us focus on the following two orbits: the one obtained from the wrapping numbers $(1,0,1,1)$, denoted $\mathcal O_N$ and the one obtained from the wrapping numbers $(1,1,1,0)$, denoted $\mathcal O_M$.

The amplitudes below should be read in two complementary ways. In the loop channel they determine the open-string spectrum on the corresponding probe orbit, including the Chan-Paton projection induced by the invariant group of the brane. In the tree channel they display the closed-string modes sourced by the boundary state. This second form is the one that will be most directly comparable with the tadpole analysis of the orientifold.

\subsection{Annulus amplitudes for \texorpdfstring{$Dp/D(p+4)$}{Dp/D(p+4)} orbits at the most symmetric loci}\label{sec:brane-orbit-origin}

We construct the annulus amplitudes using the standard orbifold boundary-state and open-descendant formalism~\cite{Cardy:1989ir,Bill__2001,Angelantonj:2002ct}.

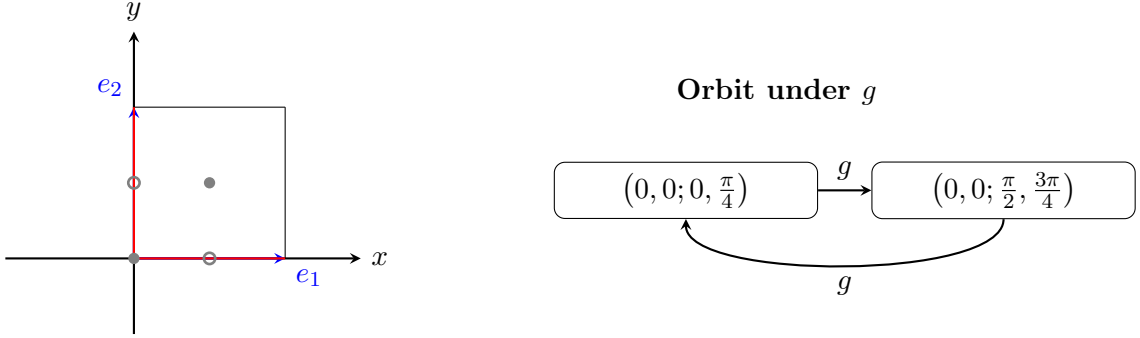
\begin{figure}[htb]
\centering
\begin{tikzpicture}[>=stealth]

    %%%%%%%%%%%%%%%%%%%%%%%%%
    % Left panel: torus plot
    %%%%%%%%%%%%%%%%%%%%%%%%%
    \begin{scope}[scale=2]
        % Axes
        \draw[->, thick] (-.85, 0) -- (1.5, 0) node[right] {$x$};
        \draw[->, thick] (0, -0.5) -- (0, 1.5) node[above] {$y$};

        % Basis vectors
        \draw[->, thick, blue] (0,0) -- (1,0) node[below right] {$e_1$};
        \draw[-] (1,1) -- (1,0);

        \draw[->, thick, blue] (0,0) -- (0,1) node[above left] {$e_2$};
        \draw[-] (0,1) -- (1,1);

        % Branes
        \draw[-, thick, red] (0,0) -- (1,0);
        \draw[-, thick, red] (0,0) -- (0,1);

        % Fixed points
        \node[circle, fill=gray, inner sep=1.5pt] at (0,0) {};
        \node[circle, fill=gray, inner sep=1.5pt] at (1/2,1/2) {};
        \node[circle, draw=gray, fill=none, inner sep=1.5pt, line width=1pt] at (1/2,0) {};
        \node[circle, draw=gray, fill=none, inner sep=1.5pt, line width=1pt] at (0,1/2) {};
    \end{scope}

    %%%%%%%%%%%%%%%%%%%%%%%%%%%%%%%%%%%%%%%
    % Right panel: orbit under g
    %%%%%%%%%%%%%%%%%%%%%%%%%%%%%%%%%%%%%%%
    \begin{scope}[xshift=5.5cm]
        \node[font=\bfseries] at (3.0,2.2) {Orbit under $g$};

        \node[draw, rounded corners, align=center, text width=3.2cm, inner sep=4pt] (s1) at (1.8,0.9)
        {$\left(0,0;0,\frac{\pi}{4}\right)$};

        \node[draw, rounded corners, align=center, text width=3.2cm, inner sep=4pt] (s2) at (6.0,0.9)
        {$\left(0,0;\frac{\pi}{2},\frac{3\pi}{4}\right)$};

        \draw[->, thick] (s1.east) -- node[above] {$g$} (s2.west);
        \draw[->, thick] (s2.south) to[out=-90,in=-90, looseness=0.5] node[below] {$g$} (s1.south);
    \end{scope}
\end{tikzpicture}
\caption{Left: example of size-two brane orbits on one of the two-tori. Right: action of the orbifold generator $g$ on the corresponding orbit in terms of $(u_1,u_2;\theta_1,\theta_2)$, where $u_i$ and $\theta_i$ denote respectively the moduli and the angle of the branes.}\label{fig:brane-orbit-origin}
\end{figure}
We first consider the orbit corresponding to $u=0$ on both tori with the wrapping numbers mentioned earlier as represented in \Cref{fig:brane-orbit-origin}. The orbits have size two and the Chan-Paton action is naturally diagonalised with respect to this $\Z_6$ invariant group. The annulus amplitude can then be written in terms of the corresponding traces $N_\beta$ and $M_\beta$ as
\begin{align}
   \mathcal{A} = \frac16&\Big[T[^0_0](
    |N_0|^2\Lambda^{\mathcal{A} (1,0)}_{m_1,w_1}\Lambda^{\mathcal{A} (1,1)}_{m_2,w_2}(T,U)+
    |M_0|^2\Lambda^{\mathcal{A} (1,1)}_{m_1,w_1}\Lambda^{\mathcal{A} (1,0)}_{m_2,w_2}(T,U)  ) \\
    &+\sum_{\beta=1}^5 (|N_\beta|^2+|M_\beta|^2)T[^0_{2\beta}] \notag\\
   &+ \sum_{\beta=0}^5 (N_\beta\overline{M}_\beta+\overline{N}_\beta M_\beta)(T[^3_{2\beta}]+T[^9_{2\beta}]) \notag\\
   &+ 2(|N_0|^2+|M_0|^2)T[^6_{0}] + 2(|N_3|^2+|M_3|^2)T[^6_{6}]-\sum_{\beta=1,2,4,5}(|N_\beta|^2+|M_\beta|^2)T[^6_{2\beta}]\notag\Big],
\end{align}
where the lattices\footnote{All lattices involving extended objects, boundaries or crosscaps, are expressed in the frame where they correspond to objects intersecting at angles.} can be found in \Cref{app:lattices} and the Chan-Paton traces $N_\beta$ and $M_\beta$ are defined by
\begin{align}
    N_\beta &= \Tr(\gamma^\beta_N)\,, \\
    M_\beta &= \Tr(\gamma^\beta_M)\,,
\end{align}
where
\begin{align}
    \gamma_N &= \operatorname{diag}\!\left(
        \Un_{n_0},
        \e{2i\pi/6}\Un_{n_1},
        \e{4i\pi/6}\Un_{n_2},
        \e{6i\pi/6}\Un_{n_3},
        \e{8i\pi/6}\Un_{n_4},
        \e{10i\pi/6}\Un_{n_5}
    \right)\,,\\
    \gamma_M &= \operatorname{diag}\!\left(
        \Un_{m_0},
        \e{2i\pi/6}\Un_{m_1},
        \e{4i\pi/6}\Un_{m_2},
        \e{6i\pi/6}\Un_{m_3},
        \e{8i\pi/6}\Un_{m_4},
        \e{10i\pi/6}\Un_{m_5}
    \right)\,.
\end{align}
All indices on the multiplicities $n_\gamma$ and $m_\gamma$ are understood modulo $6$ in this subsection. The massless contribution to this amplitude is
\begin{align}
    \mathcal{A}_0 \sim& \sum_{\gamma=0}^5\Big[(n_\gamma\overline{n}_\gamma+m_\gamma\overline{m}_\gamma)\chi[^0_0]+(n_\gamma\overline{m}_\gamma+\overline{n}_\gamma m_\gamma)(\chi[^3_0]+\chi[^9_0])\\
    &+ (n_\gamma\overline{n}_{\gamma+1}+m_\gamma\overline{m}_{\gamma+1})\chi[^0_1]+(n_\gamma\overline{n}_{\gamma-1}+m_\gamma\overline{m}_{\gamma-1})\chi[^0_{11}]\notag\\
    &+ (n_\gamma(\overline{n}_{\gamma+2}+\overline{n}_{\gamma-2})+m_\gamma(\overline{m}_{\gamma+2}+\overline{m}_{\gamma-2}))\chi[^6_0]\Big]\,.\notag
\end{align}
This leads to a $\prod_{\gamma=0}^5 U(n_\gamma)\times\prod_{\gamma=0}^5 U(m_\gamma)$ gauge theory.
The tree channel form displays the closed-string modes sourced by this boundary orbit. Written in terms of the $\xi_4$ combinations introduced above, it takes the form\footnote{Recall that the prime for the lattices, e.g. $\Lambda'$, means that the zero momentum and winding contribution has been removed.}
\begin{align}
    \widetilde{\mathcal{A}} = \frac{2^{-(p+1)/2}}{3}&\Big[\frac14T[^0_0](
        |N_0|^2\widetilde{\Lambda}^{(1,0)}_{0,0}\widetilde{\Lambda}^{(1,1)}_{0,0} +
        |M_0|^2\widetilde{\Lambda}^{(1,1)}_{0,0}\widetilde{\Lambda}^{(1,0)}_{0,0})' \\
        &+\frac12(|N_0+M_0|^2\xi_4[^0_0]+|N_0-M_0|^2\xi_4[^0_2]) \notag\\
    &+\sum_{\beta=1}^5(|N_\beta-M_\beta|^2\xi_4[^{2\beta}_0]+|N_\beta+M_\beta|^2\xi_4[^{2\beta}_2]) \notag\\
    &+\sum_{\beta\in\{2,4\}}(|N_\beta|^2+|M_\beta|^2)(\xi_4[^{2\beta}_0]+\xi_4[^{2\beta}_1]+\xi_4[^{2\beta}_2]+\xi_4[^{2\beta}_3]) \notag\\
    &+ 2(|N_3|^2+|M_3|^2)(\xi_4[^6_1]+\xi_4[^6_3]) \Big]\,.\notag
\end{align}
From the tree channel, one can read how the branes couple to the different tensors. While the two orbits both couple to the untwisted tensor and the $\Z_2$, $\Z_3$ and $\Z_6$ singlet tensors with different signs, they do couple to different quadruplets and couple neither to the triplets not to the sextupletsnor to the $\Z_4$ and $\Z_{12}$ singlets. Coupling to these fields would require at least a $\Z_4$ symmetric configuration which is not possible for this class of branes. The spectrum is summarised in \Cref{tab:Dp-orbit2-spectrum}.
\begin{table}[htb]
\centering
\begin{tabular}{c|c|c}
    Sector & Representation & Field content \\\hline
    $\alpha=0$ &
    $n_\gamma\overline{n}_\gamma+m_\gamma\overline{m}_\gamma$ &
    $A^\mu+(4-2k)\phi+2^{2-k}(\psi_L+\psi_R)$ \\\hline
    $\alpha=0$ &
    $n_\gamma(\overline{n}_{\gamma+1}+\overline{n}_{\gamma-1})+m_\gamma(\overline{m}_{\gamma+1}+\overline{m}_{\gamma-1})$ &
    $\Phi+2^{1-k}(\psi_L+\psi_R)$ \\\hline
    $\alpha\in\{3,9\}$ &
    $n_\gamma\overline{m}_\gamma+\overline{n}_\gamma m_\gamma$ &
    $2(2\varphi+2^{1-k}(\psi_L+\psi_R))$ \\\hline
    $\alpha=6$ &
    $n_\gamma(\overline{n}_{\gamma+2}+\overline{n}_{\gamma-2})+m_\gamma(\overline{m}_{\gamma+2}+\overline{m}_{\gamma-2})$ &
    $2\varphi+2^{1-k}(\psi_L+\psi_R)$ \\\hline
\end{tabular}
\caption{Massless spectrum of the $Dp/D(p+4)$ orbit at the origin. The first column indicates the sector, i.e.\ the states are the excitations from branes at relative angle $\frac{2\pi}{12}\alpha$. The second column indicates the representation under the gauge group $\prod_{\gamma=0}^5 U(n_\gamma)\times\prod_{\gamma=0}^5 U(m_\gamma)$, and the third column indicates the field content. The fields are the gauge bosons $A^\mu$, the transverse position real scalars in the non-compact space $\phi$, the transverse position complex scalars in the torus $\Phi$, other real scalars $\varphi$ and the worldvolume fermions $\psi_L$ and $\psi_R$. The number of scalars and fermions depends on the dimension of the brane in the non-compact space, which is $p=2k+1$.}\label{tab:Dp-orbit2-spectrum}
\end{table}
Starting from this special position, one can move the branes to a generic point $u$ in the bulk by turning on a Higgs-branch vev. A brane which can leave the fixed point must contain the regular representation of this invariant group, so on the branch leading to a generic orbit one takes $n_\gamma=n$ and $m_\gamma=m$ for all $\gamma$, i.e.\ exactly six of these fractional branes are needed to recombine into a bulk brane~\cite{DouglasMoore:1996,nagaoka2012higgsingintersectingbranemodels}. The relevant Higgs fields are the internal transverse complex scalars $\Phi$ in the second line of \Cref{tab:Dp-orbit2-spectrum}, which map neighbouring Chan-Paton components into one another. Giving equal vevs
\begin{equation}
\langle\Phi^N_\gamma\rangle=u_N\mathbf 1_n,\qquad \langle\Phi^M_\gamma\rangle=u_M\mathbf 1_m,
\end{equation}
identifies the six fractional components and breaks the gauge group to its diagonal subgroup,
\begin{equation}
\prod_{\gamma=0}^5 U(n_\gamma)\times\prod_{\gamma=0}^5 U(m_\gamma)\longrightarrow U(n)_{\rm diag}\times U(m)_{\rm diag}.
\end{equation}
The parameters $u_N$ and $u_M$ are the positions of the resulting bulk brane orbits in the internal torus. Around this Higgsed vacuum, the fields should be reorganised into diagonal and non-diagonal combinations with respect to the six Chan-Paton components. In the first line of \Cref{tab:Dp-orbit2-spectrum}, only the diagonal combinations of $A^\mu$, $\phi$, $\psi_L$ and $\psi_R$ remain massless. The non-diagonal combinations are lifted: the corresponding gauge bosons become massive, while the associated scalars and fermions combine with them into massive multiplets. The fields in the second line, namely the internal transverse complex scalars $\Phi$, are the Higgs fields. Their diagonal component is the physical position modulus of the generic bulk orbit, while their non-diagonal gauge-orbit fluctuations are the Goldstone modes eaten by the massive gauge bosons. The remaining fields in the sectors $\alpha\in\{3,6,9\}$ are not Goldstone modes for this displacement. Instead, after the breaking to $U(n)_{\rm diag}\times U(m)_{\rm diag}$, their Chan-Paton labels are recombined into representations of the diagonal gauge group, with the multiplicities shown in \Cref{tab:Dp-orbitgeneric-spectrum}. Note that at a generic position, the branes do not couple to the twisted sectors anymore, since they are no longer at fixed points of the orbifold action.
\begin{table}[htb]
\centering
\begin{tabular}{c|c|c}
    Sector & Representation & Field content \\\hline
    $\alpha=0$ &
    $n\overline{n}+m\overline{m}$ &
    $A^\mu+(4-2k)\phi+2^{2-k}(\psi_L+\psi_R)$ \\\hline
    $\alpha=0$ &
    $n\overline{n}+m\overline{m}$ &
    $2(\Phi+2^{1-k}(\psi_L+\psi_R))$ \\\hline
    $\alpha\in\{3,9\}$ &
    $n\overline{m}+\overline{n} m$ &
    $12(2\varphi+2^{1-k}(\psi_L+\psi_R))$ \\\hline
    $\alpha=6$ &
    $n\overline{n}+m\overline{m}$ &
    $12(2\varphi+2^{1-k}(\psi_L+\psi_R))$ \\\hline
\end{tabular}
\caption{Massless spectrum of the $Dp/D(p+4)$ orbit at a generic position. The first column indicates the sector, i.e.\ the states are the excitations from branes at relative angle $\frac{2\pi}{12}\alpha$. The second column indicates the representation under the gauge group $U(n)_{\rm diag}\times U(m)_{\rm diag}$, and the third column indicates the field content. The fields are the gauge bosons $A^\mu$, the transverse position real scalars in the non-compact space $\phi$, the transverse position complex scalars in the torus $\Phi$, other real scalars $\varphi$ and the worldvolume fermions $\psi_L$ and $\psi_R$. The number of scalars and fermions depends on the dimension of the brane in the non-compact space, which is $p=2k+1$.}\label{tab:Dp-orbitgeneric-spectrum}
\end{table}

The remaining shortened boundary orbits occur when the representative boundary is invariant under the $\Z_3$, $\Z_2$ or $\Z_4$ subgroups. Their annulus amplitudes, Chan-Paton decompositions and massless spectra are collected in \Cref{app:probe-branes}. These sectors are useful for characterising the possible fractional charges and Higgs branches, but the orientifold vacuum constructed below only requires the origin orbit described above.

\section{Orientifold descendants}\label{sec:orientifold-descendants}

We now turn to the main subject of this work, namely the construction of the unoriented descendants of the six-dimensional quasicrystalline background introduced in the previous section. Contrary to the parent closed-string orbifold, whose consistency is essentially fixed by the worldsheet orbifold construction together with modular invariance, the orientifold requires additional dynamical input~\cite{Sagnotti:1987tw,Bianchi:1990tb,Gimon_1996,Angelantonj:2002ct}. One must specify an involution compatible with the asymmetric $\Z_{12}$ action, determine the corresponding projection on the closed-string spectrum, extract the R-R and NS-NS tadpoles from the Klein bottle amplitude, and then identify the open-string sectors required for consistency. Six-dimensional K3 orientifolds with non-standard projections and enhanced tensor sectors provide closely related examples \cite{Dabholkar_1996,Polchinski_1997}.

This problem is non-trivial already in ordinary six-dimensional antisymmetric orientifolds where $\phi_L=-\phi_R$~\cite{Bianchi_2000,Angelantonj:2000xf,Blumenhagen:1999md,Blumenhagen:2000fp}. In known crystallographic examples, the orientifold projection may leave twisted sectors invariant and therefore generate non-vanishing twisted contributions in the Klein bottle amplitude. The resulting tadpoles cannot in general be cancelled by a standard open sector alone, and their cancellation requires more elaborate brane configurations, admitting in suitable \Tp-dual frames an interpretation in terms of D-branes at angles or, equivalently, magnetised branes. In the present case the problem is further sharpened by the fact that the parent orbifold is quasicrystalline, so that the quotient is not associated with an ordinary geometric action on the bulk torus.

The strategy of this section is therefore the following. We first analyse the unoriented closed-string sector and determine the Klein bottle projection associated with the quasicrystalline $\Z_{12}$ orbifold. This fixes the spectrum of crosscap charges and the structure of the closed-string tadpoles. We then introduce the open-string sector needed to compensate these charges. Finally, combining Klein bottle, annulus and Möbius strip amplitudes, we derive the tadpole cancellation conditions, determine the resulting gauge group and massless spectrum, and confront them with the six-dimensional anomaly constraints.

In this way, the orientifold construction provides a genuine open-string completion of the quasicrystalline orbifold background. It also offers a perturbative setting in which the interplay between non-geometric closed-string data and orientifold consistency conditions can be analysed explicitly.

In the main text we focus on the orientifold projection $\Omega\sigma$ where
\begin{align}
    \Omega:&\, X_L^i\leftrightarrow X_R^i \,,\\
    \sigma:&\,
    \begin{cases}
        Z^1\mapsto Z^1 \\
        (Z^2_L,Z^2_R)\mapsto (\e{2i\pi/4} Z^2_L, \e{-2i\pi/4} Z^2_R) \,.
    \end{cases}
\end{align}
This choice is convenient for two reasons. First, it gives a slightly richer structure as we shall see later on. Secondly, it fixes simple degeneracies and coefficients in the loop and tree channels. Yet other projections can be obtained quite easily from this one and can be found in the appendices, for instance for the standard uniform $\Omega$ projection. We recall that on the \Tp-dual picture, the orientifold projection will be $\Omega\widehat{\sigma}\mathcal I_{79}(-1)^{F_L}$ with
\begin{align}
    \widehat{\sigma}:&\,
    \begin{cases}
        Z^1\mapsto Z^1 \\
        Z^2\mapsto \e{2i\pi/4} Z^2 \,.
    \end{cases}
\end{align}
One can verify that this orientifold projection satisfies
\begin{equation}
    (\Omega\sigma)^{-1}.g.(\Omega\sigma)=g^7\,,
\end{equation}
which ensures this projection is really an automorphism of the orbifold group~\cite{Blumenhagen:1999md,Angelantonj:2000xf,
Bianchi:1999}. Other orientifold projections are analysed in \Cref{app:other_vacua}.

\subsection{Klein bottle projection, closed spectrum and O-plane cycles}

The first step in the construction of the orientifold vacuum is the computation of the unoriented closed-string spectrum. This is encoded in the Klein bottle amplitude, which can be computed by applying the orientifold projection to the parent closed-string orbifold. The resulting amplitude can be expressed in terms of the characters of the $\Z_{12}$ orbifold and of the lattice sums associated with the toroidal compactification. From this expression, one can read off the spectrum of crosscap states and determine the structure of the R-R and NS-NS tadpoles. On the orbifold the orientifold projection sends the twisted sector Hilbert space $\mathcal{H}_{\alpha}$ to $\mathcal{H}_{5\alpha}$ such that the only invariant states come from the twisted sectors $\alpha=0,3,6,9$. From an amplitude point of view, the orientifold projection selects all the diagonal characters of the torus amplitude. Moreover, the tree channel expression must be interpreted as the tree-level exchange of closed-string states between two crosscaps. This imposes a non-trivial consistency requirement: the coefficients of propagating characters must be compatible with squares of crosscap one-point functions~\cite{FioravantiPradisiSagnotti:1993,Pradisi_1995,Pradisi_1995_open}. This criterion fixes the relative signs in the loop channel Klein bottle amplitude and determines which closed-string tadpoles have to be cancelled by the open-string sector. It turns out that the choice compatible with supersymmetry and with a positive tree channel factorization is the following Klein bottle projection
\begin{align}
    \mathcal{K} =& \frac13\Big[T[^0_0]\left(\frac{
        \Lambda^{\mathcal K,(1,0)}_1\Lambda^{\mathcal K,(1,1)}_2(T,U)+
        \Lambda^{\mathcal K,(1,1)}_2\Lambda^{\mathcal K,(1,0)}_1(T,U)
        }{2}\right) + \sum_{\beta=1}^2 T[^0_{4\beta}] \\
    &+ 4T[^3_0] + \sum_{\beta=1}^2 T[^3_{4\beta}]
    + 8T[^6_0] -  \sum_{\beta=1}^2 T[^6_{4\beta}]
    + 4T[^9_0]  + \sum_{\beta=1}^2 T[^9_{4\beta}]\Big]\notag\\
    =&\frac16T[^0_0]\left(
        \Lambda^{\mathcal K,(1,0)}_1\Lambda^{\mathcal K,(1,1)}_2+
        \Lambda^{\mathcal K,(1,1)}_2\Lambda^{\mathcal K,(1,0)}_1
    \right)' +\xi_3[^0_0]+\xi_3[^3_0]-\xi_3[^6_0]+\xi_3[^9_0]\notag\\
    &+ \sum_{\beta=0}^3(\xi_4[^3_\beta]+\xi_4[^6_\beta] +\xi_4[^9_\beta] ) + 2\xi_2[^6_0] +2\xi_2[^6_1] \,,\notag
\end{align}
whose tree channel reads
\begin{align}
    \widetilde{\mathcal{K}} = \frac{2^3}{3} & \Big[T[^0_0]
        \begin{aligned}[t]
        \bigl(\mathcal{C}_+(m_1,w_1)^2\widetilde{\Lambda}^{(1,0)}_{0,0}\mathcal{C}_-(m_2,w_2)^2\widetilde{\Lambda}^{(1,1)}_{0,0} +\\
        \mathcal{C}_-(m_1,w_1)^2\widetilde{\Lambda}^{(1,1)}_{0,0}\mathcal{C}_+(m_2,w_2)^2\widetilde{\Lambda}^{(1,0)}_{0,0} \bigr)'
        \end{aligned}
        \label{eq:Kleintransverse} \\
        &+8\xi_4[^0_0]
        +\sum_{\alpha\in\{4,8\}}(4\xi_4[^{\alpha}_2]+2\xi_4[^{\alpha}_0]+2\xi_4[^{\alpha}_1]+2\xi_4[^{\alpha}_2]+2\xi_4[^{\alpha}_3]) \Big] \,.\notag
\end{align}
The lattices involved here are again defined in \Cref{app:lattices}, where the phases display the orientifold-plane structure and will be important for understanding the Möbius strip amplitude.
From the latter amplitude, one can read the following tadpole contributions
\begin{align}
    \widetilde{\mathcal{K}}_0 =& \frac{2^3}{3}\Big[8\chi[^0_0] + 2\chi[^4_0]+2\chi[^8_0] \Big] \,.
    \label{eq:Kleintadpole}
\end{align}
It shows that the orientifold planes contribute to an untwisted tadpole and to a twisted tadpole for the sectors $\alpha=4,8$ located at the singlet. The structure is similar to the standard geometric $\Z_3$ orbifold~\cite{Gimon_1996_K3} which is consistent with the fact that it is a subgroup of the $\Z_{12}$ action. While the first line of \Cref{eq:Kleintransverse} displays the coupling of the planes to the untwisted closed strings, the second line shows the coupling to the $\Z_3$ twisted sectors. Indeed the first term "$4\xi_4[^\alpha_2]$" shows a coupling to the sector living at the quadruplets and the rest of the terms show a coupling to the sector living at the singlet.

After the orientifold projection, the unoriented closed-string spectrum consists of the six-dimensional $\mathcal N=(1,0)$ supergravity multiplet, together with $7$ tensor multiplets and $14$ hypermultiplets. The projection acts differently on the different $\Z_{12}$ orbits. In the untwisted sector it keeps the $\mathcal N=(1,0)$ supergravity multiplet and one neutral hypermultiplet. In the twisted sectors, the sectors paired by the orientifold action give one hypermultiplet and one tensor multiplet for each paired orbit. Thus the singlet orbits receive $(H+T)$ contributions from the pairs $(1,5)$, $(7,11)$, $(2,10)$ and $(4,8)$. The same mechanism applies to the two quadruplet orbits arising from the paired $\alpha=4,8$ sectors, giving one $(H+T)$ contribution for each quadruplet orbit.

The sectors invariant under the orientifold action contribute according to their Klein bottle eigenvalue. Even, or symmetrized, combinations give hypermultiplets, whereas odd, or antisymmetrized, combinations give tensor multiplets. The sectors $\alpha=3$ and $\alpha=9$ are symmetrized, giving two hypermultiplets associated with the singlet orbits and two further hypermultiplets associated with the triplet orbits. The sector $\alpha=6$ is antisymmetrized on the singlet orbit, giving one tensor multiplet, while its triplet and two sextuplet orbits are symmetrized and therefore give three additional hypermultiplets. The unoriented closed-string spectrum is summarised in \Cref{tab:Z12-unoriented-closed-spectrum}.

In the untwisted sector, the orientifold projection removes one of the R-R scalars of the oriented theory and leaves invariant the dilaton together with three R-R scalars, including the axion. These four real scalars form the neutral hypermultiplet of the untwisted $\mathcal N=(1,0)$ spectrum.
\begin{table}[htb]
\centering
\small
\renewcommand{\arraystretch}{1.15}
\begin{tabularx}{\textwidth}{p{0.19\textwidth}|p{0.2\textwidth}|p{0.25\textwidth}|>{\raggedright\arraybackslash}X}
Sector(s) & Orbit type & Orientifold projection & Contribution \\
\hline
Untwisted sector & bulk sector & symmetrized & $\mathcal N=(1,0)$ SUGRA $+\,1H$ \\
\hline
$\alpha=(1,5)$, $(7,11)$, $(2,10)$ and $(4,8)$ & $1$ & paired sectors & $4(H+T)$ \\
\hline
$\alpha=(4,8)$ & $4+4$ & paired sectors & $2(H+T)$ \\
\hline
$\alpha=3,9$ & $1$ & symmetrized & $2H$ \\
\hline
$\alpha=3,9$ & $3$ & symmetrized & $2H$ \\
\hline
$\alpha=6$ & $1$ & antisymmetrized & $1T$ \\
\hline
$\alpha=6$ & $3+6+6$ & symmetrized & $3H$ \\
\hline
Total twisted & -- & -- & $13H+7T$ \\
\hline
Total closed spectrum & -- & -- & $\mathcal N=(1,0)$ SUGRA $+\,14H+7T$ \\
\hline
\end{tabularx}
\caption{Closed-string spectrum after the orientifold projection. Paired twisted sectors contribute one hypermultiplet ($H$) and one tensor multiplet ($T$) for each paired orbit. Sectors invariant under the orientifold action contribute according to their Klein bottle eigenvalue: symmetrized combinations give hypermultiplets, while antisymmetrized combinations give tensor multiplets.}\label{tab:Z12-unoriented-closed-spectrum}
\end{table}

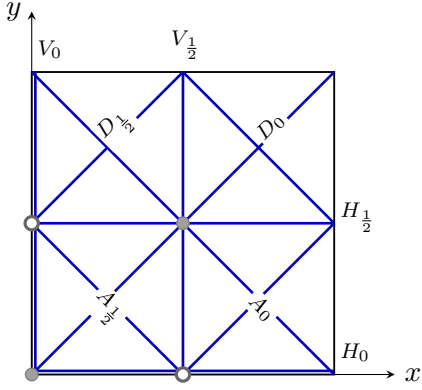
\begin{figure}[htb]
    \centering
    \begin{minipage}[t]{0.475\textwidth}
    \centering
    \vspace{0pt}
    \begin{tikzpicture}[scale=4,>=stealth]

        % Small offset used only to display boundary cycles inside the cell
        \def\eps{0.012}

        \tikzset{
            cell/.style={black, line width=0.7pt},
            ocycle/.style={
                blue!75!black,
                line width=1pt
            },
            fpfull/.style={
                circle,
                fill=gray!75,
                draw=black!50,
                line width=0.5pt,
                inner sep=0pt,
                minimum size=4.7pt
            },
            fpopen/.style={
                circle,
                draw=gray!80!black,
                fill=white,
                line width=1.2pt,
                inner sep=0pt,
                minimum size=5pt
            },
            lab/.style={
                font=\scriptsize,
                fill=white,
                fill opacity=1,
                text opacity=1,
                inner sep=1pt
            }
        }

        % Fundamental cell
        \draw[cell] (0,0) rectangle (1,1);
        \draw[->] (0,0) -- (1.2,0) node[right] {$x$};
        \draw[->] (0,0) -- (0,1.2) node[left] {$y$};

        %------------------------------------------------
        % Horizontal supports H_0 and H_{1/2}
        %------------------------------------------------
        \draw[ocycle] (0,\eps) -- (1,\eps);
        \draw[ocycle] (0,0.5) -- (1,0.5);
        
        %------------------------------------------------
        % Vertical supports V_0 and V_{1/2}
        %------------------------------------------------
        \draw[ocycle] (\eps,0) -- (\eps,1);
        \draw[ocycle] (0.5,0) -- (0.5,1);
        
        %------------------------------------------------
        % Diagonal supports D_0 and D_{1/2}
        % D_c : y = x + c mod 1
        %------------------------------------------------
        \draw[ocycle] (0,0) -- (1,1);
        \draw[ocycle] (0,0.5) -- (0.5,1);
        \draw[ocycle] (0.5,0) -- (1,0.5);
        
        %------------------------------------------------
        % Anti-diagonal supports A_0 and A_{1/2}
        % A_c : y = -x + c mod 1
        %------------------------------------------------
        \draw[ocycle] (0,1) -- (1,0);
        \draw[ocycle] (0,0.5) -- (0.5,0);
        \draw[ocycle] (0.5,1) -- (1,0.5);
        
        %------------------------------------------------
        % Fixed points
        %------------------------------------------------
        \node[fpfull] at (0,0) {};
        \node[fpopen] at (0,0.5) {};
        \node[fpopen] at (0.5,0) {};
        \node[fpfull] at (0.5,0.5) {};

        %------------------------------------------------
        % Labels
        %------------------------------------------------
        \node[lab, anchor=west] at (1.01,0.075) {$H_0$};
        \node[lab, anchor=west] at (1.01,0.52) {$H_{\frac12}$};
        \node[lab, anchor=west] at (0.01,1.075) {$V_0$};
        \node[lab, anchor=west] at (0.45,1.09) {$V_{\frac12}$};
        \node[lab, rotate=45] at (0.79,0.82) {$D_0$};
        \node[lab, rotate=45] at (0.28,0.82) {$D_{\frac12}$};
        \node[lab, rotate=-45] at (0.76,0.22) {$A_0$};
        \node[lab, rotate=-45] at (0.25,0.22) {$A_{\frac12}$};

    \end{tikzpicture}
    \end{minipage}
    \hfill
    \begin{minipage}[t]{0.515\textwidth}
    \centering
    \vspace{0pt}
    \renewcommand{\arraystretch}{1.35}
    \setlength{\extrarowheight}{1pt}
    \newcommand{\supp}[1]{\raisebox{0.4ex}{$#1$}}
    \begin{tabular}{|c|>{\scriptsize}c|}
        \hline
        \normalsize Label & \normalsize Crosscap components \\
        \hline
        \supp{H_0} &
        $\sigma_{(1,0)}^{(0,0)} \text{ and } \sigma_{(1,0)}^{(\frac12,0)}$
        \\
        \hline
        \supp{H_{\frac12}} &
        $\sigma_{(1,0)}^{(0,\frac12)} \text{ and } \sigma_{(1,0)}^{(\frac12,\frac12)}$
        \\
        \hline
        \supp{V_0} &
        $\sigma_{(0,1)}^{(0,0)} \text{ and } \sigma_{(0,1)}^{(0,\frac12)}$
        \\
        \hline
        \supp{V_{\frac12}} &
        $\sigma_{(0,1)}^{(\frac12,0)} \text{ and } \sigma_{(0,1)}^{(\frac12,\frac12)}$
        \\
        \hline
        \supp{D_0} &
        $\sigma_{(1,1)}^{(0,0)} \text{ and } \sigma_{(1,1)}^{(\frac12,\frac12)}$
        \\
        \hline
        \supp{D_{\frac12}} &
        $\sigma_{(1,1)}^{(0,\frac12)} \text{ and } \sigma_{(1,1)}^{(\frac12,0)}$
        \\
        \hline
        \supp{A_0} &
        $\sigma_{(1,-1)}^{(0,0)} \text{ and } \sigma_{(1,-1)}^{(\frac12,\frac12)}$
        \\
        \hline
        \supp{A_{\frac12}} &
        $\sigma_{(1,-1)}^{(0,\frac12)} \text{ and } \sigma_{(1,-1)}^{(\frac12,0)}$
        \\
        \hline
    \end{tabular}
    \end{minipage}
    \caption{Orientifold fixed one-cycles on a single $\T^2$. The labels $H,V,D,A$ denote horizontal, vertical, diagonal and anti-diagonal supports. Each label represents two crosscap components whose lattice positions differ by a half-translation along the wrapped cycle. The complete identification with the components $\sigma_{(p,q)}^{(a/2,b/2)}$ is given in the table, where $(p,q)$ denotes the wrapped cycle and the label $(a/2,b/2)$ records both the transverse half-position of the support and the half-shift along the wrapped cycle, which is invisible in the visible support but appears in the lattice phases. It includes the position normal to the cycle and the dual position or shift along the cycle.}\label{fig:single-torus-crosscap-supports}
\end{figure}

The lattice sums appearing in the tree Klein bottle amplitude determine the orientifold loci and their untwisted charges. The positions and charges can be read from the phases
\begin{align}
    \mathcal{C}_+^{m_i,w_i} &= \frac{1+(-1)^{m_i}}{2}\, \frac{1+(-1)^{w_i}}{2}\,,\\
    \mathcal{C}_-^{m_i,w_i} &= \frac{1-(-1)^{m_i}-(-1)^{w_i}-(-1)^{m_i+w_i}}{2} \,.
\end{align}
In the original frame these contributions may be interpreted in terms of $O9$- and $O5$-plane charges. However, the geometry of the fixed loci is more transparent after the \Tp-duality, where the same crosscap data is described by $O7$-planes wrapping one-cycles at different angles on the two dual two-tori~\cite{Blumenhagen:1999md,
blumenhagen2000newclasssupersymmetricorientifolds}. In this frame the orientifold background is the one shown in \Cref{fig:single-torus-crosscap-supports} and \Cref{tab:O7planesbulkcycles}. The horizontal and vertical fixed cycles form four positively charged $O7^+$ orbits and twelve negatively charged $O7^-$ orbits. In addition, the diagonal fixed cycles give four further $O7^+$ orbits and twelve further $O7^-$ orbits. One can check that these are indeed the fixed loci under the orientifold action $\Omega\sigma\mathcal I_{79}(-1)^{F_L}$.

\begin{table}[htb]
\centering
\renewcommand{\arraystretch}{1.25}
\begin{tabular}{c|c}
    $O$-planes & Wrapped cycle on $\T^2\times\T^2$ \\\hline
    $O7^-$-planes &
    $\begin{aligned}
    \Pi_H^- = & (H_0+H_{\frac12})\circ (D_{\frac12}+\sigma_{(1,1)}^{(\frac12,\frac12)})+
    (V_0+V_{\frac12})\circ (A_{\frac12}+\sigma_{(1,-1)}^{(\frac12,\frac12)}) \\
    \Pi_D^- = & (A_{\frac12}+\sigma_{(1,-1)}^{(\frac12,\frac12)})\circ (H_0+H_{\frac12})+
    (D_{\frac12}+\sigma_{(1,1)}^{(\frac12,\frac12)})\circ (V_0+V_{\frac12})
    \end{aligned}$
    \\\hline
    $O7^+$-planes &
    $\begin{aligned}
    \Pi_H^+ = & (H_0+H_{\frac12})\circ \sigma_{(1,1)}^{(0,0)}+
    (V_0+V_{\frac12})\circ \sigma_{(1,-1)}^{(0,0)} \\
    \Pi_D^+ = & \sigma_{(1,-1)}^{(0,0)}\circ (H_0+H_{\frac12})+
    \sigma_{(1,1)}^{(0,0)}\circ (V_0+V_{\frac12})
    \end{aligned}$
    \\\hline
\end{tabular}
\caption{Wrapped bulk cycles of the $O7^\pm$-plane orbits in the \Tp-dual frame. The symbols $H,V,D,A$ refer to the horizontal, vertical, diagonal and anti-diagonal one-cycle supports defined in \Cref{fig:single-torus-crosscap-supports}. The product $\circ$ denotes the product of one-cycles on the two factors of $\T^2\times\T^2$. The classes $\Pi_H^\pm$ collect the $H\circ D$ and $V\circ A$ orbit contributions, while $\Pi_D^\pm$ collect the $A\circ H$ and $D\circ V$ contributions. The superscript $\pm$ denotes the untwisted $O7^\pm$ charge.}\label{tab:O7planesbulkcycles}
\end{table}

\subsection{Open-string sector}

In order to cancel the tadpoles generated by the Klein bottle amplitude, one needs to introduce an open-string sector. More specifically, one needs to introduce space-filling branes whose R-R charges compensate those of the orientifold planes. The annulus and Möbius strip amplitudes associated with these branes must be consistent with the closed-string data encoded in the Klein bottle amplitude, and their massless spectra must satisfy the six-dimensional anomaly cancellation conditions.

The most natural choice is to put $D7$-branes parallel to the $O7$-planes in the dual picture, at the position $u=0$ on both tori (see \Cref{sec:brane-orbit-origin}). Let $N_0$ be the number of $D7$-branes parallel to the horizontal/vertical $O7$-planes whose orbit will be labelled $\mathcal O_N$, and $M_0$ be the number of $D7$-branes parallel to the diagonal $O7$-planes whose orbit will be labelled $\mathcal O_M$ (see \Cref{fig:single-torus-crosscap-supports}). The orbit $\mathcal O_N$ corresponds to the bulk cycles
\begin{equation}
    \Pi_N = \sigma_{(1,0)}^{(0,0)}\circ\sigma_{(1,1)}^{(0,0)} + \sigma_{(0,1)}^{(0,0)}\circ\sigma_{(1,-1)}^{(0,0)}\,,
\end{equation}
while the orbit $\mathcal O_M$ corresponds to the bulk cycles
\begin{equation}
    \Pi_M = \sigma_{(1,-1)}^{(0,0)}\circ\sigma_{(1,0)}^{(0,0)} + \sigma_{(1,1)}^{(0,0)}\circ\sigma_{(0,1)}^{(0,0)}\,.
\end{equation}
The annulus amplitude for these branes is given by the following expression
\begin{align}
   \mathcal{A} =& \frac16\Big[T[^0_0](
    N_0^2\Lambda^{\mathcal{A} (1,0)}_{m_1,w_1}\Lambda^{\mathcal{A} (1,1)}_{m_2,w_2}(T,U)+
    M_0^2\Lambda^{\mathcal{A} (1,1)}_{m_1,w_1}\Lambda^{\mathcal{A} (1,0)}_{m_2,w_2}(T,U)  ) \notag\\
    &+\sum_{\beta=1}^5 (N_\beta^2+M_\beta^2)T[^0_{2\beta}] \notag\\
   &+ 2\sum_{\beta=0}^5 N_\beta M_\beta(T[^3_{2\beta}]+T[^9_{2\beta}]) \notag\\
   &+ 2(N_0^2+M_0^2)T[^6_{0}] + 2(N_3^2+M_3^2)T[^6_{6}]-\sum_{\beta=1,2,4,5}(N_\beta^2+M_\beta^2)T[^6_{2\beta}] \Big],
\end{align}
where $N_\beta$ and $M_\beta$ are the Chan-Paton trace factors under the action of the orbifold.
The tree channel of this amplitude is given by the following expression
\begin{align}
    \widetilde{\mathcal{A}} = \frac{2^{-3}}{3}&\Big[\frac14T[^0_0](
        N_0^2\widetilde{\Lambda}^{(1,0)}_{0,0}\widetilde{\Lambda}^{(1,1)}_{0,0} +
        M_0^2\widetilde{\Lambda}^{(1,1)}_{0,0}\widetilde{\Lambda}^{(1,0)}_{0,0})' \notag\\
        &+\frac12((N_0+M_0)^2\xi_4[^0_0]+(N_0-M_0)^2\xi_4[^0_2]) \notag\\
    &+\sum_{\beta=1}^5((N_\beta-M_\beta)^2\xi_4[^{2\beta}_0]+(N_\beta+M_\beta)^2\xi_4[^{2\beta}_2]) \notag\\
    &+\sum_{\beta\in\{2,4\}}(N_\beta^2+M_\beta^2)(\xi_4[^{2\beta}_0]+\xi_4[^{2\beta}_1]+\xi_4[^{2\beta}_2]+\xi_4[^{2\beta}_3]) \notag\\
    &+ 3(N_3^2+M_3^2)(\xi_4[^6_1]+\xi_4[^6_3]) \Big]\,.
    \label{eq:AnnulusT}
\end{align}
The tree channel of the annulus amplitude directly encodes the closed-string couplings of the boundary states. Schematically, if $|i\rangle\!\rangle$ denotes an Ishibashi state of the preserved chiral algebra,
\begin{equation}
    |B_a\rangle=\sum_i B_a^{\,i}|i\rangle\!\rangle,
    \qquad
    \widetilde{\mathcal A}_{ab}
    =\sum_i B_a^{\,i}B_b^{\,i}\chi_i,
\end{equation}
so that the coefficients of a tree channel character measure the products of the corresponding boundary reflection coefficients. In the present construction, \Cref{eq:AnnulusT} contains only even-twist characters, and the boundary states therefore couple only to the $\Z_6=\langle g^2\rangle$ subsector of the closed theory. We have checked that, within the class of boundary states built from the orbifold eigencharacters $\chi[^\alpha_\beta]$, adding couplings to the genuinely $\Z_{12}$-twisted sectors does not lead to a consistent unoriented open-string theory: after modular transformation one finds either non-integral loop channel annulus multiplicities or Möbius coefficients incompatible with the annulus projection. This excludes such couplings within the present character ansatz, but does not constitute a general no-go theorem for $\Z_{12}$-twisted boundary states. Indeed, the $\chi[^\alpha_\beta]$ package several more elementary representations of the underlying chiral algebra. Resolving them into finer current-algebra characters, and constructing the corresponding Ishibashi states, may allow more general boundary conditions preserving a smaller chiral algebra, as occurs for non-geometric branes in asymmetric orbifolds~\cite{Cardy:1989ir,Brunner_1999,Gaberdiel:2002jr}. We leave this more general boundary-CFT problem for future work.

The tree Möbius amplitude is fixed by factorization: it is the geometric mean of the tree Klein bottle and annulus coefficients, with the appropriate hatted characters implementing the $P$ modular transformation. After transforming back to the loop channel, the resulting coefficients must be compatible with the annulus multiplicities and define a consistent orientifold projection on the open-string Hilbert space. The resulting tree channel Möbius amplitude is given by the following expression
\begin{align}
    \widetilde{\mathcal{M}}
    = \frac13\Bigg[&
    T[^0_0]
    \begin{aligned}[t]
        \bigl(N_0 \mathcal{C}_+(m_1,w_1)\widetilde{\Lambda}^{(1,0)}_{0,0} \mathcal{C}_-(m_2,w_2)\widetilde{\Lambda}^{(1,1)}_{0,0} + \\
        M_0 \mathcal{C}_-(m_1,w_1)\widetilde{\Lambda}^{(1,1)}_{0,0} \mathcal{C}_+(m_2,w_2)\widetilde{\Lambda}^{(1,0)}_{0,0}\bigr)'
    \end{aligned}
    \label{eq:MobiusT}\\
    -&4(N_0+M_0)\widehat{\xi}_4[^0_0]
    +2\sum_{\beta\in\{2,4\}}(N_\beta+M_\beta)
    \left(
        \widehat{\xi}_4[^{2\beta}_0]
        -\widehat{\xi}_4[^{2\beta}_1]
        +3\widehat{\xi}_4[^{2\beta}_2]
        -\widehat{\xi}_4[^{2\beta}_3]
    \right)
    \Bigg]\,.\notag
\end{align}

\subsection{Tadpoles, massless spectrum and consistency checks}
Thus the tadpole cancellation conditions read
\begin{align}
    N_0+M_0 &= 32\,,\\
    N_1-M_1 &= N_2-M_2 = N_3-M_3 =N_4-M_4=N_5-M_5=0\,,\\
    N_2=M_2 &= N_4=M_4=-8\,.
\end{align}
The twisted tadpoles generated by the Klein bottle amplitude require the branes to be located on top of the $O7^+$-planes and fix a real parametrisation of the Chan-Paton traces 
\begin{align}
    N_\beta &= \Tr(\gamma^\beta_N)\,, \\
    M_\beta &= \Tr(\gamma^\beta_M)\,,
\end{align}
where
\begin{align}
    \gamma_N &= \operatorname{diag}\!\left(
        \Un_{n_0},
        \e{2i\pi/6}\Un_{n_1},
        \e{4i\pi/6}\Un_{n_2},
        \e{6i\pi/6}\Un_{n_3},
        \e{8i\pi/6}\Un_{n_4},
        \e{10i\pi/6}\Un_{n_5}
    \right)\,,\\
    \gamma_M &= \operatorname{diag}\!\left(
        \Un_{m_0},
        \e{2i\pi/6}\Un_{m_1},
        \e{4i\pi/6}\Un_{m_2},
        \e{6i\pi/6}\Un_{m_3},
        \e{8i\pi/6}\Un_{m_4},
        \e{10i\pi/6}\Un_{m_5}
    \right)\,.
\end{align}
The orientifold projection further constrains the Chan-Paton matrices with the following relations
\begin{equation}
    n_5=n_1,\qquad n_4= n_2,\qquad
    m_5=m_1,\qquad m_4= m_2.
\end{equation}
The solution of the tadpole condition is given by
\begin{equation}
    n_0=m_0=n_3=m_3=0,
\end{equation}
together with
\begin{equation}
    n_1=m_1,\qquad n_2=m_2,\qquad n_1+n_2=8.
\end{equation}
Finally the Chan-Paton solution can be parametrised by a single integer $n_1$ as follows
\begin{align}
    \gamma_N
    = &\operatorname{diag}\!\left(
        \e{2i\pi/6}\Un_{n_1},
        \e{-2i\pi/6}\Un_{n_1},
        \e{4i\pi/6}\Un_{8-n_1},
        \e{-4i\pi/6}\Un_{8-n_1}
    \right)\,,\\
    \gamma_M
    = &\operatorname{diag}\!\left(
        \e{2i\pi/6}\Un_{n_1},
        \e{-2i\pi/6}\Un_{n_1},
        \e{4i\pi/6}\Un_{8-n_1},
        \e{-4i\pi/6}\Un_{8-n_1}
    \right)\,,
\end{align}

To keep the open-string spectra readable, we shall denote by $F_i$ the fundamental representation of the $i$-th gauge-group factor, by $\overline{F}_i$ its conjugate for unitary factors, by $A_i=\wedge^2 F_i$ and $S_i=\mathrm{Sym}^2 F_i$ the antisymmetric and symmetric representations, and by $\mathrm{Adj}_i$ the adjoint. We shall similarly write $\overline{A}_i=\wedge^2\overline{F}_i$ and $\overline{S}_i=\mathrm{Sym}^2 \overline{F}_i$.
Tensor products with omitted entries are understood to be singlets under the remaining gauge-group factors. For bifundamental representations, we shall use the unified notation
\begin{equation}
B_{ij}=(F_i,F_j), \qquad
B_{i\overline{\jmath}}=(F_i,\overline{F}_j), \qquad
B_{\overline{\imath}j}=(\overline{F}_i,F_j), \qquad
B_{\overline{\imath}\overline{\jmath}}=(\overline{F}_i,\overline{F}_j)\,.
\end{equation}

Finally, we end up with a gauge group ${U(n_1)_1\times U(8-n_1)_2\times U(n_1)_3\times U(8-n_1)_4}$ and a massless spectrum that comprises an $\mathcal{N}=(1,0)$ adjoint gauge vector and a hypermultiplet in the representation
\begin{equation}
B_{12}\oplus B_{34}\oplus B_{1\overline{3}}\oplus B_{\overline{2}4}\oplus B_{\overline{1}3}\oplus B_{2\overline{4}}\oplus A_1\oplus A_2\oplus A_3\oplus A_4 \,.
\end{equation}
The total rank is reduced by a factor of two compared with the standard $\T^4/\Z_2$ orientifold. Rank reduction associated with quantized background data and mixed orientifold-plane charges is a familiar phenomenon in type I compactifications~\cite{Witten:1997bs,Bianchi:1997rf,Angelantonj_2000,AngelantonjBlumenhagen:1999}. In the present model, the effective untwisted charge of each orientifold orbit is reduced by the coexistence of positively and negatively charged O7-planes. In the T-dual description, the diagonal fixed cycles provide a geometric manifestation of the discrete data responsible for this rank reduction.

As a further consistency check, one can compute the anomaly polynomial and verify that it factorizes as required by the six-dimensional Green-Schwarz-Sagnotti mechanism~\cite{Alvarez-Gaume:1983ihn,Green:1984sg,Sagnotti_1992,Erler_1994,Scrucca:1999uz,Scrucca_2000,Kumar_2010}. Let
\begin{equation}
    n\equiv n_1,\qquad m\equiv8-n.
\end{equation}

Using the conventions summarised in \Cref{app:anomaly}, the irreducible non-Abelian anomaly cancels and the remaining non-Abelian anomaly polynomial can be written as~\cite{Kumar_2010,Monnier_2019}
\begin{align}
    I_8^{\mathrm{NA}}
    &=
    \frac18\, a\cdot a\,\bigl(\tr R^2\bigr)^2
    +\frac18\sum_{i,j}
    \frac{b_i\cdot b_j}{\lambda_i\lambda_j}
    \tr F_i^2\,\tr F_j^2
    +\frac14\sum_i
    \frac{a\cdot b_i}{\lambda_i}
    \tr R^2\,\tr F_i^2
    \nonumber\\
    &=
    \frac18
    \left(
        a\,\tr R^2
        +\sum_i\frac{b_i}{\lambda_i}\tr F_i^2
    \right)^2,
\end{align}
where $F_i$ denotes the field strength of the non-Abelian part of the $i$-th gauge factor and $\lambda_i=1$ for the $SU(N)$ factors considered here. Localised anomalies in orbifold field theories, including their distribution among fixed loci, were analysed in \cite{Scrucca:2001eb}.

The anomaly lattice vectors can be read off from the tadpole factorization and are
\begin{equation}
\setlength{\arraycolsep}{3pt}
\begin{array}{
r@{\;}c@{\;}l@{}
r@{,\,}r@{,\,}r@{,\,}r@{,\,}r@{,\,}r@{,\,}r@{,\,}r
@{}l
}
a &=& \bigl(
& -2\sqrt{\tfrac{2}{3}} & 0 & 0 & \tfrac1{\sqrt{3}} & \tfrac1{\sqrt{3}} & 0 & 0 & 0
& \bigr),\\[1mm]
b_1 &=& \bigl(
& \sqrt{\tfrac23} & -\tfrac12 & -\tfrac1{\sqrt{3}} & -\tfrac1{\sqrt{3}} & 0 & -\tfrac{\sqrt{3}}{2} & 0 & 0
& \bigr),\\[1mm]
b_2 &=& \bigl(
& \sqrt{\tfrac23} & \tfrac12 & -\tfrac1{\sqrt{3}} & -\tfrac1{\sqrt{3}} & 0 & \tfrac{\sqrt{3}}{2} & 0 & 0
& \bigr),\\[1mm]
b_3 &=& \bigl(
& \sqrt{\tfrac23} & \tfrac12 & \tfrac1{\sqrt{3}} & 0 & -\tfrac1{\sqrt{3}} & \tfrac{\sqrt{3}}{2} & 0 & 0
& \bigr),\\[1mm]
b_4 &=& \bigl(
& \sqrt{\tfrac23} & -\tfrac12 & \tfrac1{\sqrt{3}} & 0 & -\tfrac1{\sqrt{3}} & -\tfrac{\sqrt{3}}{2} & 0 & 0
& \bigr).
\end{array}
\end{equation}
We arrange the tensor fields in the following order: first the gravity self-dual tensor, then the singlet tensor from the $\Z_2$ twisted sector, the singlet tensor from the $\Z_3$ twisted sector, the two tensors associated with the quadruplet orbits of the $\Z_3$ twisted sector, the singlet tensor from the $\Z_6$ twisted sector, and finally the two singlet tensors from the genuine $\Z_{12}$ twisted sectors.

The components of these vectors have a direct physical interpretation~\cite{Kumar_2010,Monnier_2019}. The relation between such anomaly coefficients, R-R couplings and D-brane/O-plane anomaly inflow is discussed in \cite{Scrucca:1999uz}. In the Green-Schwarz couplings, $a^\alpha$ determines the gravitational coupling of the tensor $B^\alpha$, while $b_i^\alpha$ determines its coupling to $\tr F_i^2$. Since the basis above keeps track of the orbifold-sector origin of each tensor, the vectors therefore identify which twisted sectors couple to each gauge factor. In particular, the last two components of all the $b_i$ vanish: the tensors originating from the genuine $\Z_{12}$ twisted sectors do not couple to the non-Abelian gauge fields of this vacuum. This agrees with the boundary-state analysis, where the corresponding genuinely $\Z_{12}$-twisted representations are absent from the tree annulus amplitudes of the boundary states used in the construction.

The diagonal Abelian factors require a slightly more detailed analysis, which is given in \Cref{app:abelian-anomalies}. For $0<n,m$, the mixed $U(1)-SU(N)^3$ anomaly matrix has rank three. Its one-dimensional kernel singles out the combination
\begin{equation}
    U(1)_X
    =
    -m\bigl(U(1)_1+U(1)_3\bigr)
    +n\bigl(U(1)_2+U(1)_4\bigr).
\end{equation}
The other three independent Abelian combinations participate in the generalized Green-Schwarz mechanism and are necessarily Stückelberg massive. Thus $U(1)_X$ is the only Abelian factor that can possibly remain massless~\cite{Scrucca:1999zh,Antoniadis:2002cs}. For $n=0$ or $m=0$, no analogous anomaly-compatible Abelian factor remains.

The explicit anomaly vectors also allow a direct test of the Kim-Shiu-Vafa string-probe constraint~\cite{Kim_2019, Angelantonj:2020pyr}. Let $Q$ denote the charge of a BPS string and define
\begin{equation}
    k_i=Q\cdot b_i\in\Z_{\geq0},
    \qquad
    s=-Q\cdot a\in\Z_{\geq0}.
\end{equation}
The vectors above satisfy
\begin{equation}
    b_1+b_3=b_2+b_4=-a,
\end{equation}
and therefore
\begin{equation}
    s=k_1+k_3=k_2+k_4.
\end{equation}
Using the conventions of \Cref{app:KSV-conventions}, the left-moving central charge and transverse $SU(2)_L$ level are
\begin{equation}
    c_L=3Q^2+9s+2,
    \qquad
    k_L=\frac12\left(Q^2-s+2\right)\geq0.
\end{equation}
For $0<n,m$, the non-Abelian current algebra satisfies
\begin{align}
    c_{\mathrm{KM}}^{\mathrm{NA}}
    &=
    \sum_i
    \frac{k_i\dim(G_i)}{k_i+h_i^\vee}
    \nonumber\\
    &\leq
    (n-1)(k_1+k_3)
    +(m-1)(k_2+k_4)
    =6s.
\end{align}
The $U(1)_X$ abelian factor contributes at most one additional unit to the current-algebra central charge~\cite{Lee:2019skh}. Hence
\begin{equation}
    c_{\mathrm{KM}}\leq6s+1.
\end{equation}
For $s\geq1$, positivity of $k_L$ gives $Q^2\geq s-2$, and therefore
\begin{equation}
    c_L\geq12s-4.
\end{equation}
It follows that
\begin{equation}
    c_{\mathrm{KM}}
    \leq6s+1
    \leq12s-4
    \leq c_L.
\end{equation}
At the endpoint ranks $n=0$ or $m=0$, only two $SU(8)$ factors remain and instead
\begin{equation}
    c_{\mathrm{KM}}\leq7s\leq12s-4\leq c_L.
\end{equation}
Finally, if $s=0$, all non-Abelian levels vanish implying directly that $c_{\mathrm{KM}}=0$. Since $a$ is characteristic, ${Q^2\equiv Q\cdot a=0\pmod2}$, while the BPS condition gives $Q^2\geq-1$. Hence $Q^2\geq0$ and
\begin{equation}
    c_L=3Q^2+2\geq2.
\end{equation}
The KSV current-algebra bound is therefore satisfied for all BPS string charges obeying the standard positivity conditions.

This vacuum is the simplest one that can be constructed for this orientifold. In principle, one may split the brane configuration in order to break the gauge group. The tadpole conditions, however, strongly constrain such splittings: the branes must be organised into orbits that are at least invariant under the $\Z_3$ subgroup. This follows from the presence of non-vanishing $\Z_3$-twisted Klein bottle tadpoles, whose cancellation requires corresponding twisted Chan-Paton charges. A generic bulk brane carries only untwisted charge and therefore cannot cancel these contributions.

Consequently, moving branes away from $u=0$ does not define an ordinary continuous position modulus. Instead, one obtains distinct disconnected vacua by redistributing the branes among the different $\Z_3$-invariant positions described in \Cref{sec:brane-orbit-Z3-Z2}. In particular, the tadpole solution contains no freely movable regular Chan-Paton block that could be continuously displaced into the bulk. While the branes at $u=0$ considered above give rise to products of unitary gauge groups, the other $\Z_3$-invariant positions lead to orthogonal gauge-group factors.

The continuous open-string deformations are therefore not ordinary translations of individual branes, but Higgs branches generated by vacuum expectation values of hypermultiplet scalars. Geometrically, these deformations describe brane recombination: several fractional constituents combine into a different admissible brane configuration while preserving the total R-R charges.

\subsection{Non-supersymmetric vacua}

Orientifold vacua also provide a natural setting for non-supersymmetric configurations with Brane Supersymmetry Breaking (BSB). In these constructions, the closed-string bulk preserves supersymmetry, while R-R tadpole cancellation requires branes whose worldvolume sector does not preserve the same supersymmetry as the orientifold background \cite{Antoniadis:1999xk,Angelantonj:1999ms}. The ten-dimensional $USp(32)$ model of Sugimoto provides the canonical example \cite{Sugimoto:1999tx}; reviews and further developments can be found in \cite{mourad2017updatebranesupersymmetrybreaking,
dudas2026supersymmetrybreakingfieldsstrings}.

A first non-supersymmetric variant of the present orientifold is obtained by reversing the R-R charges of all the $O7$-planes of the supersymmetric background, thus exchanging the $O^+$ and $O^-$ assignments in \Cref{tab:O7planesbulkcycles}. Tadpole cancellation then requires antibranes instead of branes, in direct analogy with the Sugimoto construction. The closed-string massless spectrum remains supersymmetric, whereas supersymmetry is broken in the open-string sector. In the present model, the reversal of the crosscap charges modifies the Möbius projection and exchanges the symmetric and antisymmetric Chan-Paton projections for the corresponding bosonic states. The resulting open-string spectrum is tachyon-free and satisfies the six-dimensional anomaly cancellation conditions.

More general BSB vacua can be obtained when only part of the orientifold-plane charge assignment is reversed. Such compact constructions, in which tadpole cancellation requires mutually non-supersymmetric brane and orientifold sectors, were studied in \cite{Angelantonj:1999ms,Angelantonj:2000xf}; more recently, twisted R-R tadpoles were used to construct particularly rigid BSB vacua in \cite{Angelantonj:2024iwi}. We shall implement an analogous partial charge reversal by switching only the charges of the $O7$-planes wrapping the diagonal cycles of \Cref{tab:O7planesbulkcycles}, as summarised in \Cref{tab:O7planesbulkcyclesBSB}.

The unoriented closed-string spectrum is now made of the six-dimensional $\mathcal{N}=(1,0)$ supergravity multiplet together with 10 hypermultiplets and 11 tensor multiplets (see \Cref{tab:Z12BSB-unoriented-closed-spectrum} for the detailed spectrum and \Cref{app:BSB} for more details on the BSB amplitudes). The tadpole contributions are now given by
\begin{align}
    \widetilde{\mathcal{K}}_0 =& \frac{2^3}{3}\Big[6 \chi[^4_0] + 6\chi[^8_0] \Big] \,.
\end{align}

\begin{table}[htb]
\centering
\renewcommand{\arraystretch}{1.25}
\begin{tabular}{c|c}
    $O$-planes & Wrapped cycle on $\T^2\times\T^2$ \\\hline
    $O7^-$-planes &
    $\begin{aligned}
    \Pi_H^- = & (H_0+H_{\frac12})\circ (D_{\frac12}+\sigma_{(1,1)}^{(\frac12,\frac12)})+
    (V_0+V_{\frac12})\circ (A_{\frac12}+\sigma_{(1,-1)}^{(\frac12,\frac12)}) \\
    \Pi_D^- = & \sigma_{(1,-1)}^{(0,0)}\circ (H_0+H_{\frac12})+
    \sigma_{(1,1)}^{(0,0)}\circ (V_0+V_{\frac12})
    \end{aligned}$
    \\\hline
    $O7^+$-planes &
    $\begin{aligned}
    \Pi_H^+ = & (H_0+H_{\frac12})\circ \sigma_{(1,1)}^{(0,0)}+
    (V_0+V_{\frac12})\circ \sigma_{(1,-1)}^{(0,0)} \\
    \Pi_D^+ = & (A_{\frac12}+\sigma_{(1,-1)}^{(\frac12,\frac12)})\circ (H_0+H_{\frac12})+
    (D_{\frac12}+\sigma_{(1,1)}^{(\frac12,\frac12)})\circ (V_0+V_{\frac12})
    \end{aligned}$
    \\\hline
\end{tabular}
\caption{Wrapped bulk cycles of the $O7^\pm$-plane orbits in the \Tp-dual frame of the BSB model.}\label{tab:O7planesbulkcyclesBSB}
\end{table}

Note that there are no untwisted tadpoles; only twisted ones remain~\cite{Rabad_n_2001}. In order to cancel these tadpoles, one needs to introduce $D7$-branes at the same positions as in the supersymmetric case, but now there are $N_0$ branes parallel to the horizontal/vertical $O7$-planes and $M_0$ antibranes parallel to the diagonal $O7$-planes. While the closed sector remains supersymmetric, supersymmetry is broken in the open sector. The consistency of coupling such a non-supersymmetric brane sector to the supersymmetric bulk, together with the corresponding non-linear realisation of local supersymmetry at low energies, was studied in \cite{Dudas_2001,Pradisi_2001}, including for six-dimensional $\mathcal N=(1,0)$ BSB models.

\begin{table}[htb]
\centering
\small
\renewcommand{\arraystretch}{1.15}
\begin{tabularx}{\textwidth}{p{0.19\textwidth}|p{0.2\textwidth}|p{0.25\textwidth}|>{\raggedright\arraybackslash}X}
Sector(s) & Orbit type & Orientifold projection & Contribution \\
\hline
Untwisted sector & bulk sector & symmetrized & $\mathcal N=(1,0)$ SUGRA $+\,1H$ \\
\hline
$\alpha=(1,5)$, $(7,11)$, $(2,10)$ and $(4,8)$ & $1$ & paired sectors & $4(H+T)$ \\
\hline
$\alpha=(4,8)$ & $4+4$ & paired sectors & $2(H+T)$ \\
\hline
$\alpha=3,9$ & $1$ & antisymmetrized & $2T$ \\
\hline
$\alpha=3,9$ & $3$ & antisymmetrized & $2T$ \\
\hline
$\alpha=6$ & $1$ & antisymmetrized & $1T$ \\
\hline
$\alpha=6$ & $3+6+6$ & symmetrized & $3H$ \\
\hline
Total twisted & -- & -- & $9H+11T$ \\
\hline
Total closed spectrum & -- & -- & $\mathcal N=(1,0)$ SUGRA $+\,10H+11T$ \\
\hline
\end{tabularx}
\caption{Unoriented closed-string spectrum after the BSB orientifold projection. Compare with the SUSY case in~\Cref{tab:Z12-unoriented-closed-spectrum}.}\label{tab:Z12BSB-unoriented-closed-spectrum}
\end{table}

The open-string spectrum is a non-supersymmetric gauge theory with gauge group
\begin{equation}
    \begin{aligned}
     [SO(r+2t)_1\times SO(r-2t)_2\times U(r+s-4)_3\times U(r-s-4)_4]_{D7}
     &\,\times\\
     [USp(r-2t)_5\times USp(r+2t)_6 \times U(r-s-4)_7\times U(r+s-4)_8]_{\overline{D7}}&\,,
    \end{aligned}
\end{equation} 
and with the following matter content 
\begin{itemize}
    \item Sector $D7-D7$: SUSY $\mathcal{N}=(1,0)$
    \begin{align}
        V\sim& A_1\oplus A_2\oplus \mathrm{Adj}_3\oplus \mathrm{Adj}_4 \\
        H\sim& A_3\oplus A_4\oplus B_{1\overline{3}}\oplus B_{1\overline{4}}\oplus B_{2\overline{3}}\oplus B_{2\overline{4}}\oplus B_{3\overline{4}}
    \end{align}
    \item Sector $\overline{D7}-\overline{D7}$: non-SUSY
    \begin{align}
        A_\mu \sim& S_5\oplus S_6\oplus \mathrm{Adj}_7\oplus \mathrm{Adj}_8 \\
        \lambda_L\sim& A_5\oplus A_6\oplus \mathrm{Adj}_7\oplus \mathrm{Adj}_8 \\
        4\phi\sim& S_7\oplus S_8\oplus B_{5\overline{7}}\oplus B_{5\overline{8}}\oplus B_{6\overline{7}}\oplus B_{6\overline{8}}\oplus B_{7\overline{8}}\\
        \lambda_R\sim& A_7\oplus A_8\oplus B_{5\overline{7}}\oplus B_{5\overline{8}}\oplus B_{6\overline{7}}\oplus B_{6\overline{8}}\oplus B_{7\overline{8}}
    \end{align}
    \item Sector $D7-\overline{D7}$: non-SUSY + tachyonic with $m^2=-\frac1{4}$
    \begin{align}
        \phi\sim& B_{1\overline{8}}\oplus B_{2\overline{7}}\oplus B_{37}\oplus B_{48}\oplus B_{5\overline{4}}\oplus B_{6\overline{3}}\oplus \mathrm{c.c.} \\
        \lambda_L\sim&2(B_{15}\oplus B_{26})\oplus 2(B_{3\overline{7}}\oplus B_{4\overline{8}}\oplus\mathrm{c.c.}) \\
        T\sim& B_{1\overline{7}}\oplus B_{2\overline{8}}\oplus B_{3\overline{8}}\oplus B_{4\overline{7}}\oplus B_{5\overline{3}}\oplus B_{6\overline{4}}\oplus \mathrm{c.c.}
    \end{align}
\end{itemize}

The spectrum is generically tachyonic because of the presence of branes and antibranes at angles, although it is free of irreducible gauge and gravitational anomalies. A distinguished minimal branch occurs for $s=0$ and $r=4$. This branch contains no open-string tachyons and has gauge group $[SO(4+2t)_1\times SO(4-2t)_2]_{D7}\times [USp(4-2t)_5\times USp(4+2t)_6]_{\overline{D7}}$. It is natural to interpret this branch as an endpoint of brane-antibrane recombination driven by tachyon condensation. To make this structure explicit, let us introduce the Chan-Paton charge vectors
\begin{align}
    v_N&=(n_0,n_1,n_2,n_3,n_4,n_5)\,,\\
    v_M&=(m_0,m_1,m_2,m_3,m_4,m_5)\,.
\end{align}
Using the tadpole solution, they can be parametrised as
\begin{align}
    v_N&=(r+2t,r+s-4,r-s-4,r-2t,r-s-4,r+s-4)\,,\\
    v_M&=(r-2t,r-s-4,r+s-4,r+2t,r+s-4,r-s-4)\,.
\end{align}
These charges naturally arrange into $\Z_3$ packets $c_0 =(1,0,0,1,0,0)$, $c_1=(0,1,0,0,0,1)$, $c_2=(0,0,1,0,1,0)$ and $d_0=(1,0,0,-1,0,0)$. There is also the regular brane representation $v_R=(1,1,1,1,1,1)$. Defining $r_{\mathrm{min}}=\max(4+|s|,2|t|)=r-k_R$, the Chan-Paton charges can be rewritten as
\begin{align}
    v_N=& k_R v_R + (r_{\mathrm{min}}c_0+2t d_0+s(c_1-c_2)) \,,\\
    v_M=& k_R v_R + (r_{\mathrm{min}}c_0-2t d_0-s(c_1-c_2))\,.
\end{align}
These expressions show that there are several branches of tachyon condensation which can lead to different recombination or annihilation channels for branes and antibranes. We can study one specific branch with $s=0$ and $|t|\leq2$. Then the Chan-Paton charges can be rewritten as
\begin{align}
    v_N=& k_R v_R + (4c_0+2t d_0) \,,\\
    v_M=& k_R v_R + (4c_0-2t d_0)\,.
\end{align}
This branch consists of a fractional core $(4c_0\pm2td_0)$ together with $k_R$ regular brane-antibrane pairs. The regular pairs can annihilate through open-string tachyon condensation \cite{Sen_1998,Sen:1999md}, leaving the fractional core as the tachyon-free configuration. This analysis isolates one particular condensation channel, namely $s=0$ with $|t|\leq2$. Determining the endpoint of a generic $(r,s,t)$ configuration would instead require the full tachyon potential in order to identify the dynamically preferred recombination channel. Related aspects of tachyon dynamics in non-supersymmetric string vacua are reviewed in \cite{Leone_2026,dudas2026supersymmetrybreakingfieldsstrings}. We shall not pursue this dynamical problem here.
The purpose of this subsection is instead to record how the orientifold signs and Chan-Paton projections are modified when one reverses part of the orientifold-plane charge assignment. Nevertheless, even the non-tachyonic branch carries a non-vanishing NS-NS tadpole and therefore does not solve the classical bulk equations around the original background. Explicit backreacted solutions in non-supersymmetric string models with dilaton tadpoles were studied in \cite{Dudas_2000,Blumenhagen_2001}, while the more general issue of background redefinition in the presence of tadpoles is discussed in \cite{Fischler:1986ci,Fischler:1986tb,Dudas_2005}. The resulting backreaction and dilaton potential are beyond the scope of the present work.

The final noteworthy feature of this BSB vacuum is that, at the non-tachyonic point $s=0$ and $r=4$, the open spectrum is free of any massless scalar. This branch is therefore particularly rigid as a consequence of the twisted R-R tadpole conditions, in close analogy with the rigid BSB vacua of \cite{Angelantonj:2024iwi}. In the present model, any continuous brane fluctuation would violate the tadpole conditions. 

\subsection{Freely acting five-dimensional orientifold}

We discussed the fact that the quasicrystalline orbifold vacua have a very rigid structure and consequently very few untwisted moduli. However, there are still many moduli coming from the twisted sectors.

To lift these twisted-sector moduli, one can further compactify on a circle and implement a Scherk-Schwarz reduction. The mechanism provides a standard way of breaking supersymmetry spontaneously through twisted boundary conditions along a compact direction \cite{Scherk:1978ta,Scherk:1979zr}. Its implementation in string compactifications was developed in \cite{Rohm:1983aq,Kounnas:1988ye,Ferrara:1987es,Ferrara:1988jx}. Freely acting orbifolds provide a worldsheet realisation of this mechanism, including in unoriented and open-string constructions \cite{Scrucca_2001,Antoniadis:1998ki,Scrucca:2002is}. This also gives masses to the twisted moduli. However, unlike standard freely acting shifts along lightlike lattice directions, where part of the supersymmetry is restored in one of the infinite-radius limits, supersymmetry is not restored in either radius limit in our model. Indeed, the freely acting $\Z_{12}$ orbifold preserves sixteen supercharges, while the orientifold projection leaves eight supercharges in five dimensions; both radius limits preserve the same eight supercharges after the orientifold projection. Before the orientifold projection, the freely acting construction gives five-dimensional $\mathcal N=2$ supergravity coupled to one vector multiplet. The spectrum therefore contains only the massless dilaton, four R-R scalars and the radius of the circle. While the freely acting shift of the extra circle must satisfy modular invariance and level matching conditions for the oriented theory to be consistent, the orientifold projection further constrains the possible choices of the shift. Indeed, for a geometric rotation on the torus, the shift must be along the momentum direction of the circle Narain lattice in order to be mutually compatible with the parity projection. For an antisymmetric rotation, the shift must be along the winding direction of the circle Narain lattice~\cite{Antoniadis_1998,Antoniadis:1998ep}. Thus in the case of the $\Z_{12}$ orientifold, the shift in the extra circle must combine a momentum and a winding shift in order to reproduce the structure of the asymmetric orbifold. Since $\Z_{12}$ decomposes as an order 3 geometric rotation and an order 4 antisymmetric rotation, the shift must be a combination of a momentum shift of order 3 and a winding shift of order 4. The generator $g$ of the orbifold group on $\T^4\times S^1$ is then given by
\begin{equation}
    g
    \begin{cases}
        Z^1_L \to \exp(2i\pi/12)Z^1_L\,, &\quad Z^1_R \to \exp(14i\pi/12)Z^1_R\,, \\
        Z^2_L \to \exp(-2i\pi/12)Z^2_L\,, &\quad Z^2_R \to \exp(-14i\pi/12)Z^2_R\,, \\
        Y_L \to Y_L + \frac{\pi R}{3} + \frac{\pi}{4R}\,, &\quad Y_R \to Y_R + \frac{\pi R}{3} - \frac{\pi}{4R}\,, \\
    \end{cases}
\end{equation}
where $Y$ is the coordinate of the extra circle of radius $R$. 

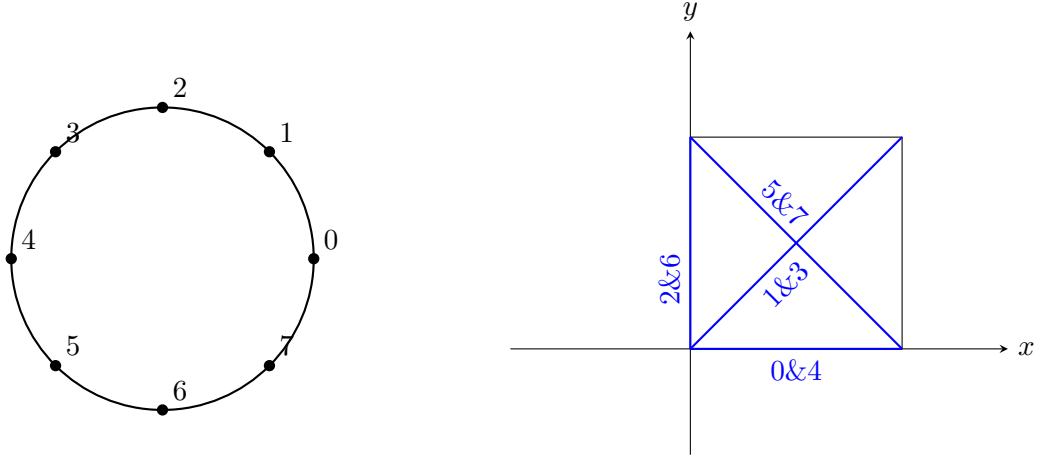
\begin{figure}
\begin{minipage}[t]{0.475\textwidth}
    \vspace{1cm}
    \centering
    \begin{tikzpicture}[scale=2,>=stealth]
        % Circle
        \draw[thick] (0,0) circle(1);

        % Angles in radians
        \foreach \angle in {0,1,2,3,4,5,6,7} {
        % Point position
        \node[circle,fill=black, inner sep=1.5pt] at ({cos(45*\angle)}, {sin(45*\angle)}) {}
        ;
        \node[above right] at ({cos(45*\angle)}, {sin(45*\angle)}) {$\angle$}
        ;}
    \end{tikzpicture}
\end{minipage}
\hfill
\begin{minipage}[t]{0.475\textwidth}
    \vspace{0cm}
    \centering
    \begin{tikzpicture}[scale=2.8,>=stealth]
        % Axes
        \draw[->] (-.85, 0) -- (1.5, 0) node[right] {$x$};
        \draw[->] (0, -0.5) -- (0, 1.5) node[above] {$y$};

        \draw[-] (0,0) -- (0,1) -- (1,1) -- (1,0) -- cycle;

        % Planes
        \draw[-, thick, blue] (0,0) -- (1,0) node[midway, below] {$0 \& 4$};
        \draw[-, thick, blue] (0,0) -- (0,1) node[midway, above left, rotate=90] {$2 \& 6$};
        \draw[-, thick, blue] (0,0) -- (1,1) node[midway, below left, rotate=45] {$1 \& 3$};
        \draw[-, thick, blue] (1,0) -- (0,1) node[midway, above left, rotate=-45] {$5 \& 7$};
    \end{tikzpicture}
\end{minipage}
    \caption{Schematic structure of the $O6$-plane configuration on the T-dual circle and the $\Tp$ torus. The plane at position $i$ on the circle corresponds to the cycle labelled by $i$ on the torus.}\label{fig:S1BG}
\end{figure}

With this construction, the resulting five-dimensional orientifold retains, along the internal $\T^4$ directions, the same orientifold-plane structure as in the six-dimensional model. The new ingredient is the freely acting shift along the additional circle: in the T-dual description, the different orientifold-plane components are localised at distinct positions along the dual circle, as illustrated in \Cref{fig:S1BG}. At a generic finite radius, the unoriented closed-string spectrum consists of five-dimensional $\mathcal N=1$ supergravity together with one neutral vector multiplet and one neutral hypermultiplet. The orientifold planes of this model carry no twisted R-R charge. Related freely acting orientifold constructions in which orientifold planes couple only to massive twisted states were recently studied in \cite{bossard2024twistedorientifoldplanessduality}.
The freely acting shift lifts the closed-string states which would generate twisted R-R tadpoles, so that only an untwisted massless R-R tadpole is present. Its cancellation fixes the total Chan-Paton rank carried by the two D-brane stacks $N$ and $M$. For the branch considered here, the branes are distributed between the two allowed boundary orbits and the resulting gauge group is
\begin{equation}
G_{\rm open}=U(4)_N\times U(4)_M\,.
\end{equation}
Each factor carries the standard five-dimensional $\mathcal N=1$ vector multiplet. At special relative boundary-crosscap configurations, brane-image strings may also remain massless and give a hypermultiplet in a two-index representation of $U(4)$, either antisymmetric or symmetric depending on the Möbius projection. For a generic displacement along the T-dual circle, these brane-image strings acquire a position-dependent mass and the corresponding two-index multiplets are lifted.

\begin{table}[htbp]
\centering
\renewcommand{\arraystretch}{1.4}
\begin{tabularx}{\textwidth}{@{}p{0.13\textwidth}XX@{}}
\hline
\textbf{Radius} & \textbf{Closed sector} & \textbf{Open sector} \\
\hline

$0<R<\infty$
&
5d $\mathcal N=1$

\par
$G+V+H$
&
5d $\mathcal N=1$

\par
$U(4)_N\times U(4)_M$

\par
$V:\ (\mathbf{16},1)+(1,\mathbf{16})$

\par
$H:\ (\mathbf{6/10},1)+(1,\mathbf{6/10})$
\\[1mm]

\hline

$R\to0$
&
Type IIA, symmetric $\Z_3$

\par
Covering bulk: 6d $\mathcal N=(1,1)$

\par
$G+20V$

\par
After $\Omega I_{\widetilde y}$: 5d $\mathcal N=1$

\par
$G+11V+11H$
&
5d $\mathcal N=1$ defect

\par
$U(4)$ locally

\par
$V:\ \mathbf{16}$

\par
$H:\ \mathbf{6/10}$
\\[1mm]

\hline

$R\to\infty$
&
Type IIB, antisymmetric $\Z_4$
\par
6d $\mathcal{N}=(1,0)$
\par
$G+T+20H$
\par
&
6d $\mathcal N=(1,0)$

\par
$G_{\rm open}=USp(8)^4$

\par
\(
\begin{aligned}
V:\quad&
(\mathbf{36},1,1,1)+(1,\mathbf{36},1,1)\\
&+(1,1,\mathbf{36},1)+(1,1,1,\mathbf{36})
\end{aligned}
\)

\par
\(
\begin{aligned}
H:\quad&
(\mathbf{28},1,1,1)+(1,\mathbf{28},1,1)\\
&+(1,1,\mathbf{28},1)+(1,1,1,\mathbf{28})\\
&+(\mathbf8,\mathbf8,1,1)+(\mathbf8,1,\mathbf8,1)\\
&+(1,\mathbf8,1,\mathbf8)+(1,1,\mathbf8,\mathbf8)
\end{aligned}
\)
\\

\hline
\end{tabularx}

\caption{Massless spectra of the five-dimensional $\Z_{12}$ orientifold and its two decompactification limits. Note that for symplectic groups, the antisymmetric representation is reducible and decomposes as $\mathbf{28}=\mathbf{27}+\mathbf{1}$.}
\label{tab:Z12_spectra_limits}
\end{table}

The two extreme-radius limits are qualitatively different and correspond to two distinct six-dimensional decompactification limits of the same five-dimensional theory. In the $R\to0$ limit, the T-dual radius diverges as $\widetilde R=\frac{1}{R}\longrightarrow\infty$, and the $g^4$- and $g^8$-twisted closed-string sectors become massless. The surviving orbifold subgroup is therefore $\langle g^4\rangle\simeq\Z_3$. Since the tower which becomes continuous is a winding tower in the original type IIB variables, the corresponding six-dimensional bulk theory is type IIA on the geometric orbifold $\T^4/\Z_3$, with spectrum $\mathcal N=(1,1)$ supergravity together with 20 vector multiplets. The orientifold projection becomes of the form $\Omega \sigma I_{\widetilde y}$ and preserves only eight supercharges, so that the O-planes and D-branes are five-dimensional defects localised along the decompactified dual direction. The orientifolded massless sector may equivalently be organised in five-dimensional $\mathcal N=1$ supergravity, eleven vector multiplets and eleven hypermultiplets. This spectrum is T-dual to the circle reduction of the ordinary type IIB symmetric $\Z_3$ orientifold spectrum made of six-dimensional $\mathcal{N}=(1,0)$ supergravity, ten tensors and eleven hypermultiplets.

An additional feature of the $R\to0$ limit is that the $g^4$- and $g^8$-twisted R-R fields, which were massive at finite radius, become massless. Their one-point functions are then interpreted in the type IIA description as sources for non-dynamical six-dimensional top-form field strengths. A generic non-compact domain-wall configuration may therefore support non-zero twisted top-form flux. In the branch considered here, however, we require the limiting background to remain on the fluxless perturbative orientifold branch. This zero twisted-flux condition fixes the relative distribution of the $N$ and $M$ stacks at the allowed orientifold loci and removes the remaining freedom which was present at finite radius. In the strict decompactification limit, the two boundary orbits are separated by a proper distance of order $\widetilde R$ and therefore become infinitely far apart. In a local patch centred on one orientifold defect, only the corresponding $U(4)$ open-string sector remains at finite distance, while the second stack is sent to infinity. Its particle states decouple locally, although its net source data can remain encoded in the asymptotic values of the top-form fluxes.

The opposite limit, $R\to\infty$, is controlled by an ordinary momentum decompactification. The sectors with $\alpha=3,6,9$ become massless and reconstruct the asymmetric $\Z_4$ orbifold generated by $g^3$. The resulting six-dimensional type IIB orientifold has closed unoriented spectrum $\mathcal N=(1,0)$ supergravity, one tensor and twenty hypermultiplets. At the same time, all open-string momentum splittings induced by the finite-radius construction vanish as $1/R$, so that the two finite-radius gauge factors enhance according to
\begin{equation}
U(4)_N\times U(4)_M
\quad\xrightarrow{R\to\infty}\quad
[USp(8)^2]_N\times [USp(8)^2]_M\,.
\end{equation}
The full open spectrum then reorganises into that of the supersymmetric symplectic branch of the six-dimensional antisymmetric $\Z_4$ orientifold.

\begin{figure}[htbp]
\centering
\begin{tikzpicture}[>=stealth,scale=1]
    % Main interpolation line
    \draw[thick,->] (0,0) -- (11,0);

    % Left endpoint
    \filldraw (0,0) circle (2pt);
    \node[above=6pt] at (0,0) {$R\to 0$};
    \node[below=6pt,align=center] at (0,0) {IIA symmetric $\Z_3$\\orientifold};

    % Middle point
    \filldraw (5.5,0) circle (2pt);
    \node[above=6pt] at (5.5,0) {generic finite $R$};
    \node[below=6pt,align=center] at (5.5,0) {$5d$ IIB freely acting $\Z_{12}$\\orientifold};

    % Right endpoint
    \filldraw (11,0) circle (2pt);
    \node[above=6pt] at (11,0) {$R\to \infty$};
    \node[below=6pt,align=center] at (11,0) {IIB antisymmetric $\Z_4$\\orientifold};
\end{tikzpicture}
\caption{Interpolation of the five-dimensional freely acting $\Z_{12}$ orientifold between its two six-dimensional endpoint theories. At small radius the bulk reorganises into the IIA symmetric $\Z_3$ model, while at large radius one recovers the IIB antisymmetric $\Z_4$ model.}\label{fig:RinterpolationZ3Z4}
\end{figure}
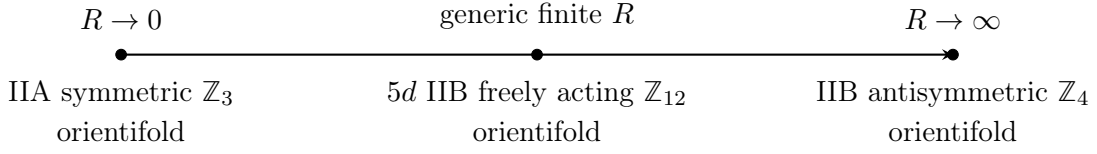

The five-dimensional construction therefore interpolates between two genuinely different six-dimensional limits, as summarised in \Cref{fig:RinterpolationZ3Z4}. At large radius one recovers a type IIB antisymmetric $\Z_4$ orientifold with space-filling open strings, whereas at small radius the winding tower reconstructs a type IIA geometric $\Z_3$ bulk with $\Omega I_{\widetilde y}$ orientifold defects. Requiring vanishing twisted top-form flux selects the perturbative $R\to0$ endpoint, while more general choices would correspond instead to non-compact type IIA domain-wall backgrounds carrying non-zero twisted flux and would require a refined study of the backreaction of these objects.

It is important that this five-dimensional construction is not continuously connected to the six-dimensional $\Z_{12}$ orientifold of the preceding sections: both radius limits approach different six-dimensional theories. Models using freely acting shifts along an extra dimension in order to interpolate between two vacua have already been considered, see for instance~\cite{Blum_1997,Blum_1998}.

\section{Discussion and conclusions}\label{sec:discussion-conclusions}

We have constructed and analysed orientifold descendants of a six-dimensional type IIB $\Z_{12}$ quasicrystalline orbifold. The parent theory is defined by an automorphism of the full Narain lattice which is not crystallographic on either the geometric torus or its T-dual frame. Its twisted sectors nevertheless admit a precise CFT description: the ground-state degeneracies organise into orbits under the full $\Z_{12}$ action, and these orbits determine the projected closed-string spectrum. The resulting oriented vacuum has $\mathcal N=(2,0)$ supersymmetry and twenty-one tensor multiplets, while the orbifold fixes the geometric torus moduli.

The orientifold projection adds a second layer of structure. The crosscap states are supported on several orientifold-plane orbits whose geometry becomes particularly transparent in the partially T-dual frame. The corresponding horizontal, vertical and diagonal cycles are permuted non-trivially by the quasicrystalline action, and their twisted R-R couplings are reflected directly in the tree Klein bottle. Tadpole cancellation therefore cannot be achieved by an arbitrary bulk D-brane configuration. It selects boundary orbits carrying the required twisted charges, which in the partially T-dual description are D7-branes at angles. This mechanism is reminiscent of crystallographic six-dimensional orientifolds with twisted Klein bottle tadpoles~\cite{Blumenhagen:1999md,Angelantonj:2000xf}, but here it is embedded in a parent CFT whose bulk orbifold action is genuinely quasicrystalline.

For the supersymmetric descendant, factorization of the Klein bottle, annulus and Möbius strip amplitudes fixes the Chan-Paton embedding and leads to the family
\begin{equation}
    U(n)_1\times U(8-n)_2\times U(n)_3\times U(8-n)_4,
    \qquad 0\leq n\leq8.
\end{equation}
The reduction of the total Chan-Paton rank is tied to the simultaneous presence of orientifold components with opposite R-R charges and has a natural interpretation in the T-dual cycle description. The massless spectrum passes the six-dimensional irreducible gravitational and gauge-anomaly tests, while the remaining anomaly polynomial factorizes through the seven tensor multiplets of the unoriented closed sector. The Abelian anomaly analysis singles out one possible anomaly-compatible combination $U(1)_X$ for generic rank, although anomaly cancellation by itself does not determine whether this combination is free of Stückelberg couplings. The same anomaly lattice also satisfies the Kim--Shiu--Vafa current-algebra bound for the allowed BPS string charges~\cite{Kim_2019, Angelantonj:2020pyr}. These checks provide complementary tests of the perturbative construction beyond the worldsheet factorization conditions.

The boundary-orbit description also clarifies the structure of deformations. Because twisted tadpoles constrain the admissible Chan-Paton charges, generic translations of individual branes are not moduli of the orientifold vacuum. Discrete redistributions among shortened boundary orbits lead instead to different tadpole solutions, while continuous deformations are naturally described as Higgs or recombination branches of the open-string theory. In this sense the rigidity of the quasicrystalline bulk is accompanied by a correspondingly constrained space of boundary configurations.

The same framework produces a non-supersymmetric Brane Supersymmetry Breaking descendant when the R-R charges of the diagonal orientifold-plane orbits are reversed. The closed sector remains supersymmetric, whereas tadpole cancellation requires D7-branes and anti-D7-branes whose open sectors do not preserve the same supersymmetry. Generic solutions contain brane-antibrane tachyons, but a minimal branch with $r=4$, $s=0$ and $|t|\leq2$ is perturbatively tachyon-free and contains no massless open-string scalars. It therefore provides a particularly rigid BSB configuration, closely related in spirit to other rigid orientifold vacua with non-trivial twisted R-R tadpole conditions~\cite{Angelantonj:2024iwi}. This statement concerns perturbative open-string stability: the configuration still carries NS-NS tadpoles, and its backreacted vacuum cannot be inferred from the one-loop spectrum alone.

A second application is the freely acting five-dimensional orientifold obtained by combining the quasicrystalline generator with momentum and winding shifts along an additional circle. The freely acting quotient realises a spontaneous reduction from thirty-two to sixteen supercharges through a stringy Scherk-Schwarz mechanism, while the orientifold projection leaves eight supercharges on the orientifold configuration. Additionally, the shift lifts the twisted sectors at generic radius. The construction has two inequivalent six-dimensional decompactification limits. For $R\to0$, the winding tower reconstructs a type IIA geometric $\Z_3$ bulk with orientifold defects localised in the decompactified dual direction. For $R\to\infty$, the momentum tower reconstructs the type IIB antisymmetric $\Z_4$ orientifold and its symplectic open-string branch. The finite-radius model should therefore be viewed as an interpolation between these two endpoint theories rather than as a circle deformation returning to the original six-dimensional $\Z_{12}$ orientifold.

Taken together, these constructions show that the absence of a conventional geometric bulk description does not obstruct a perturbative orientifold completion. The relevant geometric information is reorganised rather than lost: twisted ground-state orbits are naturally described in the Narain CFT, while boundary and crosscap sectors can retain useful geometric descriptions after suitable partial T-dualities. This separation between the intrinsically non-geometric bulk action and the more geometric organisation of the source sectors is one of the main structural lessons of the $\Z_{12}$ example.

Several questions follow directly from this picture. The first is whether boundary conditions preserving a smaller chiral algebra can couple to the genuine odd $\Z_{12}$ twisted sectors that are absent from the symmetric boundary states used here. Constructing such states would require a more refined boundary-CFT analysis, along the lines of the asymmetric-orbifold boundary states of \cite{Brunner_1999,Gaberdiel:2002jr}, and could enlarge the set of admissible quasicrystalline D-branes. A second question concerns global charge constraints: lower-dimensional probes could test discrete or K-theoretic conditions that are not visible in the perturbative R-R tadpoles~\cite{Witten_1998,Uranga_2001}. For the BSB branch, the natural next problem is instead dynamical, namely to determine the tachyon-condensation channels and the backreaction induced by the NS-NS tadpoles. Finally, it would be useful to classify compatible orientifold parities for the other cyclic quasicrystalline backgrounds and to determine which features found here, in particular the relation between twisted crosscap charges and shortened boundary orbits, persist beyond the $\Z_{12}$ example.

\acknowledgments

I am particularly grateful to Carlo Angelantonj and Emilian Dudas for their guidance throughout this work, for many illuminating discussions, and for generously sharing their knowledge and insight. I also thank Guillaume Bossard, Adrien Fiorucci, Victor Chabirand and Paul Ghiringhelli for interesting and useful discussions.

OpenAI's ChatGPT was used to assist with the writing and language editing of the manuscript and with the preparation of some figures; all scientific content and final outputs were reviewed and verified by the author.

\addtocontents{toc}{\protect\setcounter{tocdepth}{1}}
\appendix

\section{Worldsheet and amplitude conventions}\label{sec:conventions}

This appendix collects the worldsheet and one-loop amplitude conventions used throughout the paper~\cite{Ibanez:2012zz,Blumenhagen:2013fgp,Kiritsis:2019npv}. We first summarise the theta-function and character conventions, then describe the modular properties of the orbifold blocks $T[^\alpha_\beta]$ and their relation to the eigencharacters $\chi[^\alpha_\beta]$. We finally give the normalisation and loop/tree channel dictionary for the torus, Klein bottle, annulus and Möbius strip amplitudes, together with the boundary-state and crosscap interpretation used in the factorization analysis. Compact zero-mode lattices are treated separately in \Cref{app:lattices}.

\subsection{Dedekind and Jacobi functions}\label{sec:DedekindJacobi}

We use
\begin{equation}
    \q=\e{2\pi i\tau},
    \qquad
    \eta(\tau)=\q^{1/24}\prod_{n=1}^{\infty}(1-\q^n),
\end{equation}
and define the Jacobi theta function with characteristics by
\begin{equation}
    \theta[^a_b](z,\tau)
    =
    \sum_{n\in\Z}
    \q^{\frac12(n+a)^2}
    \e{2\pi i(n+a)(z+b)}.
\end{equation}
The characteristics are understood modulo integers, with
\begin{equation}
    \theta[^{a+n}_{b+m}](z,\tau)
    =
    \e{2\pi i a m}\,
    \theta[^a_b](z,\tau),
    \qquad n,m\in\Z.
\end{equation}
The modular transformations needed below are
\begin{align}
    \eta(\tau+1)
    &=\e{i\pi/12}\eta(\tau),
    \\
    \eta(-1/\tau)
    &=(-i\tau)^{1/2}\eta(\tau),\\
    \theta[^a_b](z,\tau+1)
    &=\e{-i\pi a(a-1)}
      \theta[^a_{b+a-1/2}](z,\tau),
    \\
    \theta[^a_b]\left(\frac{z}{\tau},-\frac1\tau\right)
    &=(-i\tau)^{1/2}
      \e{i\pi z^2/\tau+2\pi iab}
      \theta[^b_{-a}](z,\tau).
\end{align}

\subsection{Modular covariance of the orbifold blocks}\label{sec:modular_transformations}

The geometrical meaning of the two orbifold labels makes their modular action particularly simple. The first label specifies the twist around one worldsheet cycle and the second the insertion around the other. Consequently,
\begin{equation}
    T:\qquad
    (\alpha,\beta)\longmapsto(\alpha,\alpha+\beta),
    \qquad
    S:\qquad
    (\alpha,\beta)\longmapsto(\beta,-\alpha),
\end{equation}
with all labels understood modulo $N$.

For the explicit holomorphic blocks of our model, one must in addition keep track of their modular weight and of the normalisation $d_{\alpha,\beta}$. With the conventions used in this paper, the $T$ transformation is
\begin{equation}
    T[^\alpha_\beta](\tau+1)
    =
    i\,
    T[^\alpha_{\alpha+\beta}](\tau).
    \label{eq:Tblock-T}
\end{equation}
For every non-trivial pair $(\alpha,\beta)\neq(0,0)$, the $S$ transformation is
\begin{equation}
    T[^\alpha_\beta](-1/\tau)
    =
    -(-i\tau)^{-2}
    \frac{d_{\alpha,\beta}}{d_{\beta,-\alpha}}\,
    T[^\beta_{-\alpha}](\tau).
    \label{eq:Tblock-S}
\end{equation}
The factor $(-i\tau)^{-2}$ is the modular weight associated with the non-compact bosonic oscillators contained in our definition of $T[^\alpha_\beta]$. The ratio of $d$ factors is relevant when an untwisted sector with a non-trivial insertion is exchanged with a twisted sector. In the complete torus amplitude it combines with the twisted-ground-state degeneracies and compact zero-mode factors, so modular invariance must be checked for the complete sector rather than for the holomorphic block alone.

The universal phase $i$ in \Cref{eq:Tblock-T} and the sign in \Cref{eq:Tblock-S} cancel against their right-moving counterparts in the corresponding closed-string contribution. For most of the discussion in the main text only the induced transformation of the sector labels is therefore displayed explicitly.

The Möbius strip involves instead the transformation
\begin{equation}
    P=T^{1/2}ST^2ST^{1/2},
\end{equation}
which maps
\begin{equation}
    \tau_{\mathcal M}
    =
    \frac12+\frac{i\tau_2}{2}
    \qquad\longrightarrow\qquad
    \widetilde\tau_{\mathcal M}
    =
    \frac12+\frac{i}{2\tau_2}.
\end{equation}
Rather than applying $P$ to unhatted characters, we use the standard hatted basis. If
\begin{equation}
    \chi_i(\tau)
    =
    \q^{h_i-c/24}
    \sum_{n\geq0}a_n\q^n,
\end{equation}
the corresponding Möbius character is defined by the conventional phase
\begin{equation}
    \widehat\chi_i\left(\frac12+it\right)
    =
    \e{-i\pi(h_i-c/24)}
    \chi_i\left(\frac12+it\right).
\end{equation}
The same prescription is applied term by term to the orbifold blocks and defines $\widehat T[^\alpha_\beta]$. In this basis the loop Möbius coefficients are real and the $P$ transformation relates them directly to the boundary and crosscap reflection coefficients of the tree channel.

\subsection{Six-dimensional amplitudes}\label{sec:6d_amplitudes}

We use the following normalisation for the one-loop vacuum amplitudes in six dimensions. The non-compact volume is denoted by $V_d$ for $d$ non-compact dimensions, and the model-dependent CFT contributions, including GSO projections, orbifold sums, compact lattice sums or fixed-point degeneracies, oscillator characters and Chan-Paton traces, are collected in the dimensionless quantities $\mathcal T$, $\mathcal K$, $\mathcal A$ and $\mathcal M$. The loop-channel full amplitudes are
\begin{equation}
\begin{aligned}
\mathscr T &= \frac{V_d}{(4\pi^2\alpha')^{\frac{d}{2}}}\int_{\mathcal F}\frac{\dd^2\tau}{\tau_2^4}\,\mathcal T(\tau,\overline\tau),\\
\mathscr K &= \frac{V_d}{(4\pi^2\alpha')^{\frac{d}{2}}}\int_0^\infty \frac{\dd\tau_2}{\tau_2^4}\,\mathcal K(2i\tau_2),\\
\mathscr A_p &= \frac{V_{p+1}}{(4\pi^2\alpha')^{(p+1)/2}}\int_0^\infty \frac{\dd\tau_2}{\tau_2^{1+\frac{p+1}{2}}}\,\mathcal A_p(i \frac{\tau_2}{2}),\\
\mathscr M_p &= \frac{V_{p+1}}{(4\pi^2\alpha')^{(p+1)/2}}\int_0^\infty \frac{\dd\tau_2}{\tau_2^{1+\frac{p+1}{2}}}\,\mathcal M_p(i \frac{\tau_2}{2}+\frac12)\,,
\end{aligned}
\end{equation}
where $p+1$ is the number of non-compact spacetime dimensions wrapped by the D-brane in the open sector. The integration domain $\mathcal F$ is the fundamental domain of the modular group $SL(2,\Z)$.
The four one-loop amplitudes used below are
\begin{itemize}
    \item Torus
    \begin{equation}
        \mathcal{T}
        =
        \mathrm{Tr}_{\mathrm{closed}}
        \left(
        P_{\mathrm{GSO}}\,
        \q^{L_0-c/24}
        \qbar^{\overline L_0-c/24}
        \right).
    \end{equation}

    \item Klein bottle
    \begin{equation}
        \mathcal{K}
        =
        \mathrm{Tr}_{\mathrm{closed}}
        \left(
        P_{\mathrm{GSO}}\,
        \Omega\,
        \q^{L_0-c/24}
        \qbar^{\overline L_0-c/24}
        \right).
    \end{equation}

    \item Annulus
    \begin{equation}
        \mathcal{A}_p
        =
        \mathrm{Tr}_{\mathrm{open}}
        \left(
        P_{\mathrm{GSO}}\,
        \q^{\frac12(L_0-c/24)}
        \right).
    \end{equation}

    \item Möbius strip
    \begin{equation}
        \mathcal{M}_p
        =
        \mathrm{Tr}_{\mathrm{open}}
        \left(
        P_{\mathrm{GSO}}\,
        \Omega\,
        \q^{\frac12(L_0-c/24)}
        \right).
    \end{equation}
\end{itemize}

In the loop channel, in $\mathcal A, \mathcal K \text{ and } \mathcal M$, we use $\q=\e{-2\pi\tau_2}$ for the one-loop modulus. With these conventions, the loop channel Klein bottle characters are evaluated at $\q^2$, the annulus characters at $\sqrt{\q}$, and the Möbius strip characters at $\sqrt{\q}$ with an additional phase encoded in the hatted characters $\widehat\chi$. The tree channel amplitudes are obtained by the $S$ transformation for the Klein bottle and annulus, and by the $P=T^{1/2}ST^2ST^{1/2}$ transformation for the Möbius strip. We use the tree channel proper times $\ell$ such that the tree channel amplitudes are written with $\q=\e{-2\pi \ell}$, where $\ell=\frac{1}{2\tau_2}$ for the Klein bottle and Möbius strip, while $\ell=\frac{2}{\tau_2}$ for the annulus. In the tree channel all characters are evaluated at $\q$, again with hatted characters in the Möbius amplitude. The torus amplitude is instead written with the modular parameter $\tau$ and $\q=\e{2\pi i\tau}$. The tree-channel full amplitudes are
\begin{equation}
\begin{aligned}
\mathscr K &= \frac{V_d}{(4\pi^2\alpha')^{\frac{d}{2}}}\int_0^\infty \dd\ell\,\widetilde{\mathcal K}(i\ell),\\
\mathscr A_p &= \frac{V_{p+1}}{(4\pi^2\alpha')^{(p+1)/2}}\int_0^\infty \dd\ell\,\ell^{\frac{p+1-d}{2}}\widetilde{\mathcal A}_p(i \ell),\\
\mathscr M_p &= \frac{V_{p+1}}{(4\pi^2\alpha')^{(p+1)/2}}\int_0^\infty \dd\ell\,\ell^{\frac{p+1-d}{2}}\widetilde{\mathcal M}_p(i \ell+\frac12)\,.
\end{aligned}
\end{equation}

\begin{equation}
\begin{array}{c|c|c|c}
\text{surface} & \text{loop modulus} & \text{transformation} & \text{tree channel proper time}\\
\hline
\mathcal K & 2i\tau_2 & S & \ell=\dfrac{1}{2\tau_2}\\[2mm]
\mathcal A & \dfrac{i\tau_2}{2} & S & \ell=\dfrac{2}{\tau_2}\\[3mm]
\mathcal M & \dfrac12+\dfrac{i\tau_2}{2} & P & \ell=\dfrac{1}{2\tau_2}
\end{array}
\label{tab:channel-transformations}
\end{equation}

\subsection{Boundary and crosscap interpretation}\label{sec:boundary-crosscap-factorization}

The tree amplitudes admit a direct closed-string interpretation in terms of boundary and crosscap states~\cite{Pradisi_1995,Pradisi_1995_open,FioravantiPradisiSagnotti:1993,Angelantonj:2002ct}. Let $|i\rangle\!\rangle$ denote an Ishibashi state of the chiral algebra preserved by the boundary condition. A boundary state $|B_a\rangle$ and a crosscap state $|C\rangle$ can be expanded as
\begin{equation}
    |B_a\rangle
    =
    \sum_i B_a^{\,i}|i\rangle\!\rangle,
    \qquad
    |C\rangle
    =
    \sum_i\Gamma^i|i\rangle\!\rangle.
\end{equation}
Their closed-string overlaps give
\begin{align}
    \widetilde{\mathcal K}
    &=
    \langle C|
    \e{-2\pi\ell H_{\rm cl}}
    |C\rangle
    =
    \sum_i(\Gamma^i)^2\chi_i,
    \\
    \widetilde{\mathcal A}_{ab}
    &=
    \langle B_a|
    \e{-2\pi\ell H_{\rm cl}}
    |B_b\rangle
    =
    \sum_i B_a^{\,i}B_b^{\,i}\chi_i,
    \\
    \widetilde{\mathcal M}_a
    &=
    \langle B_a|
    \e{-2\pi\ell H_{\rm cl}}
    |C\rangle
    =
    \sum_i B_a^{\,i}\Gamma^i\widehat\chi_i.
\end{align}
Accordingly, the coefficient of a tree channel character in the Klein bottle is the square of the corresponding crosscap reflection coefficient, the annulus coefficient is a product of boundary reflection coefficients and the Möbius coefficient is their boundary and crosscap geometric mean, up to the signs fixed by the orientifold projection.

This factorization is used repeatedly in the main text. In particular, once the tree channel Klein bottle determines the R-R representations sourced by the crosscaps, the annulus must contain boundary states with compatible couplings. The Möbius strip then fixes the relative signs between boundary and crosscap reflection coefficients. Transforming the result back to the loop channel must yield non-negative integer annulus multiplicities and a Möbius projection compatible with
\begin{equation}
    \frac12\left(\mathcal A+\mathcal M\right).
\end{equation}
These requirements provide a stronger consistency condition than the cancellation of the massless R-R tadpoles alone.

In the present $\Z_{12}$ construction, the characters appearing in the tree channel amplitudes should be interpreted as characters of the chiral algebra preserved by the corresponding symmetric boundary ansatz. The absence of a given $\chi[^\alpha_\beta]$ therefore means that the associated boundary reflection coefficient vanishes within this class of boundary states. As discussed in the main text, this statement need not exclude more general boundary conditions preserving a smaller chiral algebra, for which the characters can decompose into finer representations and additional Ishibashi states may become available.

\section{Compact lattice sums}
\label{app:lattices}

In this appendix we collect the conventions for the compact lattice sums entering the annulus, Klein bottle and Möbius strip amplitudes.

For an object wrapping the one-cycle $(a,b)$ of a two-torus with complex structure $U$ and Kähler structure $T$, we define the tree channel closed-string lattice
\begin{equation}
    \widetilde{\Lambda}^{(a,b)}_{x,y}(U,T)
    =
    \sum_{m,w\in\Z}
    \e{2\pi i(wx+my)}
    \q^{\frac14\frac{|a+bU|^2}{\Im(T)\Im(U)}|m+Tw|^2}\,,
\end{equation}
where, for a boundary, $(x,y)$ denote respectively the Wilson line along the wrapped cycle and the transverse position, while for a crosscap they denote respectively its position along the T-dual cycle and its geometrical transverse position, both fixed by the orientifold projection. In an overlap between two sources, only the corresponding relative values of $(x,y)$ enter. The integers $(m,w)$ denote the closed-string momentum and winding numbers.
Brane or orientifold images belonging to the same orbifold orbit have identical conformal weights, since the orbifold generator acts as an automorphism of the Narain CFT. Consequently, their lattice sums differ only by the induced shifts and tree channel phases~\cite{Blumenhagen:1999md,Blumenhagen_2007}.

The Poisson resummation formula used throughout is 
\begin{equation}
    \sum_{m\in\Lambda} \e{- \pi (m+y)^T A (m+y)} \e {2i\pi x^T (m+y)} = \frac{1}{\text{Vol}(\Lambda)\sqrt{\text{det}(A)}}\sum_{n\in\Lambda^*} \e{-\pi (n+x)^T A^{-1} (n+x)-2i\pi n^T y}\,.
\end{equation}

\subsection{Annulus}

For two parallel boundary components with relative Wilson line and position $(x,y)$, the loop annulus lattice is
\begin{equation}
    \Lambda^{\mathcal A,(a,b)}_{m+x,w+y}(U,T)
    =
    \sum_{m,w\in\Z}
    \q^{\frac12\frac{\Im(U)}{\Im(T)}
    \frac{|m+x-T(w+y)|^2}{|a+bU|^2}}\,.
\end{equation}
With $l=2/\tau_2$, Poisson resummation gives
\begin{equation}
    \Lambda^{\mathcal A,(a,b)}_{m+x,w+y}(U,T)
    \xrightarrow{S}
    \frac{l}{2}\frac{|a+bU|^2}{\Im(U)}\,
    \widetilde{\Lambda}^{(a,b)}_{x,y}(U,T)\,.
\end{equation}
The loop expression displays the open-string zero-mode mass,
\begin{equation}
    M^2_{\mathcal A}
    =
    \frac{\Im(U)}{\Im(T)}
    \frac{|m+x-T(w+y)|^2}{|a+bU|^2}\,.
\end{equation}

\subsection{Klein bottle}

For the reflection-type orientifold projections relevant below, let the crosscap wrap the primitive one-cycle $(a,b)$. The corresponding primitive reflected cycle $(c,d)$ is fixed, up to an overall sign, by metric orthogonality,
\begin{equation}
    \Re\!\left[(a+bU)\overline{(c+dU)}\right]=0\,,
\end{equation}
and we define
\begin{equation}
    \Delta=|ad-bc|\,.
\end{equation}
For a crystallographic reflection one has $\Delta=1$ or $2$. Equivalently, in a lattice basis adapted to the wrapped cycle $(a,b)$, these two cases correspond respectively to $\Re(U)=0$ and $\Re(U)=1/2$ modulo integer shifts.

The states invariant in the loop Klein bottle have momentum along the wrapped cycle and winding along the reflected cycle. In terms of the original two-dimensional Narain integers this amounts to
\begin{equation}
    \boldsymbol m=m(d,-c)\,,\qquad \boldsymbol w=w(c,d)\,,\qquad m,w\in\Z\,.
\end{equation}
Using
\begin{equation}
    |a+bU|\,|c+dU|=\Delta\,\Im(U)\,,
\end{equation}
the corresponding loop channel lattice can be written universally as
\begin{equation}
    \Lambda^{\mathcal K,(a,b)}_{\Delta}(U,T)
    =\sum_{m,w\in\Z}\q^{\frac{\Delta^2}{2}\frac{\Im(U)}{\Im(T)}
    \frac{|m-Tw|^2}{|a+bU|^2}}\,.
\end{equation}

Under the $S$ transformation, with $l=1/(2\tau_2)$, the exchanged closed-string has winding along $(a,b)$ and momentum transverse to it,
\begin{equation}
    \boldsymbol w_{\rm tr}=\frac{2}{\Delta}w(a,b)\,,\qquad
    \boldsymbol m_{\rm tr}=\frac{2}{\Delta}m(b,-a)\,.
\end{equation}
The factor $2/\Delta$ determines the crosscap phases. For $\Delta=1$ the tree channel momentum and winding integers are even, whereas for $\Delta=2$ the tree channel lattice is unrestricted. It is therefore convenient to introduce
\begin{align}
    \mathcal C_{+}(m,w)&=\frac{1+(-1)^m}{2}\frac{1+(-1)^w}{2}\,,\\
    \mathcal C_{-}(m,w)&=\frac{1-(-1)^m-(-1)^w-(-1)^{m+w}}{2}\,,\notag
\end{align}
with
\begin{equation}
    \mathcal C_{+}^2=\mathcal C_{+}\,,\quad \mathcal C_{-}^2=1\,.
\end{equation}
The $S$ transformation of the loop channel lattice then takes the compact form
\begin{equation}
    \Lambda^{\mathcal K,(a,b)}_{\Delta}(U,T)
    \xrightarrow{S}
    \frac{2l}{\Delta^2}\frac{|a+bU|^2}{\Im(U)}
    \mathcal C_{\Delta}(m,w)^2\,
    \widetilde{\Lambda}^{(a,b)}_{0,0}(U,T)\,,
\end{equation}
where $\mathcal C_{\Delta=1}\equiv\mathcal C_{+}$ and $\mathcal C_{\Delta=2}\equiv\mathcal C_{-}$. Thus the $\Delta=1$ case projects the tree channel lattice onto even momentum and winding integers, while for $\Delta=2$ the Klein bottle contains the full underlying tree channel lattice. The non-trivial relative signs contained in $\mathcal C_{-}$ disappear after squaring in the Klein bottle but remain visible in the Möbius strip.

The wrapped cycle $(a,b)$ together with the torus geometry therefore fixes the entire Klein bottle lattice: the primitive orthogonal cycle $(c,d)$ determines $\Delta$, which in turn determines both the loop channel invariant lattice and the tree channel crosscap structure.

\subsection{Möbius strip}

The Möbius strip is a boundary-crosscap overlap. In the tree channel the crosscap structure therefore appears linearly, so that the lattice contribution is proportional to
\begin{equation}
    \mathcal C_{\Delta}(m,w)\,\widetilde{\Lambda}^{(a,b)}_{x,y}(U,T)\,,
\end{equation}
where $(x,y)$ are the boundary Wilson line and position relative to the origin used to describe the crosscap components.

In the loop channel the same information is encoded in the even and odd sectors of the open-string lattice. For $\rho,\sigma\in\{0,1\}$ we define
\begin{equation}
    \Lambda^{\mathcal M,(a,b)}_{\rho,\sigma;x,y}(U,T)
    =\sum_{m,w\in\Z}\q^{\frac12\frac{\Im(U)}{\Im(T)}
    \frac{|2m+\rho+2x-T(2w+\sigma+2y)|^2}{|a+bU|^2}}\,.
\end{equation}
With $l=1/(2\tau_2)$, the $P$ transformation gives
\begin{equation}
    \Lambda^{\mathcal M,(a,b)}_{\rho,\sigma;x,y}(U,T)
    \xrightarrow{P}
    \frac{l}{2}\frac{|a+bU|^2}{\Im(U)}
    (-1)^{\rho w+\sigma m}\,
    \widetilde{\Lambda}^{(a,b)}_{x,y}(U,T)\,.
\end{equation}
Consequently, the two crosscap structures correspond respectively to the loop channel parity combinations
\begin{align}
    \Delta=1:\qquad&
    \Lambda^{\mathcal M}_{0,0}
    +\Lambda^{\mathcal M}_{1,0}
    +\Lambda^{\mathcal M}_{0,1}
    +\Lambda^{\mathcal M}_{1,1}
    \quad\longleftrightarrow\quad
    4\mathcal C_{+}(m,w)\,\widetilde{\Lambda}^{(a,b)}_{x,y}\,,
    \\
    \Delta=2:\qquad&
    \Lambda^{\mathcal M}_{0,0}
    -\Lambda^{\mathcal M}_{1,0}
    -\Lambda^{\mathcal M}_{0,1}
    -\Lambda^{\mathcal M}_{1,1}
    \quad\longleftrightarrow\quad
    2\mathcal C_{-}(m,w)\,\widetilde{\Lambda}^{(a,b)}_{x,y}\,,
\end{align}
where the common factor $\frac{l}{2}\frac{|a+bU|^2}{\Im(U)}$ on the tree channel side is understood. The overall normalisation of these parity combinations is absorbed into the coefficient of the complete Möbius amplitude and is fixed by the corresponding boundary and crosscap reflection coefficients.

\section{Additional probe-brane orbits}\label{app:probe-branes}

\subsection{\texorpdfstring{$Dp/D(p+4)$}{Dp/D(p+4)}-branes at \texorpdfstring{$\Z_3$}{Z3} and \texorpdfstring{$\Z_2$}{Z2} points}\label{sec:brane-orbit-Z3-Z2}

We now consider $Dp/D(p+4)$ boundary orbits, described as $D(p+2)$-brane orbits in the \Tp-dual frame, whose representative is no longer at the origin, but still lies at a fixed point of a non-trivial subgroup of the orbifold group. The invariant group is then smaller than the $\Z_6$ invariant group at the origin, and the Chan-Paton action is diagonalised only with respect to this smaller group. This gives intermediate fractional configurations, between the orbit of size two at the origin and a completely generic orbit of size twelve.

\begin{figure}[htb]
\centering
\begin{tikzpicture}[>=stealth]

    %%%%%%%%%%%%%%%%%%%%%%%%%
    % Left panel: torus plot
    %%%%%%%%%%%%%%%%%%%%%%%%%
    \begin{scope}[scale=2]
        % Axes
        \draw[->, thick] (-.85, 0) -- (1.5, 0) node[right] {$x$};
        \draw[->, thick] (0, -0.5) -- (0, 1.5) node[above] {$y$};

        % Basis vectors
        \draw[->, thick, blue] (0,0) -- (1,0) node[below right] {$e_1$};
        \draw[-] (1,1) -- (1,0);

        \draw[->, thick, blue] (0,0) -- (0,1) node[above left] {$e_2$};
        \draw[-] (0,1) -- (1,1);

        % Branes
        \draw[-, thick, red] (0,1/3) -- (1,1/3);
        \draw[-, thick, red] (1/3,0) -- (1/3,1);
        \draw[-, thick, red] (0,2/3) -- (1,2/3);
        \draw[-, thick, red] (2/3,0) -- (2/3,1);

        % Fixed points
        \node[circle, fill=gray, inner sep=1.5pt] at (0,0) {};
        \node[circle, fill=gray, inner sep=1.5pt] at (1/2,1/2) {};
        \node[circle, draw=gray, fill=none, inner sep=1.5pt, line width=1pt] at (1/2,0) {};
        \node[circle, draw=gray, fill=none, inner sep=1.5pt, line width=1pt] at (0,1/2) {};
    \end{scope}

    %%%%%%%%%%%%%%%%%%%%%%%%%%%%%%%%%%%%%%%
    % Right panel: orbit under g
    %%%%%%%%%%%%%%%%%%%%%%%%%%%%%%%%%%%%%%%
    \begin{scope}[xshift=5.4cm,yshift=-1cm]
        \node[font=\bfseries] at (2.6,5.4) {Orbit under $g$};

        \node[draw, rounded corners, align=center, text width=5.2cm, inner sep=4pt] (s1) at (2.6,4.5)
        {$\left(\frac13+\frac{U}{3},0;0,\frac{\pi}{4}\right)$};

        \node[draw, rounded corners, align=center, text width=5.2cm, inner sep=4pt] (s2) at (2.6,3.3)
        {$\left(\frac13+\frac{U}{3},0;\frac{\pi}{2},\frac{3\pi}{4}\right)$};

        \node[draw, rounded corners, align=center, text width=5.2cm, inner sep=4pt] (s3) at (2.6,2.1)
        {$\left(\frac23+\frac{2U}{3},0;0,\frac{\pi}{4}\right)$};

        \node[draw, rounded corners, align=center, text width=5.2cm, inner sep=4pt] (s4) at (2.6,0.9)
        {$\left(\frac23+\frac{2U}{3},0;\frac{\pi}{2},\frac{3\pi}{4}\right)$};

        \draw[->, thick] (s1.south) -- node[right] {$g$} (s2.north);
        \draw[->, thick] (s2.south) -- node[right] {$g$} (s3.north);
        \draw[->, thick] (s3.south) -- node[right] {$g$} (s4.north);
        \draw[->, thick] (s4.east) to[bend right=45] node[right] {$g$} (s1.east);
    \end{scope}

\end{tikzpicture}
\caption{Left: example of size-four brane orbits on one of the two-tori. Right: action of the orbifold generator $g$ on the corresponding orbit.}\label{fig:brane-orbit-z3-example}
\end{figure}
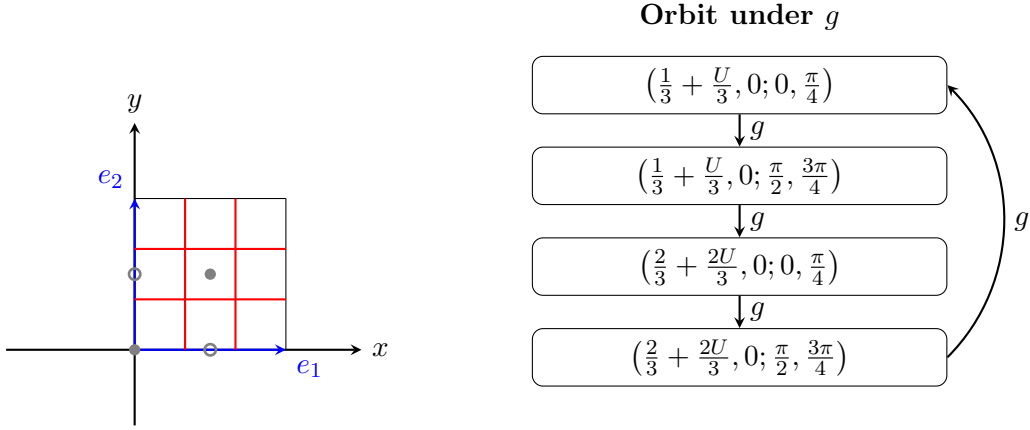

\begin{table}[htb]
\centering
\begin{tabular}{c|c|c}
Sector & Representation & Field content \\\hline
$\alpha=0$ &
$n_\gamma\overline{n}_\gamma+m_\gamma\overline{m}_\gamma$ &
$A^\mu+(4-2k)\phi+2^{2-k}(\psi_L+\psi_R)$ \\\hline
$\alpha=0$ &
$n_\gamma(\overline{n}_{\gamma+1}+\overline{n}_{\gamma-1})+m_\gamma(\overline{m}_{\gamma+1}+\overline{m}_{\gamma-1})$ &
$\Phi+2^{1-k}(\psi_L+\psi_R)$ \\\hline
$\alpha\in\{3,9\}$ &
$n_\gamma\overline{m}_\gamma+\overline{n}_\gamma m_\gamma$ &
$4(2\varphi+2^{1-k}(\psi_L+\psi_R))$ \\\hline
$\alpha=6$ &
$n_\gamma(\overline{n}_{\gamma+1}+\overline{n}_{\gamma-1})+m_\gamma(\overline{m}_{\gamma+1}+\overline{m}_{\gamma-1})$ &
$2(2\varphi+2^{1-k}(\psi_L+\psi_R))$ \\\hline
\end{tabular}
\caption{Massless spectrum of the $Dp/D(p+4)$ orbit at a $\Z_3$ fixed point. The second column gives the representations under the gauge group $\prod_{\gamma=0}^2 U(n_\gamma)\times\prod_{\gamma=0}^2 U(m_\gamma)$, with $\gamma$ understood modulo $3$.}\label{tab:Dp-orbitZ3-spectrum}
\end{table}

Let us first place the representative brane at a $\Z_3$ fixed point. This corresponds to the positions $u=\frac{\alpha}{3}(1+U)$ on the first torus and $u=\frac{\beta}{3}(1+U)$ on the second torus, with $\alpha,\beta=0,1,2$ and $(\alpha,\beta)\neq(0,0)$, as represented in \Cref{fig:brane-orbit-z3-example}. The invariant group is then $\Z_3$.

The corresponding gauge group is $\prod_{\gamma=0}^2 U(n_\gamma)\times\prod_{\gamma=0}^2 U(m_\gamma)$. The spectrum is summarised in \Cref{tab:Dp-orbitZ3-spectrum}.

The tree channel expression contains the untwisted exchange and the twisted exchanges in the $\alpha=4,8$ sectors, with phase factors determined by the displaced lattice sums. Thus these boundary states couple to the untwisted tensor and to the $\Z_3$ twisted fields localised at the corresponding fixed-point orbit. They do not couple to the $\Z_2$ twisted fields, nor to the $\Z_6$ and $\Z_{12}$ singlets, because the representative brane is not located at the corresponding fixed loci.

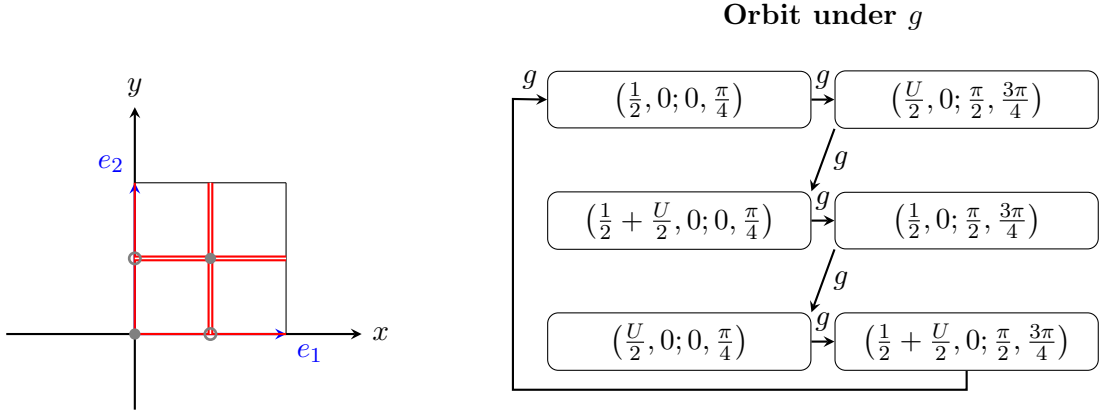
\begin{figure}[htb]
\centering
\begin{tikzpicture}[>=stealth]

    %%%%%%%%%%%%%%%%%%%%%%%%%
    % Left panel: torus plot
    %%%%%%%%%%%%%%%%%%%%%%%%%
    \begin{scope}[scale=2]
        % Axes
        \draw[->, thick] (-.85, 0) -- (1.5, 0) node[right] {$x$};
        \draw[->, thick] (0, -0.5) -- (0, 1.5) node[above] {$y$};

        % Basis vectors
        \draw[->, thick, blue] (0,0) -- (1,0) node[below right] {$e_1$};
        \draw[-] (1,1) -- (1,0);

        \draw[->, thick, blue] (0,0) -- (0,1) node[above left] {$e_2$};
        \draw[-] (0,1) -- (1,1);

        % Branes
        \draw[-, thick, red] (0,0) -- (1,0);
        \draw[-, thick, red] (0,0) -- (0,1);
        \draw[double, thick, red] (0,1/2) -- (1,1/2);
        \draw[double, thick, red] (1/2,0) -- (1/2,1);

        % Fixed points
        \node[circle, fill=gray, inner sep=1.5pt] at (0,0) {};
        \node[circle, fill=gray, inner sep=1.5pt] at (1/2,1/2) {};
        \node[circle, draw=gray, fill=none, inner sep=1.5pt, line width=1pt] at (1/2,0) {};
        \node[circle, draw=gray, fill=none, inner sep=1.5pt, line width=1pt] at (0,1/2) {};
    \end{scope}

    %%%%%%%%%%%%%%%%%%%%%%%%%%%%%%%%%%%%%%%
    % Right panel: orbit under g
    %%%%%%%%%%%%%%%%%%%%%%%%%%%%%%%%%%%%%%%
    \begin{scope}[xshift=5.8cm, yshift=-1cm]
        \node[font=\bfseries] at (3.3,5.2) {Orbit under $g$};

        % Left column: theta1 = 0
        \node[draw, rounded corners, align=center, text width=3.2cm, inner sep=4pt] (L1) at (1.4,4.1)
        {$\left(\frac12,0;0,\frac{\pi}{4}\right)$};

        \node[draw, rounded corners, align=center, text width=3.2cm, inner sep=4pt] (L2) at (1.4,2.5)
        {$\left(\frac12+\frac{U}{2},0;0,\frac{\pi}{4}\right)$};

        \node[draw, rounded corners, align=center, text width=3.2cm, inner sep=4pt] (L3) at (1.4,0.9)
        {$\left(\frac{U}{2},0;0,\frac{\pi}{4}\right)$};

        % Right column: theta1 = pi/2
        \node[draw, rounded corners, align=center, text width=3.2cm, inner sep=4pt] (R1) at (5.2,4.1)
        {$\left(\frac{U}{2},0;\frac{\pi}{2},\frac{3\pi}{4}\right)$};

        \node[draw, rounded corners, align=center, text width=3.2cm, inner sep=4pt] (R2) at (5.2,2.5)
        {$\left(\frac12,0;\frac{\pi}{2},\frac{3\pi}{4}\right)$};

        \node[draw, rounded corners, align=center, text width=3.2cm, inner sep=4pt] (R3) at (5.2,0.9)
        {$\left(\frac12+\frac{U}{2},0;\frac{\pi}{2},\frac{3\pi}{4}\right)$};

        % Orbit arrows
        \draw[->, thick] (L1.east) -- node[above] {$g$} (R1.west);
        \draw[->, thick] (R1.south west) -- node[right] {$g$} (L2.north east);
        \draw[->, thick] (L2.east) -- node[above] {$g$} (R2.west);
        \draw[->, thick] (R2.south west) -- node[right] {$g$} (L3.north east);
        \draw[->, thick] (L3.east) -- node[above] {$g$} (R3.west);
        \draw[->, thick] (R3.south)-- ++(0,-0.25)-- ++(-6,0) -- ++(0,3.845)-- node[above] {$g$} (L1.west);
    \end{scope}
\end{tikzpicture}
\caption{Left: example of size-six brane orbits on one of the two-tori. Right: action of the orbifold generator $g$ on the corresponding orbit.}\label{fig:brane-orbit-z2-example}
\end{figure}

Let us now consider a representative brane at a $\Z_2$ fixed point. This corresponds to the positions $u=\frac12(\alpha_1+\alpha_2 T)$ on the first torus and $u=\frac12(\beta_1+\beta_2 T)$ on the second torus, with $\alpha_i,\beta_i=0,1$ and $(\alpha_1,\alpha_2,\beta_1,\beta_2)\neq(0,0,0,0)$, as represented in \Cref{fig:brane-orbit-z2-example}. The invariant group is now $\Z_2$.

The corresponding gauge group is $U(n_0)\times U(n_1)\times U(m_0)\times U(m_1)$. The spectrum is summarised in \Cref{tab:Dp-orbitZ2-spectrum}.
\begin{table}[htb]
\centering
\begin{tabular}{c|c|c}
Sector & Representation & Field content \\\hline
$\alpha=0$ &
$n_0\overline{n}_0+n_1\overline{n}_1+m_0\overline{m}_0+m_1\overline{m}_1$ &
$A^\mu+(4-2k)\phi+2^{2-k}(\psi_L+\psi_R)$ \\\hline
$\alpha=0$ &
$n_0\overline{n}_1+n_1\overline{n}_0+m_0\overline{m}_1+m_1\overline{m}_0$ &
$2(\Phi+2^{1-k}(\psi_L+\psi_R))$ \\\hline
$\alpha\in\{3,9\}$ &
$n_0\overline{m}_0+\overline{n}_0m_0+n_1\overline{m}_1+\overline{n}_1m_1$ &
$6(2\varphi+2^{1-k}(\psi_L+\psi_R))$ \\\hline
$\alpha=6$ &
$n_0\overline{n}_0+n_1\overline{n}_1+m_0\overline{m}_0+m_1\overline{m}_1$ &
$4(2\varphi+2^{1-k}(\psi_L+\psi_R))$ \\\hline
$\alpha=6$ &
$n_0\overline{n}_1+n_1\overline{n}_0+m_0\overline{m}_1+m_1\overline{m}_0$ &
$2(2\varphi+2^{1-k}(\psi_L+\psi_R))$ \\\hline
\end{tabular}
\caption{Massless spectrum of the $Dp/D(p+4)$ orbit at a $\Z_2$ fixed point. The second column gives the representations under the gauge group $U(n_0)\times U(n_1)\times U(m_0)\times U(m_1)$.}\label{tab:Dp-orbitZ2-spectrum}
\end{table}

In the tree channel, the boundary state at a $\Z_2$ fixed point couples to the untwisted fields and to the $\Z_2$ twisted fields supported on the orbit of that fixed point. Since the representative is neither at a $\Z_3$ nor at a $\Z_4$ fixed point, it does not source the $\Z_3$, $\Z_4$, $\Z_6$ or $\Z_{12}$ twisted fields.

One can move from these $\Z_3$ or $\Z_2$ fixed points to a generic bulk point by turning on a Higgs-branch vev.
The process is similar to the one described before for the higgsing of the size twelve orbit at the origin.
The resulting massless spectrum is the generic one displayed in \Cref{tab:Dp-orbitgeneric-spectrum}.

\subsection{Annulus amplitudes for genuine \texorpdfstring{$D(p+2)$}{D(p+2)}-brane orbits}

\begin{figure}[htb]
    \centering

    % --- Subfigure A ---
    \begin{minipage}[t]{0.48\textwidth}
    \vspace{0pt}
        \centering
        \begin{tikzpicture}[scale=2,>=stealth]
            % Axes
            \draw[->, thick] (-.85, 0) -- (1.5, 0) node[right] {$x$};
            \draw[->, thick] (0, -0.5) -- (0, 1.5) node[above] {$y$};

            % Vectors
            \draw[->, thick, blue] (0,0) -- (1,0) node[below right] {$e_1$};
            \draw[-] ({cos(60)}, {sin(60)}) -- ({1+cos(60)}, {sin(60)});

            \draw[->, thick, blue] (0,0) -- ({cos(60)}, {sin(60)}) node[above left] {$e_2$};
            \draw[-] (1,0) -- ({1+cos(60)}, {sin(60)});

            % Fixed points
            \node[circle, fill=gray, inner sep=1.5pt] at (0,0) {};
            \node[star, star points=5, fill=gray, inner sep=1.5pt] at ({1/3+cos(60)/3}, {sin(60)/3}) {};
            \node[star, star points=5, fill=gray, inner sep=1.5pt] at ({2/3+2*cos(60)/3}, {2*sin(60)/3}) {};
            \node[circle, draw=gray, fill=none, inner sep=1.5pt, line width=1pt] at ({1/2+cos(60)/2}, {sin(60)/2}) {};
            \node[circle, draw=gray, fill=none, inner sep=1.5pt, line width=1pt] at ({cos(60)/2}, {sin(60)/2}) {};
            \node[circle, draw=gray, fill=none, inner sep=1.5pt, line width=1pt] at (1/2,0) {};

            % Branes
            \draw[-, thick, red] (0,0) -- (1,0);
            \draw[-, thick, red] (0,0) -- ({cos(60)}, {sin(60)});
            \draw[-, thick, red] (1,0) -- ({cos(60)}, {sin(60)});
        \end{tikzpicture}
        \subcaption{Configuration passing through a $\Z_4$ fixed point.\label{fig:Z12Dpp2A}}
    \end{minipage}
    \hfill
    % --- Subfigure B ---
    \begin{minipage}[t]{0.48\textwidth}
    \vspace{0pt}
        \centering
        \begin{tikzpicture}[scale=2,>=stealth]
            % Axes
            \draw[->, thick] (-.85, 0) -- (1.5, 0) node[right] {$x$};
            \draw[->, thick] (0, -0.5) -- (0, 1.5) node[above] {$y$};

            % Vectors
            \draw[->, thick, blue] (0,0) -- (1,0) node[below right] {$e_1$};
            \draw[-] ({cos(60)}, {sin(60)}) -- ({1+cos(60)}, {sin(60)});

            \draw[->, thick, blue] (0,0) -- ({cos(60)}, {sin(60)}) node[above left] {$e_2$};
            \draw[-] (1,0) -- ({1+cos(60)}, {sin(60)});

            % Fixed points
            \node[circle, fill=gray, inner sep=1.5pt] at (0,0) {};
            \node[star, star points=5, fill=gray, inner sep=1.5pt] at ({1/3+cos(60)/3}, {sin(60)/3}) {};
            \node[star, star points=5, fill=gray, inner sep=1.5pt] at ({2/3+2*cos(60)/3}, {2*sin(60)/3}) {};
            \node[circle, draw=gray, fill=none, inner sep=1.5pt, line width=1pt] at ({1/2+cos(60)/2}, {sin(60)/2}) {};
            \node[circle, draw=gray, fill=none, inner sep=1.5pt, line width=1pt] at ({cos(60)/2}, {sin(60)/2}) {};
            \node[circle, draw=gray, fill=none, inner sep=1.5pt, line width=1pt] at (1/2,0) {};

            % Parameter: vertical shift of the horizontal brane
            \def\eps{0.15}

            % Fundamental cell corners
            \coordinate (A) at (0,0);
            \coordinate (B) at (1,0);
            \coordinate (C) at ({1+cos(60)}, {sin(60)});
            \coordinate (D) at ({cos(60)}, {sin(60)});

            % Useful points
            \coordinate (O)   at (0,0);
            \coordinate (E1)  at (1,0);
            \coordinate (E2)  at ({cos(60)}, {sin(60)});
            \coordinate (E12) at ({1+cos(60)}, {sin(60)});

            \coordinate (M1)  at (1/2,0);
            \coordinate (M2)  at ({1/2+cos(60)}, {sin(60)});
            \coordinate (P1)  at ({1+cos(60)/2}, {sin(60)/2});
            \coordinate (P2)  at ({cos(60)/2}, {sin(60)/2});

            % Extra points for the last oblique brane
            \coordinate (Q1)  at ({1+cos(60)/2}, {sin(60)/2});
            \coordinate (Q2)  at ({cos(60)/2}, {sin(60)/2});

            % Helper macro:
            % #1 start, #2 end, #3 angle, #4 lattice shift along e1, #5 lattice shift along e2, #6 sign = +1 or -1, #7 color
            \newcommand{\ShiftedBraneTranslated}[7]{%
                \pgfmathsetmacro{\Tx}{#4 + #5*cos(60)}
                \pgfmathsetmacro{\Ty}{#5*sin(60)}
                \draw[-, thick, #7]
                ($ (#1) + ({-#6*\eps*sin(#3)},{#6*\eps*cos(#3)}) + (\Tx,\Ty) $)
                --
                ($ (#2) + ({-#6*\eps*sin(#3)},{#6*\eps*cos(#3)}) + (\Tx,\Ty) $);
            }

            % Draw the wrapped orbit and its nearby copies, then clip to the cell
            \begin{scope}
                \clip (A) -- (B) -- (C) -- (D) -- cycle;

                \foreach \m in {-1,0,1} {
                    \foreach \n in {-1,0,1} {

                        % First copy at +eps
                        \ShiftedBraneTranslated{O}{E1}{0}{\m}{\n}{1}{red}
                        \ShiftedBraneTranslated{O}{E2}{60}{\m}{\n}{1}{red}
                        \ShiftedBraneTranslated{E1}{E2}{120}{\m}{\n}{1}{red}

                        % Second copy at -eps
                        \ShiftedBraneTranslated{O}{E1}{0}{\m}{\n}{-1}{red}
                        \ShiftedBraneTranslated{O}{E2}{60}{\m}{\n}{-1}{red}
                        \ShiftedBraneTranslated{E1}{E2}{120}{\m}{\n}{-1}{red}

                    }
                }
            
                \draw[double, thick, red] (O) -- (E1);
                \draw[double, thick, red] (O) -- (E2);
                \draw[double, thick, red] (E1) -- (E2);
            \end{scope}
        \end{tikzpicture}
        \subcaption{Generic displaced configuration.\label{fig:Z12Dpp2B}}
    \end{minipage}

    \caption{Examples of genuine $D(p+2)$-brane configurations in the original picture. The red orbit is generated from a brane wrapping the $(1,0)$ one-cycle on the displayed two-torus.
    Panel~(\Cref{fig:Z12Dpp2A}) shows a configuration passing through a $\Z_4$ fixed point; panel~(\Cref{fig:Z12Dpp2B}) shows a generic displaced configuration. The full internal brane wraps the product of the corresponding one-cycles on the two two-tori. Here is displayed only the first $\T^2$, the second one behaving similarly.}\label{fig:Z12Dpp2}
\end{figure}
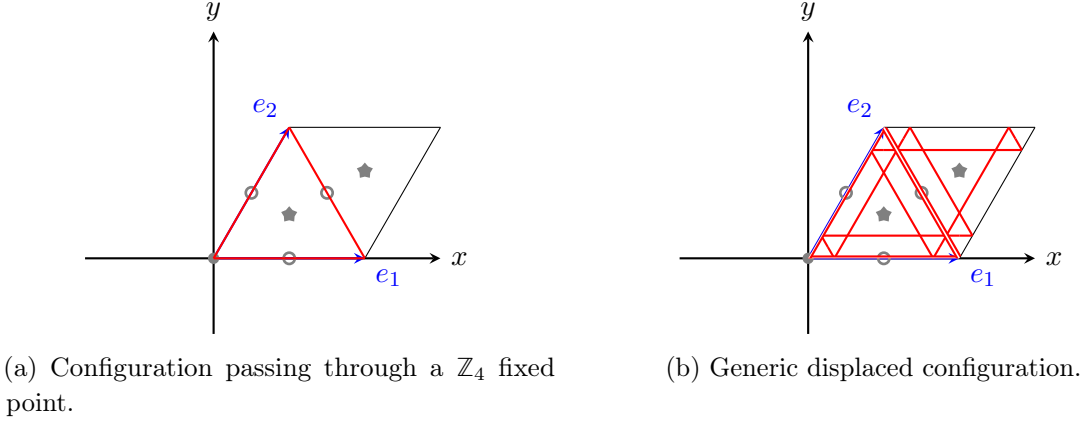

We now consider genuine $D(p+2)$-brane probes in the original frame. These branes wrap a one-cycle on each two-torus. If the representative cycle passes through a $\Z_4$ fixed point, the orbit is shortened to size 3 and is invariant under $\Z_4$. If the representative cycle passes through a $\Z_2$ fixed point, the orbit is shortened to size 6 and is invariant under $\Z_2$. Otherwise, for a generic displacement, the orbit has maximal size 12. Examples of the corresponding configurations are shown in \Cref{fig:Z12Dpp2} while details on the orbits can be found in~\Cref{tab:Dpp2-orbits}. Let us focus on the following two orbits: the one obtained from the wrapping numbers $(1,0,1,1)$, denoted $\mathcal O_N$, and the one obtained from $(1,1,1,0)$, denoted $\mathcal O_M$.

\begin{table}[htb]
\centering
\begin{tabular}{c|c|c|c}
Initial position & Invariant group & Orbit size \\
\hline
generic point & trivial & $12$ \\
$\Z_2$ fixed point & $\Z_2$ & $6$ \\
$\Z_4$ fixed point & $\Z_4$ & $3$
\end{tabular}
\caption{Orbit sizes for localised $D(p+2)$-brane probes under the $\Z_{12}$ action.}\label{tab:Dpp2-orbits}
\end{table}

\subsubsection{Branes through a \texorpdfstring{$\Z_4$}{Z4} fixed point}

\begin{figure}[htb]
\centering
\begin{tikzpicture}[>=stealth]

    %%%%%%%%%%%%%%%%%%%%%%%%%
    % Left panel: torus plot
    %%%%%%%%%%%%%%%%%%%%%%%%%
    \begin{scope}[scale=2]
        % Axes
        \draw[->, thick] (-.85, 0) -- (1.5, 0) node[right] {$x$};
        \draw[->, thick] (0, -0.5) -- (0, 1.2) node[above] {$y$};

        % Basis vectors
        \draw[->, thick, blue] (0,0) -- (1,0) node[below right] {$e_1$};
        \draw[->, thick, blue] (0,0) -- ({cos(60)}, {sin(60)}) node[above left] {$e_2$};

        % Fundamental cell
        \draw[-] ({cos(60)}, {sin(60)}) -- ({1+cos(60)}, {sin(60)});
        \draw[-] (1,0) -- ({1+cos(60)}, {sin(60)});

        % Fixed points
        \node[circle, fill=gray, inner sep=1.5pt] at (0,0) {};
        \node[star, star points=5, fill=gray, inner sep=1.5pt] at ({1/3+cos(60)/3}, {sin(60)/3}) {};
        \node[star, star points=5, fill=gray, inner sep=1.5pt] at ({2/3+2*cos(60)/3}, {2*sin(60)/3}) {};
        \node[circle, draw=gray, fill=none, inner sep=1.5pt, line width=1pt] at ({1/2+cos(60)/2}, {sin(60)/2}) {};
        \node[circle, draw=gray, fill=none, inner sep=1.5pt, line width=1pt] at ({cos(60)/2}, {sin(60)/2}) {};
        \node[circle, draw=gray, fill=none, inner sep=1.5pt, line width=1pt] at (1/2,0) {};

        % Branes
        \draw[-, thick, red] (0,0) -- (1,0);
        \draw[-, thick, red] (0,0) -- ({cos(60)}, {sin(60)});
        \draw[-, thick, red] (1,0) -- ({cos(60)}, {sin(60)});
    \end{scope}

    %%%%%%%%%%%%%%%%%%%%%%%%%%%%%%%%%%%%%%%
    % Right panel: orbit under g
    %%%%%%%%%%%%%%%%%%%%%%%%%%%%%%%%%%%%%%%
    \begin{scope}[xshift=5.8cm, yshift=-0.75cm]
        \node[font=\bfseries] at (2.4,4.0) {Orbit under $g$};

        % Nodes
        \node[draw, rounded corners, align=center, text width=2.2cm, inner sep=4pt] (A) at (2.4,3.0)
        {$\left(0,0;0,\frac{\pi}{6}\right)$};

        \node[draw, rounded corners, align=center, text width=2.2cm, inner sep=4pt] (B) at (0.6,1.0)
        {$\left(0,0;\frac{2\pi}{3},\frac{\pi}{2}\right)$};

        \node[draw, rounded corners, align=center, text width=2.2cm, inner sep=4pt] (C) at (4.2,1.0)
        {$\left(0,0;\frac{\pi}{3},\frac{5\pi}{6}\right)$};

        % Arrows
        \draw[->, thick] (A.south west) -- node[left] {$g$} (B.north);
        \draw[->, thick] (B.east) -- node[below] {$g$} (C.west);
        \draw[->, thick] (C.north) -- node[right] {$g$} (A.south east);
    \end{scope}

\end{tikzpicture}
\caption{Left: example of a size-3 brane orbit on one of the two-tori. Right: action of the orbifold generator $g$ on the corresponding orbit.}\label{fig:brane-orbit-z4-example}
\end{figure}
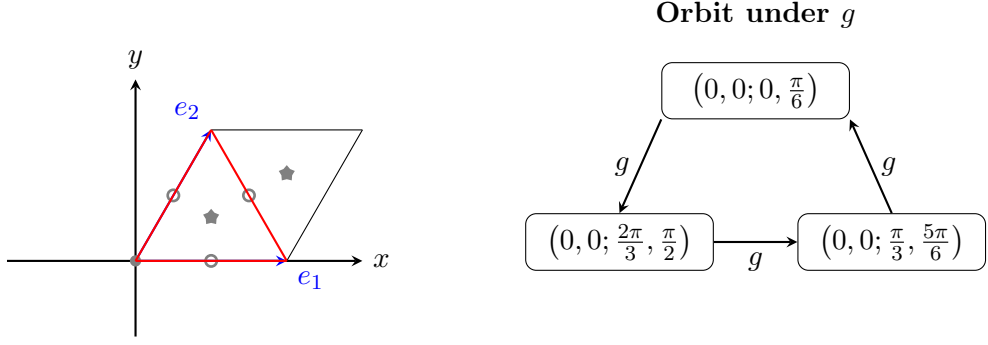

When the representative $D(p+2)$-brane passes through a $\Z_4$ fixed point, corresponding to $u=0$ or $u=\tfrac{1+T}{2}$ on both tori, the resulting boundary orbit is of size 3 (see \Cref{fig:brane-orbit-z4-example}) and is naturally diagonalised by four Chan-Paton sectors. The annulus amplitude is
\begin{align}
    \mathcal{A} = \frac14&\Big[T[^0_0](
     |N_0|^2\Lambda^{\mathcal{A} (1,0)}_{m_1,w_1}\Lambda^{\mathcal{A} (1,1)}_{m_2,w_2}(U,T)+
    |M_0|^2\Lambda^{\mathcal{A} (1,1)}_{m_1,w_1}\Lambda^{\mathcal{A} (1,0)}_{m_2,w_2}(U,T)  ) \notag\\
   &+ \sum_{\beta=1}^3 (|N_\beta|^2+|M_\beta|^2)T[^0_{3\beta}] \notag\\ 
    &+ \sum_{\beta=0}^3 (N_\beta \overline{M_\beta}+\overline{N_\beta} M_\beta)(T[^2_{3\beta}]+T[^{10}_{3\beta}]) \\
    &+ 3(|N_0|^2+|M_0|^2)(T[^4_0]+T[^8_0])-\sum_{\beta=1}^3 (|N_\beta|^2+|M_\beta|^2)(T[^4_{3\beta}]+T[^8_{3\beta}]) \notag\\
    &+4 \sum_{\beta=0,2} (N_\beta \overline{M_\beta}+\overline{N_\beta} M_\beta)T[^6_{3\beta}]+2\sum_{\beta=1,3} (N_\beta \overline{M_\beta}+\overline{N_\beta} M_\beta)T[^6_{3\beta}]\Big]\,,\notag
\end{align}
where the Chan-Paton traces $N_\beta$ and $M_\beta$ are defined by
\begin{align}
    N_\beta &= \Tr(\gamma^\beta_N)\,, \\
    M_\beta &= \Tr(\gamma^\beta_M)\,,
\end{align}
where
\begin{align}
    \gamma_N &= \operatorname{diag}\!\left(
        \Un_{n_0},
        \e{2i\pi/4}\Un_{n_1},
        \e{4i\pi/4}\Un_{n_2},
        \e{6i\pi/4}\Un_{n_3}
    \right)\,,\\
    \gamma_M &= \operatorname{diag}\!\left(
        \Un_{m_0},
        \e{2i\pi/4}\Un_{m_1},
        \e{4i\pi/4}\Un_{m_2},
        \e{6i\pi/4}\Un_{m_3}
    \right)\,.
\end{align}
The indices on $n_\gamma$ and $m_\gamma$ are understood modulo $4$. The massless contribution is
\begin{align}
    \mathcal{A}_0 \sim& \sum_{\gamma=0}^3\Big[(n_\gamma\overline{n}_\gamma+m_\gamma\overline{m}_\gamma)\chi[^0_0]+ (n_\gamma\overline{n}_{\gamma+1}+m_\gamma\overline{m}_{\gamma+1})\chi[^0_1]+( n_\gamma\overline{n}_{\gamma-1}+m_\gamma\overline{m}_{\gamma-1})\chi[^0_{11}] \notag\\
    &+(n_\gamma\overline{m}_\gamma+\overline{n}_\gamma m_\gamma)(\chi[^2_0]+\chi[^{10}_0])+(n_\gamma(3\overline{m}_{\gamma+2}+\overline{m}_{\gamma})+m_\gamma(3\overline{n}_{\gamma+2}+\overline{n}_{\gamma}))\chi[^6_0] \\
    &+(n_\gamma\overline{n}_{\gamma+1}+n_\gamma\overline{n}_{\gamma+2}+n_\gamma\overline{n}_{\gamma+3}+m_\gamma\overline{m}_{\gamma+1}+m_\gamma\overline{m}_{\gamma+2}+m_\gamma\overline{m}_{\gamma+3})(\chi[^4_0]+\chi[^8_0])\Big]\,.\notag
\end{align}

The tree channel annulus is
\begin{align}
    \widetilde{\mathcal{A}} = \frac{2^{-(p+1)/2}}{4}&\Big[ T[^0_0](
        |N_0|^2\widetilde{\Lambda}^{(1,0)}_{0,0}\widetilde{\Lambda}^{(1,1)}_{0,0} +
        |M_0|^2\widetilde{\Lambda}^{(1,1)}_{0,0}\widetilde{\Lambda}^{(1,0)}_{0,0})' \\
    &+3|N_0+M_0|^2 \xi_6[^0_0]+3|N_0-M_0|^2 \xi_6[^0_3] \notag\\
    &+ 3|N_2-M_2|^2 (2\xi_6[^6_0]+\xi_6[^6_2]+\xi_6[^6_4])+ 3|N_2+M_2|^2 (2\xi_6[^6_3]+\xi_6[^6_1]+\xi_6[^6_5]) \notag\\
    &+ |N_1-M_1|^2 (4\xi_6[^3_0]+\xi_6[^3_2]+\xi_6[^3_4])+ |N_1+M_1|^2 (4\xi_6[^3_3]+\xi_6[^3_1]+\xi_6[^3_5]) \notag\\
    &+ |N_3-M_3|^2 (4\xi_6[^9_0]+\xi_6[^9_2]+\xi_6[^9_4])+ |N_3+M_3|^2 (4\xi_6[^9_3]+\xi_6[^9_1]+\xi_6[^9_5])  \Big]\,.\notag
\end{align}
For these boundary conditions, one can see that both orbits couple to all the tensors from the $\Z_4$ twisted sectors as well as to those from the $\Z_2$ twisted sector but with different signs. However, these branes couple neither to the $\Z_3$ twisted sectors nor to the $\Z_6$ or $\Z_{12}$ twisted sectors. The gauge group is $\prod_{\gamma=0}^3 U(n_\gamma)\times\prod_{\gamma=0}^3 U(m_\gamma)$. The spectrum is summarised in \Cref{tab:Dpp2-orbitZ4-spectrum}.
\begin{table}[htb]
\centering
\begin{tabular}{c|c|c}
Sector & Representation & Field content \\\hline
$\alpha=0$ &
$n_\gamma\overline{n}_\gamma+m_\gamma\overline{m}_\gamma$ &
$A^\mu+(4-2k)\phi+2^{2-k}(\psi_L+\psi_R)$ \\\hline
$\alpha=0$ &
$n_\gamma(\overline{n}_{\gamma+1}+\overline{n}_{\gamma-1})+m_\gamma(\overline{m}_{\gamma+1}+\overline{m}_{\gamma-1})$ &
$\Phi+2^{1-k}(\psi_L+\psi_R)$ \\\hline
$\alpha\in\{2,10\}$ &
$n_\gamma\overline{m}_\gamma+\overline{n}_\gamma m_\gamma$ &
$2(2\varphi+2^{1-k}(\psi_L+\psi_R))$ \\\hline
$\alpha=6$ &
$n_\gamma\overline{m}_{\gamma+2}+m_\gamma\overline{n}_{\gamma+2}$ &
$3(2\varphi+2^{1-k}(\psi_L+\psi_R))$ \\\hline
$\alpha=6$ &
$n_\gamma\overline{m}_{\gamma}+m_\gamma\overline{n}_{\gamma}$ &
$2\varphi+2^{1-k}(\psi_L+\psi_R)$ \\\hline
$\alpha\in\{4,8\}$ &
$n_\gamma \sum_{r=1}^3\overline{n}_{\gamma+r}+m_\gamma\sum_{r=1}^3\overline{m}_{\gamma+r}$ &
$2(2\varphi+2^{1-k}(\psi_L+\psi_R))$ \\\hline
\end{tabular}
\caption{Massless spectrum of the genuine $D(p+2)$-brane orbit passing through a $\Z_4$ fixed point. The second column gives the representations under $\prod_{\gamma=0}^3 U(n_\gamma)\times\prod_{\gamma=0}^3 U(m_\gamma)$.}\label{tab:Dpp2-orbitZ4-spectrum}
\end{table}

Starting from the $\Z_4$ fixed configuration, one can move the representative brane to a generic position by turning on a Higgs-branch vev. A brane which can leave the fixed locus must contain the regular representation of the $\Z_4$ invariant group. Thus, on the branch leading to a generic bulk orbit, one takes
\begin{equation}
n_0=n_1=n_2=n_3=n,\qquad m_0=m_1=m_2=m_3=m.
\end{equation}
The relevant Higgs fields are the internal transverse complex scalars $\Phi$ in the second line of \Cref{tab:Dpp2-orbitZ4-spectrum}. They map neighbouring Chan-Paton components into one another. Giving equal vevs
\begin{equation}
\langle\Phi^N_\gamma\rangle=u_N\mathbf 1_n,\qquad \langle\Phi^M_\gamma\rangle=u_M\mathbf 1_m,\qquad \gamma=0,\ldots,3,
\end{equation}
identifies the four fractional components and breaks the gauge group to its diagonal subgroup,
\begin{equation}
\prod_{\gamma=0}^3 U(n)_\gamma\times\prod_{\gamma=0}^3 U(m)_\gamma\longrightarrow U(n)_{\rm diag}\times U(m)_{\rm diag}.
\end{equation}
For these genuine $D(p+2)$-branes, the parameters $u_N$ and $u_M$ are the transverse displacement moduli of the two internal one-cycle orbits. Equivalently, they specify the positions of the displaced representative branes away from the $\Z_4$ fixed locus.

Around this Higgsed vacuum, the fields reorganise into diagonal and non-diagonal combinations with respect to the four Chan-Paton components. In the first line of \Cref{tab:Dpp2-orbitZ4-spectrum}, only the diagonal combinations of $A^\mu$, $\phi$, $\psi_L$ and $\psi_R$ remain massless. The non-diagonal combinations are lifted: the corresponding gauge bosons become massive, while the associated scalars and fermions combine with them into massive multiplets. The fields in the second line, namely the internal transverse complex scalars $\Phi$, are the Higgs fields. Their diagonal component is the physical displacement modulus of the generic bulk orbit, while their non-diagonal gauge-orbit fluctuations are the Goldstone modes eaten by the massive gauge bosons. The remaining fields in the sectors $\alpha\in\{2,4,6,8,10\}$ are not Goldstone modes for this displacement. Instead, after the breaking to $U(n)_{\rm diag}\times U(m)_{\rm diag}$, their Chan-Paton labels are recombined into representations of the diagonal gauge group. The resulting generic-position spectrum is summarised in \Cref{tab:Dpp2-orbitgeneric-spectrum}.
\begin{table}[htb]
\centering
\begin{tabular}{c|c|c}
Sector & Representation & Field content \\\hline
$\alpha=0$ &
$n\overline{n}+m\overline{m}$ &
$A^\mu+(4-2k)\phi+2^{2-k}(\psi_L+\psi_R)$ \\\hline
$\alpha=0$ &
$n\overline{n}+m\overline{m}$ &
$2(\Phi+2^{1-k}(\psi_L+\psi_R))$ \\\hline
$\alpha\in\{2,10\}$ &
$n\overline{m}+\overline{n}m$ &
$8(2\varphi+2^{1-k}(\psi_L+\psi_R))$ \\\hline
$\alpha=6$ &
$n\overline{n}+m\overline{m}$ &
$16(2\varphi+2^{1-k}(\psi_L+\psi_R))$ \\\hline
$\alpha\in\{4,8\}$ &
$n\overline{n}+m\overline{m}$ &
$24(2\varphi+2^{1-k}(\psi_L+\psi_R))$ \\\hline
\end{tabular}
\caption{Massless spectrum of the genuine $D(p+2)$-brane orbit at a generic position. The second column gives the representations under the diagonal gauge group $U(n)_{\rm diag}\times U(m)_{\rm diag}$.}\label{tab:Dpp2-orbitgeneric-spectrum}
\end{table}

In the tree channel, the interpretation is simpler. At the $\Z_4$ fixed point the boundary state couples to the untwisted fields, to the $\Z_4$ twisted tensors localised at the corresponding fixed point orbit, and to the appropriate $\Z_2$ twisted fields, as displayed by the $\xi_6$ terms in the tree annulus. After the displacement to a generic position, the brane no longer lies on any orbifold fixed locus so it does not source any twisted fields.

\subsubsection{Branes at \texorpdfstring{$\Z_2$}{Z2} points}

We now consider boundary orbits whose representative is at a $\Z_2$ fixed point. The invariant group is then smaller than the $\Z_4$ invariant group and the Chan-Paton action is diagonalised only with respect to this smaller group. This gives intermediate fractional configurations, between the orbit of size two at the origin and a completely generic orbit of size twelve.

\begin{figure}[htb]
\centering

\begin{tikzpicture}[>=stealth]

    %%%%%%%%%%%%%%%%%%%%%%%%%
    % Left panel: torus plot
    %%%%%%%%%%%%%%%%%%%%%%%%%
    
    \begin{scope}[scale=2]
        % Axes
        \draw[->, thick] (-.85, 0) -- (1.5, 0) node[right] {$x$};
        \draw[->, thick] (0, -0.5) -- (0, 1.2) node[above] {$y$};

        % Basis vectors
        \draw[->, thick, blue] (0,0) -- (1,0) node[below right] {$e_1$};
        \draw[->, thick, blue] (0,0) -- ({cos(60)}, {sin(60)}) node[above left] {$e_2$};

        % Fundamental cell
        \draw[-] ({cos(60)}, {sin(60)}) -- ({1+cos(60)}, {sin(60)});
        \draw[-] (1,0) -- ({1+cos(60)}, {sin(60)});

        % Fixed points
        \node[circle, fill=gray, inner sep=1.5pt] at (0,0) {};
        \node[star, star points=5, fill=gray, inner sep=1.5pt] at ({1/3+cos(60)/3}, {sin(60)/3}) {};
        \node[star, star points=5, fill=gray, inner sep=1.5pt] at ({2/3+2*cos(60)/3}, {2*sin(60)/3}) {};
        \node[circle, draw=gray, fill=none, inner sep=1.5pt, line width=1pt] at ({1/2+cos(60)/2}, {sin(60)/2}) {};
        \node[circle, draw=gray, fill=none, inner sep=1.5pt, line width=1pt] at ({cos(60)/2}, {sin(60)/2}) {};
        \node[circle, draw=gray, fill=none, inner sep=1.5pt, line width=1pt] at (1/2,0) {};

        % Branes
        \draw[-, thick, red] (0,0) -- (1,0);
        \draw[-, thick, red] (0,0) -- ({cos(60)}, {sin(60)});
        \draw[-, thick, red] (1,0) -- ({cos(60)}, {sin(60)});
        \draw[-, thick, red] ({cos(60)/2}, {sin(60)/2}) -- ({cos(60)/2+1}, {sin(60)/2});
        \draw[-, thick, red] (1/2,0) -- ({cos(60)+1/2}, {sin(60)});
        \draw[-, thick, red] ({cos(60)/2}, {sin(60)/2}) -- (1/2,0);
        \draw[-, thick, red] ({cos(60)/2+1}, {sin(60)/2}) -- ({cos(60)+1/2}, {sin(60)});
    \end{scope}

    %%%%%%%%%%%%%%%%%%%%%%%%%%%%%%%%%%%%%%%
    % Right panel: orbit under g
    %%%%%%%%%%%%%%%%%%%%%%%%%%%%%%%%%%%%%%%
    \begin{scope}[xshift=5.8cm, yshift=-1cm]
        \node[font=\bfseries] at (3.3,5.2) {Orbit under $g$};

        % Left column
        \node[draw, rounded corners, align=center, text width=2.8cm, inner sep=4pt] (L1) at (1.4,4.1)
        {$\left(\frac12,0;0,\frac{\pi}{6}\right)$};

        \node[draw, rounded corners, align=center, text width=2.8cm, inner sep=4pt] (L2) at (1.4,2.5)
        {$\left(\frac12,0;\frac{\pi}{3},\frac{5\pi}{6}\right)$};

        \node[draw, rounded corners, align=center, text width=2.8cm, inner sep=4pt] (L3) at (1.4,0.9)
        {$\left(\frac12,0;\frac{2\pi}{3},\frac{\pi}{2}\right)$};

        % Right column
        \node[draw, rounded corners, align=center, text width=2.8cm, inner sep=4pt] (R1) at (5.2,4.1)
        {$\left(\frac{T}{2},0;\frac{2\pi}{3},\frac{\pi}{2}\right)$};

        \node[draw, rounded corners, align=center, text width=2.8cm, inner sep=4pt] (R2) at (5.2,2.5)
        {$\left(\frac{T}{2},0;0,\frac{\pi}{6}\right)$};

        \node[draw, rounded corners, align=center, text width=2.8cm, inner sep=4pt] (R3) at (5.2,0.9)
        {$\left(\frac{T}{2},0;\frac{\pi}{3},\frac{5\pi}{6}\right)$};

        % Orbit arrows
        \draw[->, thick] (L1.east) -- node[above] {$g$} (R1.west);
        \draw[->, thick] (R1.south west) -- node[right] {$g$} (L2.north east);
        \draw[->, thick] (L2.east) -- node[above] {$g$} (R2.west);
        \draw[->, thick] (R2.south west) -- node[right] {$g$} (L3.north east);
        \draw[->, thick] (L3.east) -- node[above] {$g$} (R3.west);
        \draw[->, thick] (R3.south)-- ++(0,-0.25)-- ++(-6,0) -- ++(0,3.845)-- node[above] {$g$} (L1.west);
    \end{scope}

\end{tikzpicture}
\caption{Left: example of a size-6 brane orbit on one of the two-tori. Right: action of the orbifold generator $g$ on the corresponding orbit.}\label{fig:brane-orbit-z2-2}
\end{figure}
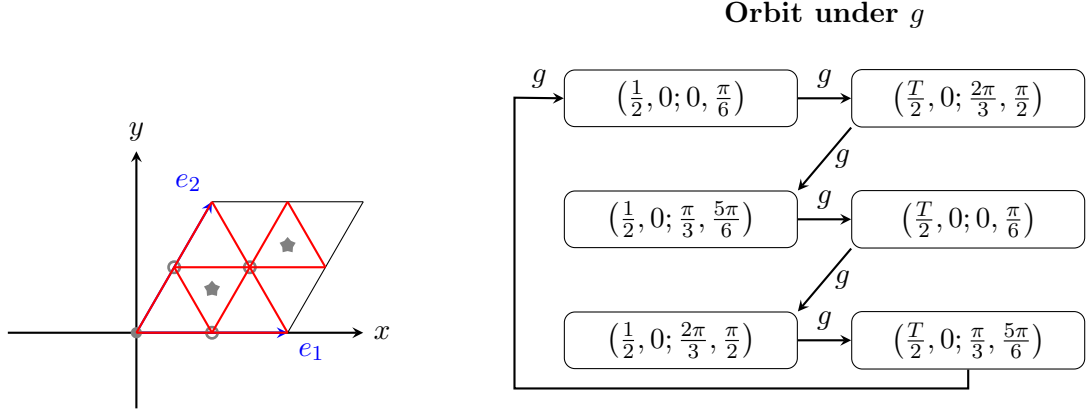
Let us first place the representative brane at a $\Z_2$ fixed point. This corresponds to the positions $u=\frac{\alpha+\beta U}{2}$ on the first torus and $u=\frac{\lambda+\mu U}{2}$ on the second torus, with $\alpha,\beta,\lambda,\mu\in\Z_2$ such that the fixed point is not a $\Z_4$ fixed point, as represented in \Cref{fig:brane-orbit-z2-2}. The invariant group is then $\Z_2$.

The corresponding gauge group is $U(n_0)\times U(n_1)\times U(m_0)\times U(m_1)$. The spectrum is summarised in \Cref{tab:Dpp2-orbitZ2-spectrum}.
\begin{table}[htb]
\centering
\begin{tabular}{c|c|c}
Sector & Representation & Field content \\\hline
$\alpha=0$ &
$n_0\overline{n}_0+n_1\overline{n}_1+m_0\overline{m}_0+m_1\overline{m}_1$ &
$A^\mu+(4-2k)\phi+2^{2-k}(\psi_L+\psi_R)$ \\\hline
$\alpha=0$ &
$n_0\overline{n}_1+n_1\overline{n}_0+m_0\overline{m}_1+m_1\overline{m}_0$ &
$2(\Phi+2^{1-k}(\psi_L+\psi_R))$ \\\hline
$\alpha\in\{2,10\}$ &
$n_0\overline{m}_0+n_1\overline{m}_1+m_0\overline{n}_0+m_1\overline{n}_1$ &
$4(2\varphi+2^{1-k}(\psi_L+\psi_R))$ \\\hline
$\alpha=6$ &
$n_0\overline{m}_0+n_1\overline{m}_1+m_0\overline{n}_0+m_1\overline{n}_1$ &
$6(2\varphi+2^{1-k}(\psi_L+\psi_R))$ \\\hline
$\alpha=6$ &
$n_0\overline{m}_1+n_1\overline{m}_0+m_0\overline{n}_1+m_1\overline{n}_0$ &
$2(2\varphi+2^{1-k}(\psi_L+\psi_R))$ \\\hline
$\alpha\in\{4,8\}$ &
$n_0\overline{n}_0+n_1\overline{n}_1+m_0\overline{m}_0+m_1\overline{m}_1$ &
$8(2\varphi+2^{1-k}(\psi_L+\psi_R))$ \\\hline
$\alpha\in\{4,8\}$ &
$n_0\overline{n}_1+n_1\overline{n}_0+m_0\overline{m}_1+m_1\overline{m}_0$ &
$4(2\varphi+2^{1-k}(\psi_L+\psi_R))$ \\\hline
\end{tabular}
\caption{Massless spectrum of the genuine $D(p+2)$-brane orbit passing through a $\Z_2$ fixed point. The second column gives the representations under $U(n_0)\times U(n_1) \times U(m_0)\times U(m_1)$.}\label{tab:Dpp2-orbitZ2-spectrum}
\end{table}

The tree channel expression contains the untwisted exchange and the twisted exchanges in the $\alpha=6$ sectors, with phase factors determined by the displaced lattice sums. Thus these boundary states couple to the untwisted tensor and to the $\Z_2$ twisted fields localised at the corresponding fixed-point orbit. They couple neither to the $\Z_3$ twisted fields, nor to the $\Z_6$ and $\Z_{12}$ singlets, because the representative brane is not located at the corresponding fixed loci.

One can again move from this $\Z_2$ fixed point to a generic bulk point by turning on a Higgs-branch vev.
The resulting massless spectrum is the generic one displayed in \Cref{tab:Dpp2-orbitgeneric-spectrum}.

\section{Untwisted R-R source conventions}\label{sec:OpDp-untwisted-couplings}

This appendix collects only the R-R source normalisations needed to interpret the tree amplitudes of the main text. We normalise the elementary D$p$-brane R-R charge to
\begin{equation}
    Q_{Dp}=+1,
    \qquad
    Q_{\overline{Dp}}=-1.
\end{equation}
For the standard orientifold planes,
\begin{equation}
    Q_{O_p^-}=-2^{p-4},
    \qquad
    Q_{O_p^+}=+2^{p-4},
\end{equation}
in D$p$-brane charge units. In particular,
\begin{equation}
    Q_{O_7^\pm}=\pm8.
\end{equation}
Changing a brane into an antibrane reverses its R-R charge but not its positive tension; this distinction is relevant for the Brane Supersymmetry Breaking descendant. Let
\begin{equation}
    v=\frac{\operatorname{Vol}(T^4)}{\alpha'^2}
\end{equation}
denote the dimensionless internal volume. For a six-dimensional space-filling source, define its wrapped internal volume by
\begin{equation}
    \mathcal V_{\parallel}^{(p)}
    =
    \begin{cases}
        v, & p=9,\\
        L_{\parallel,1}L_{\parallel,2}, & p=7,\\
        1, & p=5,
    \end{cases}
\end{equation}
where $L_{\parallel,i}$ is the dimensionless length of the wrapped one-cycle on the $i$-th two-torus. After canonical normalisation of the six-dimensional R-R field, the corresponding untwisted one-point coupling scales as
\begin{equation}
    \Gamma_p
    \propto
    Q_p\frac{\mathcal V_{\parallel}^{(p)}}{\sqrt v}.
\end{equation}
Thus
\begin{equation}
    \Gamma_p^2
    \propto
    Q_p^2\frac{\bigl(\mathcal V_{\parallel}^{(p)}\bigr)^2}{v},
\end{equation}
or explicitly
\begin{equation}
\begin{array}{c|c|c}
\text{source} & \Gamma_p/Q_p & \Gamma_p^2/Q_p^2\\
\hline
O9,D9 & \sqrt v & v\\[1mm]
O7,D7 & \dfrac{L_{\parallel,1}L_{\parallel,2}}{\sqrt v}
      & \dfrac{L_{\parallel,1}^2L_{\parallel,2}^2}{v}\\[3mm]
O5,D5 & \dfrac1{\sqrt v} & \dfrac1v
\end{array}
\end{equation}

These factors are precisely those appearing in the massless untwisted R-R part of the tree channel amplitudes
\begin{equation}
    \widetilde{\mathcal K}_{\rm RR}\sim\Gamma_C^2,
    \qquad
    \widetilde{\mathcal A}_{\rm RR}\sim\Gamma_B^2,
    \qquad
    \widetilde{\mathcal M}_{\rm RR}\sim\Gamma_C\Gamma_B.
\end{equation}
The remaining numerical coefficients depend on the orbifold projector, modular transformation and number of source images and are fixed directly by factorization of the complete amplitudes.

This explains precisely the different charges between the two orbits $N$ and $M$ in the main text and in the alternative amplitudes of \Cref{app:other_vacua} that one can see from the untwisted term in the tree channel.

If a brane or orientifold plane belongs to an orbifold orbit, its coupling to a fixed closed-string state $I$ is the coherent sum of the one-point functions of its images,
\begin{equation}
    \Gamma_I^{\rm orbit}
    =
    \sum_{a\in{\rm orbit}}
    \Gamma_{I,a}.
\end{equation}
The sum is performed before squaring in $\widetilde{\mathcal K}$ or $\widetilde{\mathcal A}$. Relative orbifold phases and signs can therefore project out particular couplings. For twisted R-R states the corresponding coefficients are read directly from the factorized twisted characters; their normalisation is not determined by the untwisted volume scaling above.

\section{Alternative orientifold projections and auxiliary amplitudes}\label{app:other_vacua}

\subsection{Second orientifold projection of the \texorpdfstring{$\Z_{12}$}{Z12}-orbifold}

In the main text we have considered the $\Z_{12}$-orbifold with the orientifold projection $\Omega\sigma$ which is not the simplest parity projection but had the advantage of making the $N$ and $M$ cycles quite symmetric. In this section we consider a second possible projection, namely the standard $\Omega$\footnote{The difference between these two projections is equivalent to the standard choices of lattices A and B.}. There is still the relation $\Omega^{-1}.g.\Omega=g^7$ such that the same twisted sectors are paired.
\begin{table}[htbp]
\centering
\small
\renewcommand{\arraystretch}{1.15}
\begin{tabularx}{\textwidth}{p{0.19\textwidth}|p{0.2\textwidth}|p{0.25\textwidth}|X}
Sector(s) & Orbit type & Orientifold projection & Contribution \\
\hline
Untwisted sector & bulk sector & symmetrized & $\mathcal N=(1,0)$ \par SUGRA $+\,1H$ \\
\hline
$\alpha=(1,5)$, $(7,11)$, $(2,10)$ and $(4,8)$ & $1$ & paired sectors & $4(H+T)$ \\
\hline
$\alpha=(4,8)$ & $4+4$ & paired sectors & $2(H+T)$ \\
\hline
$\alpha=3,9$ & $1$ & symmetrized & $2H$ \\
\hline
$\alpha=3,9$ & $3$ & symmetrized & $2H$ \\
\hline
$\alpha=6$ & $1$ & symmetrized & $1H$ \\
\hline
$\alpha=6$ & $3+6+6$ & symmetrized & $3H$ \\
\hline
Total twisted & -- & -- & $14H+6T$ \\
\hline
Total closed spectrum & -- & -- & $\mathcal N=(1,0)$ \par SUGRA $+\,15H+6T$ \\
\hline
\end{tabularx}
\caption{Closed-string spectrum after the second orientifold projection.}\label{tab:Z12AA-unoriented-closed-spectrum}
\end{table}

The Klein bottle amplitude reads
\begin{align}
    \mathcal{K} =& \frac13\Big[T[^0_0]\left(\frac{
        \Lambda^{\mathcal K,(1,0)}_1\Lambda^{\mathcal K,(1,0)}_1(T,U)+
        \Lambda^{\mathcal K,(1,1)}_2\Lambda^{\mathcal K,(1,1)}_2(T,U)
        }{2}\right) + \sum_{\beta=1}^2 T[^0_{4\beta}] \\
    &+ 4T[^3_0] + \sum_{\beta=1}^2 T[^3_{4\beta}]
    + 10T[^6_0] +  \sum_{\beta=1}^2 T[^6_{4\beta}]
    + 4T[^9_0]  + \sum_{\beta=1}^2 T[^9_{4\beta}]\Big]\notag\\
    =&\frac16T[^0_0]\left(\Lambda^{0}_{m_1,w_1,m_2,w_2}+\Lambda^{1}_{m_1,w_1,m_2,w_2}\right)' +\xi_3[^0_0]+\xi_3[^3_0]+\xi_3[^6_0]+\xi_3[^9_0] \notag\\
    &+ \sum_{\beta=0}^3(\xi_4[^3_\beta]+\xi_4[^6_\beta] +\xi_4[^9_\beta] ) + 2\xi_2[^6_0] +2\xi_2[^6_1] \,,\notag
\end{align}
whose tree channel reads
\begin{align}
    \widetilde{\mathcal{K}} =& \frac{2^3}{3}\Big[T[^0_0]
    \begin{aligned}[t]
        \bigl(\mathcal{C}_+(m_1,w_1)^2\widetilde{\Lambda}^{(1,0)}_{0,0}\mathcal{C}_+(m_2,w_2)^2\widetilde{\Lambda}^{(1,0)}_{0,0} +\\
        \mathcal{C}_-(m_1,w_1)^2\widetilde{\Lambda}^{(1,1)}_{0,0}\mathcal{C}_-(m_2,w_2)^2\widetilde{\Lambda}^{(1,1)}_{0,0} \bigr)'
        \end{aligned} \\
        &+2(\sqrt{2}+\frac1{\sqrt{2}})^2\xi_4[^0_0]+2(\sqrt{2}-\frac1{\sqrt{2}})^2\xi_4[^0_2]
    +4\sum_{\alpha\in\{4,8\}}(\xi_4[^{\alpha}_1]+\xi_4[^{\alpha}_2]+\xi_4[^{\alpha}_3]) \Big] \,.\notag
    \label{eq:KleintransverseAA}
\end{align}
One can already see, from the term $\propto (\sqrt{2}\pm\frac{1}{\sqrt{2}})^2$ that the charges of the two orbits $N$ and $M$ are now different by a factor of 2 and this will appear in the branes couplings as well (see \Cref{sec:OpDp-untwisted-couplings}).
The associated unoriented closed-string spectrum is summarised in \Cref{tab:Z12AA-unoriented-closed-spectrum}. The $O7$-planes are now of two different types, $O7^-$ and $O7^+$, with opposite R-R charges. Their wrapped bulk cycles are given in \Cref{tab:O7planesbulkcyclesAA}.

\begin{table}[htbp]
\centering
\renewcommand{\arraystretch}{1.25}
\begin{tabular}{c|c}
    $O$-planes & Wrapped cycle on $\T^2\times\T^2$ \\\hline
    $O7^-$-planes &
    $\begin{aligned}
    \Pi_H^- = & (H_0+H_{\frac12})\circ(H_0+H_{\frac12}) +
    (V_0+V_{\frac12})\circ(V_0+V_{\frac12}) \\
    \Pi_D^- = & (A_{\frac12}+\sigma_{(1,-1)}^{(\frac12,\frac12)})\circ(D_{\frac12}+\sigma_{(1,1)}^{(\frac12,\frac12)}) + \sigma_{(1,-1)}^{(0,0)}\circ \sigma_{(1,1)}^{(0,0)} \\
    +& (D_{\frac12}+\sigma_{(1,1)}^{(\frac12,\frac12)})\circ(A_{\frac12}+\sigma_{(1,-1)}^{(\frac12,\frac12)}) + \sigma_{(1,1)}^{(0,0)}\circ \sigma_{(1,-1)}^{(0,0)}
    \end{aligned}$
    \\\hline
    $O7^+$-planes &
    $\begin{aligned}
    \Pi_D^+ = & \sigma_{(1,-1)}^{(0,0)}\circ (\sigma_{(1,1)}^{(\frac12,\frac12)}+D_{\frac12})+
    \sigma_{(1,1)}^{(0,0)}\circ (\sigma_{(1,-1)}^{(\frac12,\frac12)}+A_{\frac12})\\
    +&
    (\sigma_{(1,-1)}^{(\frac12,\frac12)}+A_{\frac12})\circ \sigma_{(1,1)}^{(0,0)}+
    (\sigma_{(1,1)}^{(\frac12,\frac12)}+D_{\frac12})\circ \sigma_{(1,-1)}^{(0,0)}
    \end{aligned}$
    \\\hline
\end{tabular}
\caption{Wrapped bulk cycles of the $O7^\pm$-plane orbits in the \Tp-dual frame in the case of the second orientifold ($\Omega$) model.}\label{tab:O7planesbulkcyclesAA}
\end{table}

The most natural choice is to put $D7$-branes parallel to the $O7$-planes in the dual picture. Let $N_0$ be the number of $D7$-branes parallel to the horizontal/vertical $O7$-planes whose orbit will be labelled $\mathcal O_N$, and $M_0$ be the number of $D7$-branes parallel to the diagonal $O7$-planes whose orbit will be labelled $\mathcal O_M$. The orbits $\mathcal O_N$ and $\mathcal O_M$ correspond to the bulk cycles
\begin{align}
    \Pi_N &= \sigma_{(1,0)}^{(0,0)}\circ\sigma_{(1,0)}^{(0,0)} + \sigma_{(0,1)}^{(0,0)}\circ\sigma_{(0,1)}^{(0,0)}\,,\\
    \Pi_M &= \sigma_{(1,-1)}^{(0,0)}\circ\sigma_{(1,1)}^{(0,0)} + \sigma_{(1,1)}^{(0,0)}\circ\sigma_{(1,-1)}^{(0,0)}\,.
\end{align}
The annulus amplitude for these branes is given by the following expression
\begin{align}
   \mathcal{A} =& \frac16\Big[T[^0_0](
    N_0^2\Lambda^{\mathcal{A} (1,0)}_{m_1,w_1}\Lambda^{\mathcal{A} (1,0)}_{m_2,w_2}(T,U)+
    M_0^2\Lambda^{\mathcal{A} (1,1)}_{m_1,w_1}\Lambda^{\mathcal{A} (1,1)}_{m_2,w_2}(T,U)  ) \\
    &+\sum_{\beta=1}^5 (N_\beta^2+M_\beta^2)T[^0_{2\beta}] \notag\\
   &+ 2\sum_{\beta=0}^5 N_\beta M_\beta(T[^3_{2\beta}]+T[^9_{2\beta}]) \notag\\
   &+ (N_0^2+4M_0^2)T[^6_{0}] - (N_3^2+4M_3^2)T[^6_{6}]-\sum_{\beta=1,2,4,5}(-1)^\beta(N_\beta^2+M_\beta^2)T[^6_{2\beta}]\notag\Big],
\end{align}
where $N_\beta$ and $M_\beta$ are the Chan-Paton trace factors under the action of the orbifold.
The tree channel of this amplitude is given by the following expression
\begin{align}
    \widetilde{\mathcal{A}} = \frac{2^{-3}}{3}&\Big[\frac14T[^0_0](
        N_0^2\widetilde{\Lambda}^{(1,0)}_{0,0}\widetilde{\Lambda}^{(1,0)}_{0,0} +
        M_0^2\widetilde{\Lambda}^{(1,1)}_{0,0}\widetilde{\Lambda}^{(1,1)}_{0,0})' \notag\\
        &+\frac12((\frac{N_0}{\sqrt{2}}+\sqrt{2}M_0)^2\xi_4[^0_0]+(\frac{N_0}{\sqrt{2}}-\sqrt{2}M_0)^2\xi_4[^0_2]) \notag\\
    &+\sum_{\beta=1,2,4,5}((N_\beta-M_\beta)^2\xi_4[^{2\beta}_0]+(N_\beta+M_\beta)^2\xi_4[^{2\beta}_2]) \notag\\
    &+2\sum_{\beta\in\{2,4\}}(N_\beta^2+M_\beta^2)(\xi_4[^{2\beta}_1]+\xi_4[^{2\beta}_3]) \notag\\
    &+ (\frac{N_3}{\sqrt{2}}-\sqrt{2}M_3)^2\xi_4[^6_0]+ (\frac{N_3}{\sqrt{2}}+\sqrt{2}M_3)^2\xi_4[^6_2] \notag\\
    &+2(N_3^2+M_3^2)(\xi_4[^6_0]+\xi_4[^6_2])+\frac32 M_3^2(\xi_4[^6_1]+\xi_4[^6_3])  \Big]\,.
\end{align}
The tree channel Möbius amplitude is fixed by factorization: it is the geometric mean of the tree Klein bottle and annulus coefficients, with the appropriate hatted characters implementing the $P$ modular transformation. After transforming back to the loop channel, the resulting coefficients must be compatible with the annulus multiplicities and define a consistent orientifold projection on the open-string Hilbert space. The resulting tree channel Möbius amplitude is given by the following expression
\begin{align}
    \widetilde{\mathcal{M}}
    = \frac13\Bigg[&
    T[^0_0]
    \begin{aligned}[t]
        \bigl(N_0 \mathcal{C}_+(m_1,w_1)\widetilde{\Lambda}^{(1,0)}_{0,0} \mathcal{C}_+(m_2,w_2)\widetilde{\Lambda}^{(1,0)}_{0,0} + \\
        M_0 \mathcal{C}_-(m_1,w_1)\widetilde{\Lambda}^{(1,1)}_{0,0} \mathcal{C}_-(m_2,w_2)\widetilde{\Lambda}^{(1,1)}_{0,0}\bigr)'
    \end{aligned}
    \label{eq:MobiusTAA}\\
    -&2(\sqrt{2}M_0+\frac{N_0}{\sqrt{2}})(\sqrt{2}+\frac1{\sqrt{2}})\widehat{\xi}_4[^0_0]-2(\sqrt{2}M_0-\frac{N_0}{\sqrt{2}})(\sqrt{2}-\frac1{\sqrt{2}})\widehat{\xi}_4[^0_2] \notag\\
    -&4\sum_{\beta\in\{2,4\}}(N_\beta+M_\beta)
    \left(
        \widehat{\xi}_4[^{2\beta}_1]
        +\widehat{\xi}_4[^{2\beta}_2]
        +\widehat{\xi}_4[^{2\beta}_3]
    \right)
    \Bigg]\,.\notag
\end{align}
The tadpole condition reads as follows
\begin{align}
    N_0 = M_0 &= 16\,,\\
    N_1-M_1 &= N_2-M_2 = N_4-M_4 = N_5-M_5 = 0\,,\\
    N_3=M_3 &=0\,.
\end{align}
The parametrisation of the Chan-Paton traces is given by
\begin{align}
    N_\beta &= \Tr(\gamma^\beta_N)\,, \\
    M_\beta &= \Tr(\gamma^\beta_M)\,,
\end{align}
where
\begin{align}
    \gamma_N &= \operatorname{diag}\!\left(
        \Un_{n_0},
        \e{2i\pi/6}\Un_{n_1},
        \e{4i\pi/6}\Un_{n_2},
        \e{6i\pi/6}\Un_{n_3},
        \e{8i\pi/6}\Un_{n_4},
        \e{10i\pi/6}\Un_{n_5}
    \right)\,,\\
    \gamma_M &= \operatorname{diag}\!\left(
        \Un_{m_0},
        \e{2i\pi/6}\Un_{m_1},
        \e{4i\pi/6}\Un_{m_2},
        \e{6i\pi/6}\Un_{m_3},
        \e{8i\pi/6}\Un_{m_4},
        \e{10i\pi/6}\Un_{m_5}
    \right)\,.
\end{align}
The orientifold projection further constrains the Chan-Paton matrices with the following relations
\begin{equation}
    n_2= n_1,\qquad n_3= n_0,\qquad n_5= n_4,\qquad
    m_2= m_1,\qquad m_3= m_0,\qquad m_5= m_4.
\end{equation}
The solution of the tadpole condition is given by
\begin{equation}
    n_4=8-n_1-n_0,
\end{equation}
together with
\begin{equation}
    n_0=m_0,\qquad n_1=m_1,\qquad n_4=m_4.
\end{equation}

Finally, we obtain the gauge group
$${U(n_0)_1\times U(n_1)_2\times U(n_4)_3\times U(m_0)_4\times U(m_1)_5\times U(m_4)_6}$$
and a massless spectrum that comprises an $\mathcal{N}=(1,0)$ gauge vector and a hypermultiplet in the representation
\begin{align}
 &A_1\oplus \overline{A}_1 \oplus A_2\oplus \overline{A}_2 \oplus A_3 \oplus \overline{A}_3\oplus A_4\oplus A_5\oplus A_6 \notag \\
 &\oplus B_{1\overline{2}}\oplus B_{\overline{1}2} \oplus B_{1\overline{3}} \oplus B_{\overline{1}3} \oplus B_{2\overline{3}}\oplus B_{\overline{2}3} \oplus B_{4\overline{5}}\oplus B_{4\overline{6}} \oplus B_{5\overline{6}}\notag\\
 &\oplus B_{14}\oplus B_{\overline{1}\overline{4}} \oplus B_{25} \oplus B_{\overline{2}\overline{5}} \oplus B_{36}\oplus B_{\overline{3}\overline{6}}   \,.
\end{align}

The non-Abelian anomaly polynomial reads as follows
\begin{align}
    I_8 = \frac18 \Big[&\big(\sqrt{3} \tr R^2 - \frac2{\sqrt{3}} \tr F_1^2- \frac2{\sqrt{3}} \tr F_2^2- \frac2{\sqrt{3}} \tr F_3^2- \frac1{\sqrt{3}} \tr F_4^2- \frac1{\sqrt{3}} \tr F_5^2- \frac1{\sqrt{3}} \tr F_6^2 \big)^2 \notag\\
    &-\big(- \frac2{\sqrt{3}} \tr F_1^2+ \frac1{\sqrt{3}} \tr F_2^2+ \frac1{\sqrt{3}} \tr F_3^2+ \frac2{\sqrt{3}} \tr F_4^2- \frac1{\sqrt{3}} \tr F_5^2- \frac1{\sqrt{3}} \tr F_6^2  \big)^2 \notag\\
    &-(-  \tr F_2^2+  \tr F_3^2+  \tr F_5^2-  \tr F_6^2)^2\Big]\,.
\end{align}
A representative of the anomaly vectors is
\begin{equation}
\setlength{\arraycolsep}{3pt}
\begin{array}{
r@{\;}c@{\;}l@{}
r@{,\,}r@{,\,}r@{,\,}r@{,\,}r@{,\,}r@{,\,}r
@{}l
}

a &=& \bigl( &\sqrt{3} & 0 & 0 & 0 & 0 & 0 & 0 &\bigr),\\[1mm]
b_1 &=& \bigl(& -\tfrac{2}{\sqrt{3}} & -\tfrac{2}{\sqrt{3}} & 0 & 0 & 0 & 0 & 0 &\bigr), \\[1mm]
b_2 &=& \bigl(& -\tfrac{2}{\sqrt{3}} & \tfrac{1}{\sqrt{3}} & -1 & 0 & 0 & 0 & 0 &\bigr), \\[1mm]
b_3 &=& \bigl(& -\tfrac{2}{\sqrt{3}} & \tfrac{1}{\sqrt{3}} & 1 & 0 & 0 & 0 & 0 &\bigr), \\[1mm]
b_4 &=& \bigl(& -\tfrac{1}{\sqrt{3}} & \tfrac{2}{\sqrt{3}} & 0 & 0 & 0 & 0 & 0 &\bigr), \\[1mm]
b_5 &=& \bigl(& -\tfrac{1}{\sqrt{3}} & -\tfrac{1}{\sqrt{3}} & 1 & 0 & 0 & 0 & 0 &\bigr), \\[1mm]
b_6 &=& \bigl(& -\tfrac{1}{\sqrt{3}} & -\tfrac{1}{\sqrt{3}} & -1 & 0 & 0 & 0 & 0 &\bigr). \\
\end{array}
\end{equation}
We arrange the tensor fields in the following order: first the gravity self-dual tensor, then the singlet tensor from the $\Z_6$ twisted sector, then the singlet from the $\Z_3$ twisted sector, then the remaining tensors.

\subsection{Other \texorpdfstring{$\Z_{12}$}{Z12} vacua}

One can also consider the orientifold projection $\Omega\sigma'\mathcal{I}_{79}(-1)^{F_L}$, where
\begin{align}
    \sigma':&\,
    \begin{cases}
        Z^1\mapsto Z^1 \\
        Z^2\mapsto \e{2i\pi/3} Z^2 \,.
    \end{cases}
\end{align}
It satisfies the relation $(\Omega\sigma'\mathcal{I}_{79}(-1)^{F_L})^{-1}.g.(\Omega\sigma'\mathcal{I}_{79}(-1)^{F_L})=g^5$, and therefore it is a consistent orientifold projection.

On the orbifold the orientifold projection sends the twisted sector Hilbert space $\mathcal{H}_{\alpha}$ to $\mathcal{H}_{7\alpha}$ such that the only invariant states come from the twisted sectors $\alpha=0,2,4,6,8,10$. 
\begin{table}[htbp]
\centering
\small
\renewcommand{\arraystretch}{1.15}
\begin{tabularx}{\textwidth}{p{0.19\textwidth}|p{0.2\textwidth}|p{0.25\textwidth}|X}
Sector(s) & Orbit type & Orientifold projection & Contribution \\
\hline
Untwisted sector & bulk sector & symmetrized & $\mathcal N=(1,0)$\par SUGRA $+\,1H$ \\
\hline
$\alpha=(1,5)$, $(7,11)$ and $(3,9)$ & $1$ & paired sectors & $3(H+T)$ \\
\hline
$\alpha=2,10$ & $1$ & symmetrized & $2H$ \\
\hline
$\alpha=4,8$ & $1$ & antisymmetrized & $2T$ \\
\hline
$\alpha=4,8$ & $4+4$ & symmetrized & $2H$ \\
\hline
$\alpha=(3,9)$ & $3$ & paired sectors & $H+T$ \\
\hline
$\alpha=6$ & $1$ & symmetrized & $1H$ \\
\hline
$\alpha=6$ & $3+6+6$ & symmetrized & $3H$ \\
\hline
Total twisted & -- & -- & $15H+5T$ \\
\hline
Total closed spectrum & -- & -- & $\mathcal N=(1,0)$\par SUGRA $+\,16H+5T$ \\
\hline
\end{tabularx}
\caption{Closed-string spectrum after the other orientifold projection.}\label{tab:Z12AB-unoriented-closed-spectrum}
\end{table}

The Klein bottle amplitude reads
\begin{align}
    \mathcal{K} =& \frac12\Big[T[^0_0]\left(\frac{
        \Lambda^{\mathcal K,(1,0)}_2\Lambda^{\mathcal K,(1,1)}_2(U,T)+
        \Lambda^{\mathcal K,(1,1)}_2\Lambda^{\mathcal K,(1,0)}_2(U,T)
        }{2}\right) + T[^0_6] \\
    &+ T[^2_0] + T[^2_6]+ T[^{10}_0] + T[^{10}_6] \notag\\
    &+ 3T[^4_0] - T[^4_6]+ 3T[^8_0] - T[^8_6] \notag\\
    &+ 4T[^6_0] + 4T[^6_6]\Big]\notag\\
    =&\frac14T[^0_0]\left(
        \Lambda^{\mathcal K,(1,0)}_2\Lambda^{\mathcal K,(1,1)}_2+
        \Lambda^{\mathcal K,(1,1)}_2\Lambda^{\mathcal K,(1,0)}_2\right)' +\xi_2[^0_0] \notag\\
    &+\xi_2[^2_0]-\xi_2[^4_0]+\xi_2[^6_0]-\xi_2[^8_0]+\xi_2[^{10}_0] \notag\\
    &+ 2\sum_{\beta=0}^2(\xi_3[^4_\beta]+\xi_4[^8_\beta])+\sum_{\beta=0}^1\xi_4[^6_{2\beta}] + 2\xi_2[^6_0] \,,\notag
\end{align}
whose tree channel reads
\begin{align}
    \widetilde{\mathcal{K}} =& \frac{2^3}{2}\Big[T[^0_0]
        \begin{aligned}[t]
        \bigl(\mathcal{C}_-(m_1,w_1)^2\widetilde{\Lambda}^{(1,0)}_{0,0}\mathcal{C}_-(m_2,w_2)^2\widetilde{\Lambda}^{(1,1)}_{0,0} +\\
        \mathcal{C}_-(m_1,w_1)^2\widetilde{\Lambda}^{(1,1)}_{0,0}\mathcal{C}_-(m_2,w_2)^2\widetilde{\Lambda}^{(1,0)}_{0,0} \bigr)'
        \end{aligned} \\
        &+6\xi_6[^0_0]+6(\xi_6[^6_1]+\xi_6[^6_5]+2\xi_6[^6_3]) \Big] \,.\notag
\end{align}
The associated unoriented closed-string spectrum is summarised in \Cref{tab:Z12AB-unoriented-closed-spectrum}.

The annulus amplitude required to cancel the tadpoles is given by the following expression
\begin{align}
   \mathcal{A} =& \frac14\Big[T[^0_0](
    N_0^2\Lambda^{\mathcal{A} (1,0)}_{m_1,w_1}\Lambda^{\mathcal{A} (1,1)}_{m_2,w_2}(U,T)+
    M_0^2\Lambda^{\mathcal{A} (1,1)}_{m_1,w_1}\Lambda^{\mathcal{A} (1,0)}_{m_2,w_2}(U,T)  ) \\
   &+ \sum_{\beta=1}^3 (N_\beta^2+M_\beta^2)T[^0_{3\beta}] \notag\\
   &+ 2\sum_{\beta=0}^3 N_\beta M_\beta(-1)^\beta(T[^2_{3\beta}]+T[^{10}_{3\beta}]) \notag\\
   &+ 3(N_0^2+M_0^2)(T[^4_{0}]+T[^8_0]) + (N_1^2+M_1^2)T[^6_3] + (N_3^2+M_3^2)T[^6_9]-(N_2^2+M_2^2)T[^6_6]\notag\\
   &+ 4(2N_0M_0 T[^6_0]+N_1M_1 T[^6_3]+2N_2M_2 T[^6_6]+N_3M_3 T[^6_9]) \Big],\notag
\end{align}
where $N_\beta$ and $M_\beta$ are the Chan-Paton trace factors under the action of the orbifold.
The tree channel of this amplitude is given by the following expression
\begin{align}
    \widetilde{\mathcal{A}} = \frac{2^{-3}}{4}&\Big[T[^0_0](
        N_0^2\widetilde{\Lambda}^{(1,0)}_{0,0}\widetilde{\Lambda}^{(1,1)}_{0,0} +
        M_0^2\widetilde{\Lambda}^{(1,1)}_{0,0}\widetilde{\Lambda}^{(1,0)}_{0,0})' \notag\\
    &+3((N_0+M_0)^2\xi_6[^0_0]+(N_0-M_0)^2\xi_6[^0_3]) \notag\\
    &+3(N_1+M_1)^2(\xi_6[^3_1]+\xi_6[^3_5])+3(N_1-M_1)^2(\xi_6[^3_2]+\xi_6[^3_4]) \notag\\
    &+(N_2+M_2)^2(\xi_6[^6_1]+3\xi_6[^6_3]+\xi_6[^6_5])+(N_2-M_2)^2(3\xi_6[^6_0]+\xi_6[^6_2]+\xi_6[^6_4]) \notag\\
    &+3(N_3+M_3)^2(\xi_6[^9_1]+\xi_6[^9_5])+3(N_3-M_3)^2(\xi_6[^9_2]+\xi_6[^9_4]) \Big]\,.
\end{align}
The tree Möbius amplitude is fixed by factorization: it is the geometric mean of the tree Klein bottle and annulus coefficients, with the appropriate hatted characters implementing the $P$ modular transformation. After transforming back to the loop channel, the resulting coefficients must be compatible with the annulus multiplicities and define a consistent orientifold projection on the open-string Hilbert space. The resulting tree channel Möbius amplitude is given by the following expression
\begin{align}
    \widetilde{\mathcal{M}}
    = -\frac12\Bigg[&
    T[^0_0]
    \begin{aligned}[t]
        \bigl(N_0 \mathcal{C}_-(m_1,w_1)\widetilde{\Lambda}^{(1,0)}_{0,0} \mathcal{C}_-(m_2,w_2)\widetilde{\Lambda}^{(1,1)}_{0,0} + \\
        M_0 \mathcal{C}_-(m_1,w_1)\widetilde{\Lambda}^{(1,1)}_{0,0} \mathcal{C}_-(m_2,w_2)\widetilde{\Lambda}^{(1,0)}_{0,0}\bigr)'
    \end{aligned}
    \\
    +&6(N_0+M_0)\widehat{\xi}_6[^0_0]+(N_2+M_2)(\widehat{\xi}_6[^6_1]+\widehat{\xi}_6[^6_3]+\widehat{\xi}_6[^6_5]) \Bigg]\,.\notag
\end{align}
The tadpole conditions are
\begin{align}
    N_0 + M_0 &= 16\,,\\
    N_2 - M_2 &= 0\,.
\end{align}
The parametrisation of the Chan-Paton traces is given by
\begin{align}
    N_\beta &= \Tr(\gamma^\beta_N)\,, \\
    M_\beta &= \Tr(\gamma^\beta_M)\,,
\end{align}
where
\begin{align}
    \gamma_N &= \operatorname{diag}\!\left(
        \Un_{n_0},
        \e{2i\pi/4}\Un_{n_1},
        \e{4i\pi/4}\Un_{n_2},
        \e{6i\pi/4}\Un_{n_3}
    \right)\,,\\
    \gamma_M &= \operatorname{diag}\!\left(
        \Un_{m_0},
        \e{2i\pi/4}\Un_{m_1},
        \e{4i\pi/4}\Un_{m_2},
        \e{6i\pi/4}\Un_{m_3}
    \right)\,.
\end{align}
The orientifold projection further constrains the Chan-Paton matrices with the following relations
\begin{equation}
    n_2= n_1,\qquad n_3= n_0,\qquad
    m_2= m_1,\qquad m_3= m_0.
\end{equation}
The solution of the tadpole condition is given by
\begin{equation}
    m_1=8-m_0-n_0-n_1.
\end{equation}

Finally, we obtain the gauge group ${U(n_0)_1\times U(n_1)_2\times U(m_0)_3\times U(m_1)_4}$ and a massless spectrum that comprises an $\mathcal{N}=(1,0)$ gauge vector and a hypermultiplet in the representation
\begin{align}
    &3(A_1\oplus A_2 \oplus A_3 \oplus A_4) \\
    \oplus
    &3(B_{1\overline{2}}\oplus B_{\overline{1}2}\oplus B_{3\overline{4}}\oplus B_{\overline{3}4})  \notag\\
    \oplus
    &3(B_{14}\oplus B_{\overline{1}\overline{4}}\oplus B_{23}\oplus B_{\overline{2}\overline{3}})  \notag\\
    \oplus
    &(\text{Adj}_1\oplus \text{Adj}_2\oplus \text{Adj}_3\oplus \text{Adj}_4)\,.
\end{align}

The anomaly polynomial is given by
\begin{align}
    I_8 = \frac38 (-\tr R^2 + \tr F_1^2 + \tr F_2^2 + \tr F_3^2 + \tr F_4^2)^2\,.
\end{align}

One can also recombine these branes into a single stack of $D7$-branes wrapping all the cycles previously wrapped by the two stacks. The resulting theory has a gauge group $U(4)$ and a massless spectrum that comprises an $\mathcal{N}=(1,0)$ gauge vector and a hypermultiplet in the representation $4\text{Adj}\oplus 3(A+\overline{A})$. The anomaly polynomial is given by
\begin{align}
    I_8 = \frac38 (-\tr R^2 + 2\tr F^2)^2\,.
\end{align}
The recombined solution can be interpreted as a Higgs branch generated by bifundamental hypermultiplets connecting the two boundary orbits. Maximal-rank expectation values identify the corresponding Chan-Paton factors and break the product gauge group to the diagonal $U(4)$. From the boundary-state viewpoint, the two fractional orbit sums are replaced by a single configuration carrying the same total R-R source data, so the tadpole conditions remain unchanged. The relative vector multiplets become massive by eating the appropriate hypermultiplet degrees of freedom, while the remaining massless fields reorganise into
\begin{equation}
    4\,\mathrm{Adj}\oplus3\left(A\oplus\overline A\right),
\end{equation}
in agreement with the anomaly factorization of this branch.

For both vacua, the massless brane sector couples only to the untwisted closed-string sector, since the coefficients of all twisted characters vanish identically.

\subsection{Brane Supersymmetry Breaking model}\label{app:BSB}

We will use the following conventions for convenience. We denote by $F_{a,b}$ the contribution of the spin structure $[^a_b]$ to a given $\Z_N$ block. The standard type IIB block can then be decomposed as
\begin{equation}
    T = T_B-T_F = F_{0,0}-F_{0,1/2}-F_{1/2,0}-F_{1/2,1/2}\,,
\end{equation}
where
\begin{equation}
    T_B = F_{0,0}-F_{0,1/2}\,, 
    \qquad
    T_F = F_{1/2,0}+F_{1/2,1/2}\,.
\end{equation}
This corresponds to the standard GSO signs $s_{a,b}=(-1)^{2a+2b+4ab}$ introduced in \Cref{eq:ZN-blocks}.

In the presence of antibranes, a brane-antibrane overlap has the same NS-NS reflection coefficient as a brane-brane overlap, while its R-R reflection coefficient has the opposite sign. It is therefore convenient to introduce the BSB block
\begin{equation}
    \widetilde T_{\rm BSB}
    \equiv T_B+T_F
    = F_{0,0}-F_{0,1/2}+F_{1/2,0}+F_{1/2,1/2}\,,
\end{equation}
corresponding to
\begin{equation}
    \widetilde s^{\rm BSB}_{a,b}=(-1)^{2b+4ab}\,.
\end{equation}
Thus $\widetilde T_{\rm BSB}$ differs from the standard block $T$ by a sign flip of the R-sector contribution.

Under the modular $S$ transformation, $\widetilde T_{\rm BSB}$ is mapped to the loop channel brane-antibrane block
\begin{equation}
    T_{\rm BSB}
    = F_{0,0}+F_{0,1/2}-F_{1/2,0}+F_{1/2,1/2}\,,
\end{equation}
with
\begin{equation}
    s^{\rm BSB}_{a,b}=(-1)^{2a+4ab}\,.
\end{equation}
The Fourier characters of the standard blocks $T$ will be denoted by $\chi$, while those of the loop channel BSB blocks $T_{\rm BSB}$ will be denoted by $\tau$. The latter implement the reversed NS GSO projection characteristic of the brane-antibrane open-string sector.

We shall also use the shorthand
\begin{equation}
    \bigl(N+(-1)^F M\bigr)^2 \chi[^0_0]
    \equiv
    (N+M)^2\chi_B[^0_0]-(N-M)^2\chi_F[^0_0]\,,
\end{equation}
where $\chi[^0_0]=\chi_B[^0_0]-\chi_F[^0_0]$.

The Klein bottle amplitude for the BSB model is given by the following expression
\begin{align}
    \mathcal{K} =& \frac13\Big[T[^0_0]\left(\frac{
        \Lambda^{\mathcal K,(1,0)}_1\Lambda^{\mathcal K,(1,1)}_2(T,U)+
        \Lambda^{\mathcal K,(1,1)}_2\Lambda^{\mathcal K,(1,0)}_1(T,U)
        }{2}\right) + \sum_{\beta=1}^2 T[^0_{4\beta}] \\
    &- 4T[^3_0] - \sum_{\beta=1}^2 T[^3_{4\beta}]
    + 8T[^6_0] -  \sum_{\beta=1}^2 T[^6_{4\beta}]
    - 4T[^9_0]  - \sum_{\beta=1}^2 T[^9_{4\beta}]\Big]\notag\\
    =&\frac16T[^0_0]\left(\Lambda^{\mathcal K,(1,0)}_1\Lambda^{\mathcal K,(1,1)}_2(T,U) + \Lambda^{\mathcal K,(1,1)}_2\Lambda^{\mathcal K,(1,0)}_1(T,U)\right)' +\xi_3[^0_0]-\xi_3[^3_0]-\xi_3[^6_0]-\xi_3[^9_0] \notag\\
    &+ \sum_{\beta=0}^3(-\xi_4[^3_\beta]+\xi_4[^6_\beta] -\xi_4[^9_\beta] ) + 2\xi_2[^6_0] +2\xi_2[^6_1] \,,\notag
\end{align}
whose tree channel reads
\begin{align}
    \widetilde{\mathcal{K}} =& \frac{2^3}{3}\Big[T[^0_0]
        \begin{aligned}[t]
        \bigl(\mathcal{C}_+(m_1,w_1)^2\widetilde{\Lambda}^{(1,0)}_{0,0}\mathcal{C}_-(m_2,w_2)^2\widetilde{\Lambda}^{(1,1)}_{0,0} +\\
        \mathcal{C}_-(m_1,w_1)^2\widetilde{\Lambda}^{(1,1)}_{0,0}\mathcal{C}_+(m_2,w_2)^2\widetilde{\Lambda}^{(1,0)}_{0,0} \bigr)'
        \end{aligned}
        \label{eq:KleintransverseBSB} \\
        &+8\xi_4[^0_2]
        +\sum_{\alpha\in\{4,8\}}(4\xi_4[^{\alpha}_0]+2\xi_4[^{\alpha}_0]+2\xi_4[^{\alpha}_1]+2\xi_4[^{\alpha}_2]+2\xi_4[^{\alpha}_3]) \Big] \,.\notag
\end{align}

\begin{align}
    \widetilde{\mathcal{K}}_0 =& \frac{2^3}{3}\Big[(4+2)\chi[^4_0]+(4+2)\chi[^8_0] \Big] \,.
    \label{eq:KleintadpoleBSB}
\end{align}
The annulus amplitude for the BSB model is given by the following expression
\begin{align}
   \mathcal{A} =& \frac16\Big[T[^0_0](
    N_0^2\Lambda^{\mathcal{A} (1,0)}_{m_1,w_1}\Lambda^{\mathcal{A} (1,1)}_{m_2,w_2}(T,U)+
    M_0^2\Lambda^{\mathcal{A} (1,1)}_{m_1,w_1}\Lambda^{\mathcal{A} (1,0)}_{m_2,w_2}(T,U)  ) \\
   &+ \sum_{\beta=1}^5 (N_\beta^2+M_\beta^2)T[^0_{2\beta}] \notag\\
   &+ 2\sum_{\beta=0}^5 N_\beta M_\beta(T_{\rm BSB}[^3_{2\beta}]+T_{\rm BSB}[^9_{2\beta}]) \notag\\
   &+ 2(N_0^2+M_0^2)T[^6_{0}] + 2(N_3^2+M_3^2)T[^6_{6}]-\sum_{\beta=1,2,4,5}(N_\beta^2+M_\beta^2)T[^6_{2\beta}]\notag\Big],
\end{align}
where $N_\beta$ and $M_\beta$ are the Chan-Paton trace factors under the action of the orbifold.
The tree channel of this amplitude is given by the following expression
\begin{align}
    \widetilde{\mathcal{A}} = \frac{2^{-3}}{3}&\Big[\frac14T[^0_0](
        N_0^2\widetilde{\Lambda}^{(1,0)}_{0,0}\widetilde{\Lambda}^{(1,1)}_{0,0} +
        M_0^2\widetilde{\Lambda}^{(1,1)}_{0,0}\widetilde{\Lambda}^{(1,0)}_{0,0})' \notag\\
    &+\frac12((N_0+(-1)^FM_0)^2\xi_4[^0_0]+(N_0-(-1)^FM_0)^2\xi_4[^0_2]) \notag\\
    &+\sum_{\beta=1}^5((N_\beta-(-1)^FM_\beta)^2\xi_4[^{2\beta}_0]+(N_\beta+(-1)^FM_\beta)^2\xi_4[^{2\beta}_2]) \notag\\
    &+\sum_{\beta\in\{2,4\}}(N_\beta^2+M_\beta^2)(\xi_4[^{2\beta}_0]+\xi_4[^{2\beta}_1]+\xi_4[^{2\beta}_2]+\xi_4[^{2\beta}_3]) \notag\\
    &+ 3(N_3^2+M_3^2)(\xi_4[^6_1]+\xi_4[^6_3]) \Big]\,.
\end{align}
Finally, the tree Möbius amplitude is fixed by factorization: it is the geometric mean of the tree Klein bottle and annulus coefficients, with the appropriate hatted characters implementing the $P$ modular transformation. After transforming back to the loop channel, the resulting coefficients must be compatible with the annulus multiplicities and define a consistent orientifold projection on the open-string Hilbert space. The resulting tree channel Möbius amplitude is given by the following expression
\begin{align}
    \widetilde{\mathcal{M}}
    = \frac13\Bigg[&
    \,T[^0_0]\hspace{-0.75cm}
    \begin{aligned}[t]
        \bigl(\,&N_0 \mathcal{C}_+(m_1,w_1)\widetilde{\Lambda}^{(1,0)}_{0,0} \mathcal{C}_-(m_2,w_2)\widetilde{\Lambda}^{(1,1)}_{0,0} + \\
        (-1)^F&M_0 \mathcal{C}_-(m_1,w_1)\widetilde{\Lambda}^{(1,1)}_{0,0} \mathcal{C}_+(m_2,w_2)\widetilde{\Lambda}^{(1,0)}_{0,0}\,\bigr)'
    \end{aligned}
    \label{eq:MobiusTBSB}\\
    &-4(N_0-(-1)^FM_0)\widehat{\xi}_4[^0_2]
    \notag\\
    &+2\sum_{\beta\in\{2,4\}}(N_\beta-(-1)^FM_\beta)
    \left(
        3\widehat{\xi}_4[^{2\beta}_0]
        -\widehat{\xi}_4[^{2\beta}_1]
        +\widehat{\xi}_4[^{2\beta}_2]
        -\widehat{\xi}_4[^{2\beta}_3]
    \right)
    \Bigg]\,.\notag
\end{align}

The anomaly vectors are
\begin{equation}
\setlength{\arraycolsep}{3pt}
\begin{array}{
r@{\;}c@{\;}l@{}
r@{,\,}r@{,\,}r@{,\,}r@{,\,}r@{,\,}r@{,\,}
r@{,\,}r@{,\,}r@{,\,}r@{,\,}r@{,\,}r
@{}l
}
a &=& \bigl(
& 0 & 0 & 0
& \tfrac{2}{\sqrt{3}} & \tfrac1{\sqrt{3}} & \tfrac1{\sqrt{3}}
& 0 & 0 & 0 & 0 & 0 & 0
& \bigr),\\[1mm]
b_1 &=& \bigl(
& \tfrac2{\sqrt{6}} & \tfrac2{\sqrt{3}} & \tfrac2{\sqrt{6}}
& -\tfrac2{\sqrt{3}} & -\tfrac2{\sqrt{3}} & 0
& 0 & 0 & 0 & 0 & 0 & 0
& \bigr),\\[1mm]
b_2 &=& \bigl(
& \tfrac2{\sqrt{6}} & -\tfrac2{\sqrt{3}} & -\tfrac2{\sqrt{6}}
& -\tfrac2{\sqrt{3}} & -\tfrac2{\sqrt{3}} & 0
& 0 & 0 & 0 & 0 & 0 & 0
& \bigr),\\[1mm]
b_3 &=& \bigl(
& \tfrac2{\sqrt{6}} & \tfrac1{\sqrt{3}} & -\tfrac2{\sqrt{6}}
& \tfrac1{\sqrt{3}} & \tfrac1{\sqrt{3}} & 0
& 0 & 0 & 0 & 0 & 0 & 0
& \bigr),\\[1mm]
b_4 &=& \bigl(
& \tfrac2{\sqrt{6}} & -\tfrac1{\sqrt{3}} & \tfrac2{\sqrt{6}}
& \tfrac1{\sqrt{3}} & \tfrac1{\sqrt{3}} & 0
& 0 & 0 & 0 & 0 & 0 & 0
& \bigr),\\[1mm]
b_5 &=& \bigl(
& -\tfrac1{\sqrt{6}} & \tfrac1{\sqrt{3}} & \tfrac1{\sqrt{6}}
& -\tfrac1{\sqrt{3}} & 0 & -\tfrac1{\sqrt{3}}
& 0 & 0 & 0 & 0 & 0 & 0
& \bigr),\\[1mm]
b_6 &=& \bigl(
& -\tfrac1{\sqrt{6}} & -\tfrac1{\sqrt{3}} & -\tfrac1{\sqrt{6}}
& -\tfrac1{\sqrt{3}} & 0 & -\tfrac1{\sqrt{3}}
& 0 & 0 & 0 & 0 & 0 & 0
& \bigr),\\[1mm]
b_7 &=& \bigl(
& -\tfrac2{\sqrt{6}} & \tfrac1{\sqrt{3}} & -\tfrac2{\sqrt{6}}
& \tfrac1{\sqrt{3}} & 0 & \tfrac1{\sqrt{3}}
& 0 & 0 & 0 & 0 & 0 & 0
& \bigr),\\[1mm]
b_8 &=& \bigl(
& -\tfrac2{\sqrt{6}} & -\tfrac1{\sqrt{3}} & \tfrac2{\sqrt{6}}
& \tfrac1{\sqrt{3}} & 0 & \tfrac1{\sqrt{3}}
& 0 & 0 & 0 & 0 & 0 & 0
& \bigr).
\end{array}
\end{equation}
We arrange the tensor fields in the following order: first the gravity self-dual tensor, then the singlet tensor from the $\Z_6$ twisted sector, then the singlet tensor from the $\Z_2$ twisted sector, then the singlet tensor from the $\Z_3$ twisted sector, then the two quadruplet tensors from the $\Z_3$ twisted sector and finally the remaining tensors.

\subsection{Five-dimensional freely acting \texorpdfstring{$\Z_{12}$}{Z12} orientifold}

We dress the six-dimensional generator $g$ by a momentum shift of order
three and a winding shift of order four on $S^1_R$,
\begin{equation}
    g\,|m,w\rangle
    =
    e^{2\pi i(m/3+w/4)}|m,w\rangle,
    \qquad
    p_{L,R}^{(\alpha)}
    =
    \frac{m+\alpha/4}{R}
    \pm
    R
    \left(w+\frac{\alpha}{3}\right).
\end{equation}
The modular-covariant circle block is
\begin{equation}
    \Lambda[^\alpha_\beta](R)
    =
    \sum_{m,w\in\Z}
    e^{2\pi i\beta(m/3+w/4+\alpha/12)}
    q^{\frac{1}4(p_L^{(\alpha)})^2}
    \overline q^{\frac{1}4(p_R^{(\alpha)})^2},
\end{equation}
with
\begin{equation}
    \Lambda[^\alpha_\beta]
    \xrightarrow{T}
    \Lambda[^\alpha_{\alpha+\beta}],
    \qquad
    \Lambda[^\alpha_\beta]
    \xrightarrow{S}
    \Lambda[^\beta_{-\alpha}].
\end{equation}

The torus amplitude is
\begin{align}
    \mathcal T
    ={}&
    \frac1{12}
    |T[^0_0]|^2
    \left(
    \Gamma_{2,2}^{(1)}
    \Gamma_{2,2}^{(2)}
    \right)'
    \Lambda[^0_0]
    \nonumber\\
    &+
    \frac1{12}
    \Bigg[
    \sum_{\alpha,\beta=0}^{11}
    T[^\alpha_\beta]\,
    \overline T[^{-5\alpha}_{-5\beta}]\,
    \Lambda[^\alpha_\beta]
    \nonumber\\
    &\hspace{5mm}
    +8\sum_{\beta=0}^{2}
    |T[^4_{4\beta}]|^2
    \Lambda[^4_{4\beta}]
    +
    8\sum_{\beta=0}^{2}
    |T[^8_{4\beta}]|^2
    \Lambda[^8_{4\beta}]
    \nonumber\\
    &\hspace{5mm}
    +3\sum_{\beta=0}^{3}
    T[^3_{3\beta}]
    \overline T[^9_{-3\beta}]
    \Lambda[^3_{3\beta}]
    +
    3\sum_{\beta=0}^{3}
    T[^9_{3\beta}]
    \overline T[^3_{-3\beta}]
    \Lambda[^9_{3\beta}]
    \nonumber\\
    &\hspace{5mm}
    +3\sum_{\beta=0}^{3}
    |T[^6_{3\beta}]|^2
    \Lambda[^6_{3\beta}]
    +
    12\sum_{\beta=0}^{1}
    |T[^6_{6\beta}]|^2
    \Lambda[^6_{6\beta}]
    \Bigg].
\end{align}

The circle contributions can be treated uniformly by defining the momentum and winding lattices
\begin{align}
    P_{\delta,\phi}(R;t)&\equiv\sum_{m\in\Z}\e{2\pi i\phi m}\exp\left[-\pi t\frac{1}{R^2}(m+\delta)^2\right],\\
    W_{\delta,\phi}(R;\ell)&\equiv\sum_{w\in\Z}\e{2\pi i\delta(w-\phi)}\exp\left[-\pi\ell R^2(w-\phi)^2\right].
\end{align}
We denote by $W_{\delta,\phi}^{[w]}(R;\ell)$ the $w$-th term of the latter sum. Poisson resummation gives the universal relation
\begin{equation}
    P_{\delta,\phi}(R;t)\longrightarrow\frac{R}{\sqrt{t}}\,W_{\delta,\phi}\left(R;\frac1t\right).
\end{equation}
The three surfaces differ only by the loop modulus entering the momentum lattice. With the conventions used below,
\begin{align}
    \mathcal K:\qquad &P_{\delta,\phi}(R;2\tau_2)\xrightarrow{S}\frac{R}{\sqrt{2\tau_2}}W_{\delta,\phi}(R;\ell),\qquad &&\ell=\frac1{2\tau_2},\\
    \mathcal A:\qquad &P_{\delta,\phi}\left(R;\frac {\tau_2}{2}\right)\xrightarrow{S}\sqrt{\frac{2}{\tau_2}}R W_{\delta,\phi}(R;\ell),\qquad &&\ell=\frac2{\tau_2},\\
    \mathcal M:\qquad &P_{\delta,\phi}(R;2\tau_2)\xrightarrow{P}\frac{R}{\sqrt{2\tau_2}}W_{\delta,\phi}(R;\ell),\qquad &&\ell=\frac1{2\tau_2}.
\end{align}

To avoid clutter, let $\mathcal L_K$, $\mathcal L_A^{N,M}$ and $\mathcal L_M^{N,M}$ denote the internal loop-channel Klein bottle, annulus and Möbius lattice sums, while a tilde denotes their tree level counterparts. Their explicit expressions are collected in \Cref{app:lattices}; all have unit zero-mode contribution.

For the Klein bottle only $\alpha=3j$, $j=0,1,2,3$, survive, with $w=-j$. The loop channel amplitude is
\begin{align}
    \mathcal K=\frac13\Big[&
    \frac12T[^0_0]\mathcal L_KP_{0,0}+T[^0_4]P_{0,1/3}+T[^0_8]P_{0,2/3}\nonumber\\
    &+4T[^3_0]P_{3/4,0}+T[^3_4]P_{3/4,1/3}+T[^3_8]P_{3/4,2/3}\nonumber\\
    &+8T[^6_0]P_{3/2,0}-T[^6_4]P_{3/2,1/3}-T[^6_8]P_{3/2,2/3}\nonumber\\
    &+4T[^9_0]P_{9/4,0}+T[^9_4]P_{9/4,1/3}+T[^9_8]P_{9/4,2/3}\Big].
\end{align}
Applying the general Poisson resummation and reindexing the winding integer when convenient gives
\begin{align}
    \widetilde{\mathcal K}=\frac{2^3}{3}\sum_{w\in\Z}\Bigg\{&
    T[^0_0](\widetilde{\mathcal L}_K W_{0,0}^{[w]})'
    +8\xi_4[^0_{-w}]W_{0,0}^{[w]}\\
    +&\left[2\sum_{\rho=0}^{3}\xi_4[^4_\rho]+4\xi_4[^4_{-w}]\right]W_{0,-2/3}^{[w]}
    +\left[2\sum_{\rho=0}^{3}\xi_4[^8_\rho]+4\xi_4[^8_{2-w}]\right]W_{0,-4/3}^{[w]}\Bigg\}.\notag
\end{align}
The $w$-dependent character labels are the tree channel manifestation of the crosscap-position phases.

The loop channel annulus amplitude is
\begin{align}
    \mathcal A=\frac13\Bigg\{&
    T[^0_0]\left(N_0^2\mathcal L_A^N+M_0^2\mathcal L_A^M\right)P_{0,0}\nonumber\\
    &+\sum_{\beta=2,4}(N_\beta^2+M_\beta^2)T[^0_{2\beta}]P_{0,\beta/6}\nonumber\\
    &+2\sum_{\beta=0,2,4}N_\beta M_\beta\left[T[^3_{2\beta}]P_{3/4,\beta/6}
    +T[^9_{2\beta}]P_{-3/4,\beta/6}\right]\nonumber\\
    &+2(N_0^2+M_0^2)T[^6_0]P_{3/2,0}
    -\sum_{\beta=2,4}(N_\beta^2+M_\beta^2)T[^6_{2\beta}]P_{3/2,\beta/6}\Bigg\}.
\end{align}
Its tree channel form is
\begin{align}
    \widetilde{\mathcal A}=\frac16\sum_{w\in\Z}\Bigg\{&
    \frac14T[^0_0]\left(\left[N_0^2\widetilde{\mathcal L}_A^N+M_0^2\widetilde{\mathcal L}_A^M\right]W_{0,0}^{[w]}\right)' \\
    +&\frac12\left[(N_0+M_0)^2\xi_4[^0_{-w}]+(N_0-M_0)^2\xi_4[^0_{2-w}]\right]W_{0,0}^{[w]} \notag\\
    +&\left[(N_2-M_2)^2\xi_4[^4_{2-w}]+(N_2+M_2)^2\xi_4[^4_{-w}]
    +(N_2^2+M_2^2)\sum_{\rho=0}^{3}\xi_4[^4_\rho]\right]W_{0,-2/3}^{[w]}\nonumber\\
    +&\left[(N_4-M_4)^2\xi_4[^8_{-w}]+(N_4+M_4)^2\xi_4[^8_{2-w}]
    +(N_4^2+M_4^2)\sum_{\rho=0}^{3}\xi_4[^8_\rho]\right]W_{0,-4/3}^{[w]}\Bigg\}.\nonumber
\end{align}
The mixed $NM$ sector therefore corresponds to the relative circle position
\begin{equation}
    x_M-x_N=\frac38\qquad\left(\bmod\,\frac12\right).
\end{equation}

The tree level Möbius amplitude is fixed by the geometric mean of the boundary and crosscap reflection coefficients. Defining
\begin{equation}
    \widehat F_\alpha[r]\equiv\widehat\xi_4[^\alpha_r]-\widehat\xi_4[^\alpha_{r+1}]
    +3\widehat\xi_4[^\alpha_{r+2}]-\widehat\xi_4[^\alpha_{r+3}],
\end{equation}
with the second label understood modulo four, one finds
\begin{align}
    \widetilde{\mathcal M}=\frac23\sum_{w\in\Z}\Bigg\{&
    -\widehat T[^0_0]\left(\left[N_0\widetilde{\mathcal L}_M^N+M_0\widetilde{\mathcal L}_M^M\right]W_{0,0}^{[w]}\right)'
    -4(N_0+M_0)\widehat\xi_4[^0_{-w}]W_{0,0}^{[w]}\nonumber\\
    &+2(N_2+M_2)\widehat F_4[2-w]W_{0,-2/3}^{[w]}
    +2(N_4+M_4)\widehat F_8[-w]W_{0,-4/3}^{[w]}\Bigg\}.
\end{align}
The compatible loop amplitude is
\begin{align}
    \mathcal M=\frac13\Bigg\{&
    \widehat T[^0_0]\left(N_0\mathcal L_M^N+M_0\mathcal L_M^M\right)P_{0,0}\\
    +&(N_0+M_0)\left[\widehat T[^0_6]P_{1/2,0}-2\widehat T[^6_3]P_{-1/4,0}-2\widehat T[^6_9]P_{1/4,0}\right]\nonumber\\
    +&(N_2+M_2)\left[\widehat T[^0_2]P_{0,1/3}+\widehat T[^6_5]P_{1/4,1/3}
    +\widehat T[^0_8]P_{1/2,1/3}+\widehat T[^6_{11}]P_{3/4,1/3}\right]\nonumber\\
    +&(N_4+M_4)\left[\widehat T[^0_{10}]P_{0,2/3}+\widehat T[^6_1]P_{1/4,2/3}
    +\widehat T[^0_4]P_{1/2,2/3}+\widehat T[^6_7]P_{3/4,2/3}\right]\Bigg\}.\nonumber
\end{align}
The internal Möbius lattices are the same boundary-cycle lattices as in the annulus, dressed by the crosscap half-shift phases of the $O7$ component intersecting the corresponding boundary. Their signs are fixed by the crosscap phases and by requiring $\frac12(\mathcal A+\mathcal M)$ to give integral symmetric or antisymmetric Chan-Paton multiplicities.

At any finite $0<R<\infty$, the shifted $g^4$- and $g^8$-sectors are massive, so the only massless R-R tadpole is untwisted
\begin{equation}
    \widetilde{\mathcal K}_0+\widetilde{\mathcal A}_0+\widetilde{\mathcal M}_0
    =\frac1{12}(N_0+M_0-16)^2\xi_4[^0_0],\qquad {N_0+M_0=16}.
\end{equation}
In the $R\to0$ limit the $g^4$- and $g^8$-twisted towers also become massless, with source coefficients
\begin{equation}
    \mathcal Q_4^2=\frac1{12}\left[(N_2-M_2)^2+3(N_2+M_2+8)^2\right],
\end{equation}
and similarly for $\mathcal Q_8^2$. The perturbative zero-twisted-flux branch is
\begin{equation}
    N_0=M_0=8,\qquad N_2=M_2=N_4=M_4=-4.
\end{equation}
This gives the gauge group $U(4)_N\times U(4)_M$.
At finite radius the $NM$ sector is massive because of the $3/8$ relative displacement, while the massless gauge fields arise from the unshifted $NN$ and $MM$ sectors.

Finally,
\begin{equation}
    M_{\rm circ}^2\sim\frac{(m+\alpha/4)^2}{R^2}+R^2\left(w+\frac{\alpha}{3}\right)^2.
\end{equation}
Thus
\begin{equation}
    R\to0:\qquad \alpha=0,4,8,\qquad \langle g^4\rangle\simeq\Z_3,
\end{equation}
and the winding tower reconstructs the sixth dimension in the type IIA frame. On the zero-flux branch the two $U(4)$ boundary families separate to infinite proper distance on the decompactified dual circle. Thus we end up with the six-dimensional dual IIA geometric $\Z_3$ orientifold with five-dimensional defects. Conversely,
\begin{equation}
    R\to\infty:\qquad \alpha=0,3,6,9,\qquad \langle g^3\rangle\simeq\Z_4,
\end{equation}
and the momentum tower reconstructs the six-dimensional antisymmetric $\Z_4$ type IIB orientifold. The boundary splittings vanish as $1/R$, and in the six-dimensional normalisation
\begin{equation}
    N_{\Z_4}=2N_0=16,\qquad M_{\Z_4}=2M_0=16.
\end{equation}
Thus in the limit we end up in the standard symplectic branch of the six-dimensional IIB antisymmetric $\Z_4$ orientifold with gauge group $[USp(8)^2]_N\times [USp(8)^2]_M$.

\section{Multiplet, anomaly and string-defect conventions}\label{sec:SUSY}

\subsection{Supermultiplets in various dimensions}

We collect here the different multiplets in various dimensions, with their field content and number of supercharges~\cite{Wigner:1939cj,Ferrara_1998,Riccioni_1998}.
\subsubsection{d=6}
\paragraph{$\mathcal{N}=(1,0)$, N=8}~\\
G = $\{g_{\mu\nu}, B_{\mu\nu}^+, \psi_L^\mu \}$ \\
V = $\{A^\mu, \lambda_L\}$ \\
T = $\{B_{\mu\nu}^-,\phi, \lambda_R \}$ \\
H = $\{4\phi, \lambda_R \}$

\paragraph{$\mathcal{N}=(1,1)$, N=16}~\\
G = $\{g_{\mu\nu}, B_{\mu\nu}, \phi, 4 A^\mu, \psi_L^\mu, \psi_R^\mu, \lambda_L,\lambda_R \}$ \\
V = $\{A^\mu, 4\phi, \lambda_L, \lambda_R \}$
 
\paragraph{$\mathcal{N}=(2,0)$, N=16}~\\
G = $\{g_{\mu\nu}, 5B_{\mu\nu}^+, 2\psi_L^\mu \}$ \\
T = $\{B_{\mu\nu}^-,5\phi,2\lambda_R \}$

\paragraph{$\mathcal{N}=(2,2)$, N=32}~\\
G = $\{g_{\mu\nu}, 5B_{\mu\nu}, 16A^\mu,25\phi,2\psi^\mu_L,2 \psi_R^\mu, 10 \lambda_L, 10 \lambda_R \}$

\subsubsection{d=5}

\paragraph{$\mathcal{N}=1$, N=8}~\\
G = $\{g_{\mu\nu},A^\mu, \psi^\mu \}$ \\
V = $\{A^\mu,\phi, \lambda\}$ \\
H = $\{4\phi, \lambda \}$

\paragraph{$\mathcal{N}=2$, N=16}~\\
G = $\{g_{\mu\nu}, 6A^\mu,\phi, 2\psi^\mu,2\lambda \}$ \\
V = $\{A^\mu,5\phi,2\lambda \}$

\paragraph{$\mathcal{N}=4$, N=32}~\\
G = $\{g_{\mu\nu}, 27A^\mu,42\phi,4\psi^\mu, 24 \lambda \}$

\subsection{Anomaly conventions}\label{app:anomaly}

\subsubsection{Six-dimensional anomalies}\label{app:6d-anomaly-conventions}

We collect here the six-dimensional anomaly conventions used in the main text~\cite{Scrucca:2004jn}. Our chirality convention is such that the contributions of a right-handed spin-$3/2$ fermion, a right-handed spin-$1/2$ fermion in a representation $R$, and a self-dual two-form are
\begin{align}
    I_8^{\Psi^\mu_R} &= \frac{245}{360} \, \tr R^4 - \frac{43}{288} (\tr R^2)^2, \\
    I_8^{\lambda_R} &= \dim(R)[\frac{1}{360} \, \tr R^4 + \frac{1}{288} (\tr R^2)^2] + \frac{1}{24} \, \Tr_{R} F^4 - \frac{1}{24} \, \tr R^2 \, \Tr_{R} F^2, \\
    I_8^{B^+} &= -\frac{28}{360} \, \tr R^4 + \frac{8}{288} (\tr R^2)^2,
\end{align}
The contributions of left-handed fermions and anti-self-dual two-forms are obtained with the opposite sign.
For the representations used in the paper, the relevant trace identities are
\begin{align}
    \Tr_{\mathrm{S/A}}F^2
    &=
    (N\pm2)\tr F^2,\\
    \Tr_{\mathrm{S/A}}F^4
    &=
    (N\pm8)\tr F^4
    +3\bigl(\tr F^2\bigr)^2,\\
    \Tr_{(n,m)}F^2
    &=
    m\,\tr_nF_n^2+n\,\tr_mF_m^2,\\
    \Tr_{(n,m)}F^4
    &=
    m\,\tr_nF_n^4+n\,\tr_mF_m^4
    +6\,\tr_nF_n^2\,\tr_mF_m^2,
\end{align}
where the upper and lower signs correspond respectively to the symmetric and antisymmetric representations. We use $\tr$ for the fundamental trace and $\Tr_R$ for the trace in the representation $R$~\cite{Erler_1994,Scrucca_2000}.

The coefficient of $\tr R^4$ vanishes when
\begin{equation}
    H-V+29T=273,
\end{equation}
where charged hypermultiplets and vector multiplets are counted with the dimensions of their gauge representations. After imposing this condition, the anomaly polynomial can be written as
\begin{equation}
    I_8
    =
    \frac{9-T}{8}\bigl(\tr R^2\bigr)^2
    -\frac{1}{24}\tr R^2\,\Tr_{\mathrm{ch}}F^2
    +\frac{1}{24}\Tr_{\mathrm{ch}}F^4,
\end{equation}
where we define the net chiral gauge traces
\begin{equation}
    \Tr_{\mathrm{ch}}F^k
    \equiv
    \sum_{\text{hypermultiplets }H}\Tr_{R_H}F^k
    -\Tr_{\mathrm{Adj}}F^k.
\end{equation}

For a theory with $T$ tensor multiplets, the tensor charge lattice has signature $(1,T)$. We work in a basis in which
\begin{equation}
    \Omega_{\alpha\beta}
    =
    \operatorname{diag}(1,-1,\ldots,-1),
    \qquad
    x\cdot y
    =
    \Omega_{\alpha\beta}x^\alpha y^\beta.
\end{equation}
The Green-Schwarz-Sagnotti mechanism requires the reducible anomaly polynomial to factorize as
\begin{equation}
    I_8
    =
    \frac12
    \Omega_{\alpha\beta}
    X_4^\alpha X_4^\beta,
\end{equation}
with
\begin{equation}
    X_4^\alpha
    =
    \frac12a^\alpha\tr R^2
    +\frac12\sum_i
    \frac{b_i^\alpha}{\lambda_i}\tr F_i^2
    +\frac12\sum_{A,B}b_{AB}^\alpha f_Af_B.
\end{equation}
For the non-Abelian part this gives
\begin{equation}
    I_8^{\mathrm{NA}}
    =
    \frac18a\cdot a\,\bigl(\tr R^2\bigr)^2
    +\frac18\sum_{i,j}
    \frac{b_i\cdot b_j}{\lambda_i\lambda_j}
    \tr F_i^2\,\tr F_j^2
    +\frac14\sum_i
    \frac{a\cdot b_i}{\lambda_i}
    \tr R^2\,\tr F_i^2.
\end{equation}
The corresponding Green-Schwarz coupling is
\begin{equation}
    S_{\mathrm{GS}}
    =
    \int
    \Omega_{\alpha\beta}
    B_2^\alpha\wedge X_4^\beta.
\end{equation}
Thus the components of $a$, $b_i$ and $b_{AB}$ specify the couplings of the individual two-forms to the gravitational and gauge characteristic classes. In the main text we choose the tensor basis according to the untwisted and twisted sectors from which the two-forms originate, so these components can be interpreted directly in terms of the orbifold sectors.

\subsubsection{Abelian anomaly conventions and the \texorpdfstring{$U(1)_X$}{U(1)X} direction}\label{app:abelian-anomalies}

For each unitary factor we decompose the $U(N_i)$ field strength as
\begin{equation}
    \mathcal F_i
    =
    F_i+f_i\,\mathbf 1_{N_i},
    \qquad
    \tr F_i=0,
\end{equation}
where the fundamental representation has diagonal $U(1)_i$ charge $+1$. The antifundamental therefore has charge $-1$, while the two-index symmetric and antisymmetric representations carry charge $+2$ and their conjugates charge $-2$.

For a representation $R$ of the non-Abelian factor with Abelian charge $q$,
\begin{align}
    \Tr_R\mathcal F^2
    &=
    \Tr_RF^2
    +q^2\dim(R)\,f^2,\\
    \Tr_R\mathcal F^4
    &=
    \Tr_RF^4
    +4qf\,\Tr_RF^3
    +6q^2f^2\,\Tr_RF^2
    +q^4\dim(R)\,f^4.
\end{align}
For a representation carrying several Abelian charges, $qf$ is replaced by $\sum_iq_if_i$. In particular, the terms proportional to $f_i\tr F_\kappa^3$ are mixed $U(1)_i-SU(N_\kappa)^3$ anomalies that must be cancelled by the generalized Green-Schwarz mechanism.

For the spectrum of the supersymmetric $\Z_{12}$ orientifold, with
\begin{equation}
    n\equiv n_1,\qquad m\equiv8-n,
\end{equation}
these terms take the form
\begin{equation}
    I_8^{U(1)-SU^3}
    =
    \frac16
    \sum_{\kappa,i}
    D_{\kappa i}\,
    f_i\,\tr F_\kappa^3,
\end{equation}
where the rows are ordered as
\begin{equation}
    SU(n)_1,\quad SU(m)_2,\quad SU(n)_3,\quad SU(m)_4,
\end{equation}
the columns correspond to $U(1)_1,\ldots,U(1)_4$, and
\begin{equation}
    D=
    \begin{pmatrix}
        3n & m   & -2n & 0\\
        n  & 3m  & 0   & -2m\\
        -2n& 0   & 3n  & m\\
        0  & -2m & n   & 3m
    \end{pmatrix}.
\end{equation}
For $0<n,m$, the matrix has rank three and
\begin{equation}
    \ker D
    =
    \operatorname{Span}
    \left\{
        (-m,n,-m,n)
    \right\}.
\end{equation}
Up to an overall normalisation, the unique Abelian combination free of mixed $U(1)-SU(N)^3$ anomalies is therefore
\begin{equation}
    U(1)_X
    =
    -m\bigl(U(1)_1+U(1)_3\bigr)
    +n\bigl(U(1)_2+U(1)_4\bigr).
\end{equation}
A field-strength background along this direction can be parametrised as
\begin{equation}
    f_1=f_3=-m f_X,
    \qquad
    f_2=f_4=n f_X.
\end{equation}

The remaining reducible Abelian anomalies then combine with the non-Abelian polynomial into
\begin{equation}
    I_8
    =
    \frac18
    \left[
        a\left(\tr R^2-8nm\,f_X^2\right)
        +\sum_i\frac{b_i}{\lambda_i}\tr F_i^2
    \right]^2.
\end{equation}
Equivalently, defining
\begin{equation}
    b_{XX}=-8nm\,a,
\end{equation}
one may write using the anomaly vectors $a$ and $b_i$ of the main text
\begin{equation}
    I_8
    =
    \frac18
    \left[
        a\,\tr R^2
        +\sum_i\frac{b_i}{\lambda_i}\tr F_i^2
        +b_{XX}f_X^2
    \right]^2.
\end{equation}

The three linearly independent Abelian directions outside $\ker D$ require generalized Green-Schwarz cancellation. In the string construction, the corresponding R-R axions transform under the Abelian gauge transformations and generate Stückelberg masses for these gauge bosons~\cite{Scrucca_2000}. Hence a massless Abelian gauge factor~\cite{Scrucca:1999zh,Antoniadis:2002cs}, if present, must be proportional to $U(1)_X$. When $n=0$ or $m=0$, restricting the analysis to the gauge factors that remain shows that no anomaly-compatible Abelian direction analogous to $U(1)_X$ survives.

\subsubsection{BPS-string inflow and KSV conventions}\label{app:KSV-conventions}

For a BPS string of charge $Q$ in the six-dimensional tensor lattice, anomaly inflow~\cite{Scrucca:1999uz} induces a two-dimensional worldsheet anomaly polynomial of the form
\begin{equation}
    I_4
    =
    \frac12Q\cdot a\,\tr R^2
    +\frac12\sum_i
    \frac{Q\cdot b_i}{\lambda_i}\tr F_i^2
    +\frac12Q^2\chi(N),
\end{equation}
where $\chi(N)$ is the Euler class of the normal bundle of the string worldsheet. The BPS charges considered in the main text obey
\begin{equation}
    Q^2\geq-1,
    \qquad
    Q^2+Q\cdot a\geq-2,
    \qquad
    Q\cdot b_i\geq0.
\end{equation}
The anomaly vector $a$ is characteristic,
\begin{equation}
    Q^2\equiv Q\cdot a\pmod2.
\end{equation}

The interacting worldsheet SCFT has central charges and non-Abelian current-algebra levels~\cite{Kim_2019,Angelantonj:2020pyr}
\begin{equation}
    c_L=3Q^2-9Q\cdot a+2,
    \qquad
    c_R=3Q^2-3Q\cdot a,
    \qquad
    k_i=Q\cdot b_i,
\end{equation}
while the level of the transverse $SU(2)_L$ current algebra is
\begin{equation}
    k_L
    =
    \frac12\left(Q^2+Q\cdot a+2\right).
\end{equation}
Unitarity requires
\begin{equation}
    \sum_i
    \frac{k_i\dim(G_i)}{k_i+h_i^\vee}
    +c_{\mathrm{KM}}^{U(1)}
    \leq c_L,
\end{equation}
where $h_i^\vee$ is the dual Coxeter number of $G_i$. Each massless Abelian current with non-zero level contributes one unit to $c_{\mathrm{KM}}^{U(1)}$~\cite{Lee:2019skh}.

\bibliographystyle{JHEP}
\bibliography{biblio}

\end{document}